\documentclass[a4paper,11pt]{article}
\pdfoutput=1 % if your are submitting a pdflatex (i.e. if you have
\usepackage{jheppub} % for details on the use of the package, please
\title{From Six to Four and More:
Massless and Massive Maximal Super Yang-Mills Amplitudes in 6d and 4d and their Hidden Symmetries 
}
\usepackage[tight]{subfigure}

\usepackage{amsmath}
\usepackage{cleveref}
\usepackage{amsfonts}
\usepackage{amssymb}
\usepackage{slashed}
\usepackage{bm}%,bbold}
\usepackage{dsfont}
\usepackage{array}
\usepackage{longtable}

\usepackage{alltt}

\graphicspath{{figures/}}

\allowdisplaybreaks

\DeclareMathAlphabet{\mathpzc}{OT1}{pzc}{m}{it}
\DeclareMathAlphabet{\EuRoman}{U}{eur}{m}{n}
\SetMathAlphabet{\EuRoman}{bold}{U}{eur}{b}{n}
\newcommand{\nn}{\nonumber}
\newcommand{\cN}{{\cal N}}
\newcommand{\sfrac}[2]{{\textstyle\frac{#1}{#2}}}

\renewcommand{\a}{\alpha}
\newcommand{\da}{\dot{\alpha}}
\renewcommand{\b}{\beta}
\newcommand{\db}{\dot{\beta}}
\newcommand{\g}{\gamma}
\newcommand{\dg}{\dot{\gamma}}
\DeclareMathOperator{\hyp}{hyp}
\DeclareMathOperator{\ferm}{ferm}

\newcommand{\LR}{\begin{gathered}[c]\underrightarrow{\qquad\vphantom{A}\quad}\\\overleftarrow{\quad \vphantom{A}\qquad}\end{gathered}}

\newcommand{\bracket}[2]{\langle #1 \vert #2 \rangle}
\newcommand{\+}{\negthinspace+\negthinspace}

\renewcommand{\=}[1][=]{\mathrel{\phantom{#1}}}

\newcommand{\Tr}{\mathop{\mathrm{Tr}}}

\newcommand{\sign}{\mathop{\mathrm{sign}}}

\newcommand{\ang}[2]{\langle #1\;#2\rangle}
\newcommand{\sqb}[2]{[ #1\;#2 ]}

 \makeatletter
 \def\mr@ignsp#1 {\ifx\:#1\@empty\else #1\expandafter\mr@ignsp\fi}%
\newcommand{\multiref}[1]{\begingroup%\let\protect\string%
 \xdef\mr@no@sparg{\expandafter\mr@ignsp#1 \: }%
 \def\mr@comma{}%
 \@for\mr@refs:=\mr@no@sparg\do{\mr@comma\def\mr@comma{,}\ref{\mr@refs}}%
 \endgroup}
 \makeatother

\usepackage{color}
\usepackage{colordvi}

\definecolor{mygreen}{rgb}{0,0.75,0}

\author[a,b]{Jan Plefka}
\author[a]{, Theodor Schuster}
\author[a,c]{and Valentin Verschinin}
\note[$\ast$]{{\it E-mail:} \{jan.plefka,theodor.schuster\}@physik.hu-berlin.de} 
\note[$\dagger$]{{\it E-mail:} valentin.verschinin@cpht.polytechnique.fr}
\preprint{HU-EP-14/17}

\affiliation[a]{Institut f\"ur Physik and IRIS Adlershof, Humboldt-Universit\"at zu Berlin,\\
Newtonstrasse 15, 12489 Berlin, Germany}
\affiliation[b]{Institut f\"ur Theoretische Physik, Eidgen\"ossische Technische Hochschule 
Z\"urich,\\ Wolfgang-Pauli-Strasse 27, 8093 Z\"urich, Switzerland}
\affiliation[c]{Centre de Physique Th\'eorique, \'Ecole Polytechnique, CNRS,\\ 91128 Palaiseau, France\\}

\abstract{
A self-consistent exposition of 
the theory of tree-level superamplitudes of the 
4d $\cN=4$ and 6d $\cN=(1,1)$ maximally supersymmetric Yang-Mills theories 
is provided. In 4d we work in non-chiral superspace and
construct the superconformal and dual superconformal symmetry generators 
of the $\cN=4$ SYM theory using the non-chiral BCFW recursion to prove the latter.
In 6d we provide a complete derivation of the standard and hidden symmetries of the tree-level superamplitudes of $\cN=(1,1)$ SYM theory, again using the BCFW recursion to prove the dual conformal symmetry. 
Furthermore, we demonstrate that compact analytical formulae for tree-superamplitudes 
in $\cN=(1,1)$ SYM can be obtained from a numerical implementation of the supersymmetric BCFW recursion relation. We derive compact manifestly dual conformal representations of the five- and six-point superamplitudes as well as arbitrary multiplicity formulae valid for 
certain classes of superamplitudes related to ultra-helicity-violating massive amplitudes
in 4d.
We study massive tree superamplitudes on the Coulomb branch of the $\mathcal{N}=4$ SYM theory
from dimensional reduction of the massless superamplitudes of the six-dimensional $\mathcal{N}=(1,1)$ SYM theory. We exploit this correspondence to construct the super-Poincar\'e and
enhanced dual conformal symmetries of massive tree superamplitudes in $\mathcal{N}=4$ SYM theory
which are shown to close into a finite dimensional algebra of Yangian type. 
Finally, we address the fascinating possibility of uplifting massless 4d superamplitudes 
to 6d massless superamplitudes proposed by Huang. We confirm the uplift for multiplicities up to eight but show that finding the uplift is  highly
non-trivial and in fact not of a practical use for multiplicities larger than five.}

\begin{document}
\maketitle
\flushbottom
\section{Introduction}\label{section:intro}
Scattering amplitudes of maximally supersymmetric Yang-Mills theories in 3, 4, 6 and 10
dimensions possess remarkable properties. Next to their constitutional maximally extended 
super-Poincar\'e symmetries they all enjoy a hidden dual conformal symmetry -- at least 
at the tree-level \cite{Lipstein:2012kd,Drummond:2008vq,Brandhuber:2008pf,Dennen:2010dh,CaronHuot:2010rj}.
The four dimensional $\cN=4$ super
Yang-Mills (SYM) theory is distinguished in this series as it also has superconformal symmetry in the standard sense. The standard superconformal symmetry then further enhances the dual conformal symmetry to a dual superconformal 
symmetry \cite{Drummond:2008vq,Brandhuber:2008pf}. 
On top the closure of the two sets of superconformal symmetry algebras leads to an infinite
dimensional symmetry algebra of Yangian type \cite{Drummond:2009fd}. It is the manifestation
of an underlying integrable structure in planar $\cN=4$ SYM. The key to the discoveries of
these rich symmetry structures of maximally supersymmetric Yang-Mills theories in various dimensions is the use of a suitable on-shell superspace formalism
along with spinor helicity variables to package the component field amplitudes into
superamplitudes, which was pioneered in 4d in \cite{Nair:1988bq}. In this work we shall focus
on the four and six dimensional maximally theories: The 4d $\cN=4$ SYM 
and the 6d $\cN=(1,1)$ SYM models.

While the massless tree amplitudes of 4d $\cN=4$ SYM are very well studied and in fact known 
analytically \cite{Drummond:2008cr}, not so much is known about the massive amplitudes on the Coulomb branch of this theory. These amplitudes are obtained by giving a vacuum 
expectation value to the
scalar fields and yield -- arguably -- the simplest massive amplitudes in four 
dimensions. Alternatively, these massive amplitudes
 arise from the amplitudes of the maximally supersymmetric
6d $\cN=(1,1)$ SYM theory upon dimensional reduction, where the higher dimensional momenta 
yield the masses in the 4d theory.  
Indeed,  compact arbitrary multiplicity amplitudes for particular subclasses of 
Coulomb branch amplitudes have been obtained in \cite{Craig:2011ws} by making use
of modern on-shell techniques.
The massive 4d $\cN=4$ SYM amplitudes 
are invariant under a dual conformal symmetry which is inherited from the
6d $\cN=(1,1)$ SYM theory as shown in \cite{Dennen:2010dh}.
Moreover, this symmetry remains
intact also at loop-level if one restricts the loop-momentum integrations 
to a four-dimensional subspace.
% - under the assumption of cut-constructability of
%the theory. 
This prescription is equivalent to the Higgs regularization for infrared divergences in 4d proposed in \cite{Alday:2009zm},
where such an extended dual conformal invariance was conjectured and tested at the one-loop
four-point level. The dimensional reduction of 6d $\cN=(1,1)$ SYM to four dimensions
yields $\cN=4$ superamplitudes expressed on a non-chiral superspace \cite{Huang:2011um} which
is distinct to the usual chiral superspace of \cite{Nair:1988bq}.
In this work we explicitly construct all generators of the standard and dual (super) 
conformal symmetry generators acting in the non-chiral $\cN=4$ on-shell superspace 
as well as in the 
$\cN=(1,1)$ on-shell superspace. We also determine the standard and dual symmetries of
massive $\cN=4$ amplitudes as they are induced from an enhanced super-Poincar\'e and
enhanced dual conformal symmetry of the 6d $\cN=(1,1)$ SYM theory.

The most efficient method to analytically construct tree-level amplitudes is
based on an on-shell recursive technique due to Britto, Cachazo, Feng and Witten 
(BCFW) \cite{Britto:2004ap,Britto:2005fq}. 
In contrast to the earlier Berends-Giele off-shell recursion relations~\cite{Berends:1987me}, 
the BCFW relation uses only on-shell lower-point amplitudes, evaluated at complex, shifted
momenta.  The BCFW recursion relation is easily generalizable to an on-shell recursion
for superamplitudes, as was done for $\cN=4$ SYM in \cite{ArkaniHamed:2009dn}
(see also \cite{Bianchi:2008pu}).  In fact the knowledge of the dual superconformal invariance
of superamplitudes motivates an ansatz in terms of dual conformal invariants. 
Together with the super
BCFW recursion this allowed for the complete analytic solution \cite{Drummond:2008cr}.
In fact the variant of the BCFW recursion for 4d $\cN=4$ SYM in non-chiral superspace has
not been written down before and we will do so in this work. The BCFW recursion for 6d
$\cN=(1,1)$ SYM theory was established in \cite{Dennen:2009vk,Bern:2010qa} and 
tree-level amplitudes of multiplicities up to five were derived. The one loop
corrections were obtained in \cite{Brandhuber:2010mm}.
In this work we point out how a numerical implementation of the BCFW recursion for $\cN=(1,1)$
SYM amplitudes in combination with a suitable set of  dual conformal invariant basis
functions may be used to derive compact five and six-point amplitudes as well as
arbitrary multiplicity amplitudes for certain subclasses related to the 4d amplitudes with
two neighboring massive legs mentioned above \cite{Craig:2011ws}. 
In fact, the method we propose is very general and could be applied to further cases
as well.

A very tempting option to obtain massive 4d  amplitudes of $\cN=4$ SYM
was introduced by Huang in \cite{Huang:2011um}. He indicated that it should be possible 
to invert the dimensional reduction of $\cN=(1,1)$ to massive $\cN=4$  by uplifting the 
massless non-chiral superamplitudes of $\mathcal{N}=4$ SYM to six-dimensional 
superamplitudes of ${\mathcal N}=(1,1)$ SYM. Non-chiral superamplitudes of $\mathcal{N}=4$ SYM
are straightforward to obtain using the non-chiral BCFW recursion, resulting in an eminent practical relevance of a potential uplift. It is indeed
very surprising that in fact the massive Coulomb branch amplitudes or equivalently the six-dimensional amplitudes might not contain any more information than the massless four-dimensional amplitudes of $\mathcal{N}=4$ SYM.

It is the aim of this paper to provide a self-consistent and detailed exposition of 
the theory of superamplitudes for 4d $\cN=4$ SYM and 6d $\cN=(1,1)$ SYM. The paper
is organized as follows.
We discuss the needed spinor helicity formalisms in section 2. Section 3 and 4 are devoted to
the on-shell superspaces of both theories and the standard and hidden symmetries of the
associated superamplitudes. In section 5 we discuss the dimensional reduction from
massless 6d to massive 4d amplitudes and establish the inherited (hidden) symmetries of the 4d
amplitudes. Section 6 exposes the on-shell BCFW recursion relations for $\cN=4$ SYM in
non-chiral superspace as well as for $\cN=(1,1)$ SYM. We also provide a proof of dual
conformal symmetry of $\cN=(1,1)$ superamplitudes thereby correcting some minor mistakes in
the literature. Finally in section 8 we analyze in detail the proposal of Huang for 
uplifting 4d massless $\cN=4$ superamplitudes in non-chiral superspace to 6d
$\cN=(1,1)$ superamplitudes and point out why this uplift is non-trivial and in fact
not of a real practical use for multiplicities larger than five. Notational details and
extended formulae are relegated to the appendices.

\section{Spinor helicity formalism}
\subsection{General remarks}
Calculating scattering amplitudes of massless particles, the spinor helicity formalism has become a powerful tool in obtaining compact expressions for tree-level and one-loop amplitudes. The basic idea is to use a set of commuting spinor variables instead of the parton momenta  $\{p_i\}$. These spinors trivialize the on-shell conditions for the momenta 
\begin{equation}
 (p_i)^2=0\,.
\end{equation}
In what follows we will briefly review the spinor helicity formalism in four and six dimensions. Additional details and conventions can be found in \cref{appendix:Spinors}.
\subsection{Four dimensions}\label{section:spinor4d}
The starting point of the spinor helicity formalism in four dimensions 
\cite{DeCausmaecker:1981bg,Berends:1981uq,Kleiss:1985yh,Xu:1986xb}, 
which we briefly review here,
is to express all momenta by $(2\times2)$ matrices
\begin{align}
p_{\alpha\dot\alpha}&= \sigma^{\mu}_{\alpha\dot\alpha}\,p_{\mu}\,,&p^{\dot\alpha\alpha}&= \bar{\sigma}^{\mu\,\dot\alpha\alpha}\,p_{\mu},&\text{or inversely}&&p_\mu=\tfrac{1}{2}p_{\alpha\dot\alpha}\bar\sigma_\mu^{\dot\alpha\alpha}=\tfrac{1}{2}p^{\dot\alpha\alpha}\sigma_{\mu\,\alpha\dot\alpha}\,,
\end{align}
where we take $\sigma^{\mu}=(\mathbf{1},\vec{\sigma})$ and $\bar{\sigma}^{\mu}=(\mathbf{1},-\vec{\sigma})$ with
$\vec{\sigma}$ being the Pauli matrices. Raising and lowering of the $\alpha$ and $\dot\alpha$ indices may be conveniently defined by left multiplication with the antisymmetric $\epsilon$ symbol for which we choose the following conventions
\begin{align}
\epsilon_{12}&=\epsilon_{\dot{1}\dot{2}}=-\epsilon^{12}=-\epsilon^{\dot{1}\dot{2}}=1\,,&
\epsilon_{\alpha\beta}\epsilon^{\beta\gamma}&=\delta_\alpha^\gamma&\epsilon_{\dot{\alpha}\dot{\beta}}\epsilon^{\dot{\beta}\dot{\gamma}}&=\delta_{\dot{\alpha}}^{\dot{\gamma}}\,.
\end{align}
Besides being related by $p_{\alpha\dot{\alpha}}=\epsilon_{\alpha\beta}\epsilon_{\dot\alpha\dot\beta}p^{\dot\beta\beta}=p_{\dot\alpha\alpha}$, these matrices satisfy $p^2=\det(p_{\alpha\dot\alpha})=\det(p^{\alpha\dot\alpha})$, $p_{\alpha\dot\alpha}p^{\dot\alpha\beta}=p^2\delta_\alpha^\beta$ and $p^{\dot\alpha\alpha}p_{\alpha\dot\beta}=p^2\delta_{\dot\alpha}^{\dot\beta}$. Hence, the matrices $p^{\dot\alpha\alpha}$ and $p_{\alpha\dot\alpha}$ have rank one for massless momenta, implying the existence of chiral spinors $\lambda_\alpha$ and anti-chiral spinors $\tilde\lambda^{\dot\alpha}$ solving the massless Weyl equations
\begin{align}
 p_{\alpha\dot\alpha}\tilde\lambda^{\dot\alpha}&=0\,,&p^{\dot\alpha\alpha}\lambda_{\alpha}&=0\,.
\end{align}
These spinors can be normalized such that
\begin{align}\label{eq:bispinor}
p_{\alpha\dot\alpha}&=\lambda_{\alpha}\, \tilde \lambda_{\dot\alpha}\, .
\end{align}
For complex momenta the spinors $\lambda$ and
$\tilde\lambda$ are independent. However, for real momenta we have the reality condition $p_{\alpha\dot\beta}^*=p_{\dot\alpha\beta}$, implying $\tilde \lambda_{\dot\alpha}=c\, \lambda^{*}_\alpha$ for some $c\in \mathds{R}$. Hence, the spinors can be normalized such that 
\begin{equation}
\lambda_{\dot\alpha}=\pm\, \lambda^{*}_\alpha\,. 
\end{equation}
An explicit representation is 
\begin{equation}
|\lambda\rangle := \lambda_\alpha = \sfrac{\sqrt{p_0+p_3}}{p_1-ip_2}\,
\begin{pmatrix} % or pmatrix or bmatrix or Bmatrix or \dots
  p_1-ip_2 \\
  p_0-p_3 \\
\end{pmatrix} \, ,\qquad
|\tilde\lambda] := \tilde\lambda^{\dot\alpha} =
\sfrac{\sqrt{p_0+p_3}}{p_1+ip_2}\,
\begin{pmatrix} % or pmatrix or bmatrix or Bmatrix or \dots
  -p_0+p_3 \\ p_1+ip_2 \\
\end{pmatrix} \, ,
\label{eq:reducedspinors}
\end{equation}
with $\lambda_{\dot\alpha}=\sign(p_0+p_3)\, \lambda^{*}_\alpha$.

Obviously, \cref{eq:bispinor} is invariant under the $SO(2)$ little group transformations
\begin{align} \label{4D_Littlegroup}
\lambda_{\alpha} \rightarrow z \lambda_{\alpha}\,, && \tilde{\lambda}_{\dot\alpha} \rightarrow z^{-1} \tilde{\lambda}_{\dot\alpha}\,&&&\text{with}&|z|=1\,.
\end{align}
Labeling the external particles by $i$, each parton momentum is invariant under its own little group transformation $\lambda_{i} \rightarrow z_i\, \lambda_{i}$. The simplest Lorentz invariant and little group covariant objects that can be built out of the chiral and anti-chiral spinors are the anti-symmetric spinor products
\begin{align}\label{spinor_kontaktion}
\ang{i}{j}&= \ang{\lambda_{i}}{ \lambda_{j}} = \lambda_{i}{^\alpha} \lambda_{j\alpha}\,, &&[i \,j] = [\tilde{\lambda}_{i}\, \tilde{\lambda}_{j}] =\tilde{\lambda}_{i\dot\alpha} \tilde{\lambda}_{j}^{\dot\alpha}
\end{align}  
The little group invariant scalar products of massless momenta are then given by a product of two spinor brackets 
\begin{equation} \label{skalarprodukt}
2 p_i p_j =p_{i\,\alpha\dot\alpha}p_j^{\dot\alpha\alpha} = \langle i\, j\rangle [j\, i]\,.
\end{equation}
The spinor helicity formalism allows for a compact treatment of polarizations.
Each external gluon carries helicity $h_{i}=\pm 1$ and a momentum specified by the spinors $\lambda_{i}$
and $\tilde\lambda_{i}$. Given
this data the associated polarization vectors are
\begin{align}
\left(\varepsilon^{+}_{i}\right)^{\dot\alpha\alpha}
 &= \sqrt{2}\frac{\tilde\lambda_{i}^{\dot\alpha}\, \mu_{i}^\alpha}{\ang{\lambda_{i}}{\mu_{i}}}\, , &
\left(\varepsilon^{-}_{i}\right)^{\dot\alpha\alpha}
 &= \sqrt{2}\frac{\tilde\mu_{i}^{\dot\alpha}\,\lambda_{i}^{\alpha}} 
{\sqb{\tilde\mu_{i}}{\tilde\lambda_{i}}}\, ,&\left(\varepsilon^{\pm}_{i}\right)^{\mu}&=\tfrac{1}{2}\sigma_{\alpha\dot\alpha}^\mu\left(\varepsilon^{\pm}_{i}\right)^{\dot\alpha\alpha}\,,
\end{align}
where $(q_i)_{\alpha\dot\alpha}=\mu_{i}^{\alpha}\tilde\mu_{i}^{\dot\alpha}$ are auxiliary light-like
momenta reflecting the freedom of on-shell gauge transformations. It is straightforward to verify that the polarization vectors fulfill
\begin{align}
 \varepsilon_i^{\pm}\cdot p_i&=0\,,&\varepsilon_i^{\pm}\cdot q_i&=0\,,&\varepsilon_i^{\pm}\cdot\varepsilon_i^{\pm}&=0\,,&\varepsilon_i^{\pm}\cdot\varepsilon_i^{\mp}&=-1\,,&(\varepsilon_i^{+})_\mu^*&=(\varepsilon_i^{-})_\mu\,,
\end{align}
as well as the completeness relation
\begin{equation}
\sum_{h=\pm}(\varepsilon^h_{i})_\mu(\varepsilon^h_{i})_\nu^*=-\eta_{\mu\nu}+\frac{p_{i\,\mu}q_{i\,\nu}+p_{i\,\nu}q_{i\,\mu}}{p_{i}\cdot q_{i}}\,.
\end{equation}
A summary of all our conventions for four dimensional spinors can be found in \cref{appendix:Spinors}.
\subsection{Six dimensions}\label{section:spinor6d}

Similar to four dimensions, the six-dimensional spinor-helicity formalism \cite{Cheung:2009dc} provides a solution to the on-shell condition $p^2=0$ for massless momenta by expressing them in terms of spinors. As a first step one uses the six-dimensional analog of the
Pauli matrices $\Sigma^\mu$ and $\widetilde \Sigma^\mu$ to represent a six-dimensional vector by an antisymmetric $4\times 4$ matrix 
\begin{align}
p_{AB}&=p_\mu\Sigma^\mu_{AB}\,,& p^{AB}&=p_\mu\widetilde\Sigma^{\mu\,AB}\,,\,&&\text{or inversely} &p^\mu&=\tfrac{1}{4}\,p_{AB}\widetilde\Sigma^{\mu\,BA}=\tfrac{1}{4}\,p^{AB}\Sigma^\mu_{BA}\,.
\end{align}
Besides being related by $p_{AB}=\tfrac{1}{2}\,\epsilon_{ABCD}\,p^{CD}$, these matrices satisfy $p_{AB}p^{BC}=\delta_A^C p^2$ and $\det (p^{AB})=\det (p_{AB})=(p^2)^2$. Hence, for massless momenta, $p_{AB}$ and $p^{AB}$ have rank 2 and therefore the chiral and anti-chiral part of the Dirac equation
\begin{align}\label{eq:Weyl6d}
p_{AB}\lambda^{B\,a}&=0\,,& p^{AB}\tilde\lambda_{B\,\dot a}&=0
\end{align}
have two independent solutions, labeled by their little group indices $a=1,2$ and $\dot a= \dot 1,\dot 2$ respectively. Raising and lowering of the $SU(2)\times SU(2)$ little group indices may be conveniently defined by contraction with the antisymmetric tensors $\epsilon_{ab}$ and $\epsilon^{\dot a \dot b}$
\begin{align}
\lambda^{A}_{\phantom{A}\,a}&=\epsilon_{ab}\lambda^{A\,b}\,,&\tilde\lambda_{A}^{\phantom{A}\,\dot a}&=\epsilon^{\dot a \dot b}\tilde\lambda_{A\,\dot b}\,.
\end{align}
The anti-symmetry of $p_{AB}$ and $p^{AB}$ together with the on-shell condition 
$p_{AB}\, p^{BC}=0$ yields the bispinor representation
\begin{align}\label{eq:bispinor6d}
p_{AB}&=\tilde\lambda_{A\,\dot a}\tilde\lambda_B^{\phantom{B}\,\dot a}\, , \qquad  p^{AB}=\lambda^{A\,a}\lambda^{B}_{\phantom{B}\,a}\, \quad \text{and}
\quad \lambda^{A\,a}\tilde\lambda_{A\,\dot b}=0\, .
\end{align}
An explicit representation of the chiral and anti-chiral spinors is given by
\begin{align}
 \lambda^{A\, a}&=\begin{pmatrix}
                  0 &\sqrt{p_0+p_{3}}\\
                  \frac{-p_5+ip_4}{\sqrt{p_0+p_{3}}} &\frac{p_1+ip_2}{\sqrt{p_0+p_{3}}}\\
		  \frac{-p1+ip_2}{\sqrt{p_0+p_{3}}} &\frac{-p_5-ip_4}{\sqrt{p_0+p_{3}}}\\
		  \sqrt{p_0+p_{3}} &0
                 \end{pmatrix}\,,&
\tilde\lambda_{A\,\dot a}&=\begin{pmatrix}     0 &\sqrt{p_0-p_{3}}\\
                  \frac{p_5+ip_4}{\sqrt{p_0-p_{3}}} &\frac{-p_1+ip_2}{\sqrt{p_0-p_{3}}}\\
		  \frac{p_1+ip_2}{\sqrt{p_0-p_{3}}} &\frac{p_5-ip_4}{\sqrt{p_0-p_{3}}}\\
		  \sqrt{p_0-p_{3}} &0
                 \end{pmatrix}\,.
\end{align}
As a consequence of the properties of the six-dimensional Pauli matrices, the spinors are subject to the constraint
\begin{equation}\label{eq:6dspinorConstraint}
 \lambda^{A\, a}\lambda^{B}_{\,a}=\tfrac{1}{2}\epsilon^{ABCD}\tilde\lambda_{C\, \dot a}\tilde\lambda_{D}^{\dot a}\,.
\end{equation}
It is convenient to introduce the bra-ket notation
\begin{align}
 \lambda_i^a&=|p_i^a\rangle=|i^a\rangle\,,&\tilde\lambda_{i\,\dot a}&=|p_{i\,\dot a}]=|i_{\dot a}]
\end{align}
By fully contracting all $SU(4)$ Lorentz indices it is possible to construct little group covariant and Lorentz invariant objects. The simplest Lorentz invariants are the products of chiral and anti-chiral spinors 
\begin{align}
 \langle i^a|j_{\dot a}]=[j_{\dot a} |i^a\rangle=\lambda_i^{A\,a}\tilde\lambda_{j\,A\,\dot a}
\end{align}
These little group covariant spinor products are related to the little group invariant scalar products by
\begin{equation}\label{eq:invariants}
 2p_i\cdot p_j=\tfrac{1}{2}p_i^{AB}p_{j\,BA}=\det\left(\langle i|j]\right)\,.
\end{equation}
The spinor products are $2\times 2$ matrices whose inverse is
\begin{equation}
( \langle i^a|j_{\dot b}])^{-1}=-\frac{ [j^{\dot b}|i_a\rangle}{2p_i\cdot p_j}
\end{equation}
Each set of four linear independent spinors labeled by $i$, $j$, $k$, $l$ can be contracted with the antisymmetric tensor, to give the Lorentz invariant four brackets
\begin{align}
\langle i^a j^b k^c l^d \rangle&=\epsilon_{ABCD} \lambda_i^{A\,a}\lambda_j^{B\,b}\lambda_k^{C\,c}\lambda_l^{D\,d}=\det(\lambda_i^a\lambda_j^b\lambda_k^c\lambda_l^d)\,,\\
[i_{\dot a} j_{\dot b}k_{\dot c}l_{\dot d} ]&=\epsilon^{ABCD} \tilde\lambda_{i\,A\,\dot a}\tilde\lambda_{j\,B\,\dot b}\tilde\lambda_{k\,C\,\dot c}\tilde\lambda_{l\,D\,\dot d}=
\det(\tilde\lambda_{i\,\dot a}\tilde\lambda_{j\,\dot b}\tilde\lambda_{k\,\dot c}\tilde\lambda_{l\,\dot d})\,.
\end{align}
Note that in the above expressions the 4x4 matrix appearing in the determinants is 
defined  through its four columns vectors $\{\lambda_i^a\lambda_j^b\lambda_k^c\lambda_l^d\}$
and similarly for the second expression.

The four brackets are related to the spinor products by
\begin{equation}
 \langle I_1 I_2 I_3 I_4 \rangle[J_1 J_2 J_3 J_4]=\det(\langle I_i|J_j])\,,
\end{equation}
where $I_k=(i_k)^{a_k}$, $J_k=(j_k)_{\dot a_k}$ are multi indices labeling the spinors.
Finally, it is convenient to define the following Lorentz invariant objects
\begin{align}
\langle i^a|k_1 k_2 \cdots k_{2m+1}|j^b\rangle&=\lambda_i^{A_1\,a}(k_1)_{A_1 A_2}(k_2)^{A_2 A_3}\dots(k_{2m+1})_{A_{2m+1} A_{2m+2}}\lambda_j^{A_{2m+2}\,b}\,,\\
\langle i^a|k_1 k_2 \cdots k_{2m}|j_{\dot b}]&=\lambda_i^{A_1\,a}(k_1)_{A_1 A_2}(k_2)^{A_2 A_3}\dots(k_{2m})^{A_{2m} A_{2m+1}}\tilde\lambda_{j\,A_{2m+1}\,\dot b}\,,\\
[ i_{\dot a}|k_1 k_2 \cdots k_{2m+1}|j_{\dot b}]&=\tilde\lambda_{i\,A_1\,\dot a}(k_1)^{A_1 A_2}(k_2)_{A_2 A_3}\dots(k_{2m+1})^{A_{2m+1} A_{2m+2}}\tilde\lambda_{j\,A_{2m+2}\,\dot b}\,.
\end{align}

Similar to the four dimensional case, the polarization vectors of the gluons can be expressed in terms of spinors by introducing some light-like reference momentum $q$ with $q\cdot p\neq 0$, where $p$ denotes the gluon momentum. The four polarization states are labeled by $SO(4)\simeq SU(2)\times SU(2)$ little group indices and can be defined as
\begin{equation}
 \varepsilon_{a \dot a}^\mu=\frac{1}{\sqrt{2}}\langle p_a|\Sigma^\mu|q_b\rangle(\langle q_b|p^{\dot a}])^{-1}=\frac{1}{\sqrt{2}}[p_{\dot a}|\widetilde\Sigma^\mu|q_{\dot b}](\langle p^a|q_{\dot b}])^{-1}\,.
\end{equation}
It is straightforward to verify the properties
\begin{align}
\varepsilon_{a \dot a}\cdot p&=0\,,&\varepsilon_{a \dot p}\cdot q&=0\,,&
\varepsilon_{a \dot a}\cdot \varepsilon_{b \dot b} &=-\epsilon_{a b}\epsilon_{\dot a \dot b}\,,
\end{align}
as well as the completeness relation
\begin{equation}
  \varepsilon_{a \dot a}^\mu \varepsilon^{\nu\,a \dot a}=-\eta^{\mu\nu}+\frac{p^\mu q^\nu +p^\nu q^\mu}{p\cdot q}\,.
\end{equation}

\section{\texorpdfstring{Four-dimensional $\mathcal{N}=4$}{N=4} SYM theory}\label{section:superamps4d}

\subsection{On-shell superspaces and superamplitudes}

Dealing with scattering amplitudes of supersymmetric gauge theories is most conveniently done using appropriate on-shell superspaces. Most common for treating $\cN=4$ super Yang-Mills theory are \cite{Nair:1988bq,Witten:2003nn,Georgiou:2004by}
\begin{align}
\text{chiral superspace:}&\quad\{\lambda_i,\tilde\lambda_i,\eta_i^A\}\,, &&&\text{anti-chiral superspace:}&\quad\{\lambda_i,\tilde\lambda_i,\tilde\eta_{i\,A}\}\,.
\end{align}
The Grassmann variables $\eta_i^A$, $\tilde{\eta}_{i A}$ transform in the fundamental,  anti-fundamental representation of $SU(4)$ and can be assigned the helicities
\begin{align}\label{eq:helicities}
h_i \eta_i^A& = \tfrac{1}{2} \eta_i^A \,,&&&  h_i \tilde{\eta}_{i A} &= - \tfrac{1}{2} \tilde{\eta}_{i A}\,,
\end{align}
with $h_i$ denoting the helicity operator acting on leg $i$.
With their help it is possible to decode the sixteen on-shell states
\begin{align}
\mbox{gluons:}&\; G_{\pm}& \mbox{scalars:}&\; \phi_{A B} = \tfrac{1}{2} \epsilon_{ABCD} \phi^{CD}& \mbox{gluinos:}& \; \psi_{A}&  \mbox{anti-gluinos:} &\; \overline{\psi}^{A}
\end{align}
into a chiral or an anti-chiral superfield $\varPhi\left(\eta\right)$, $\overline{\varPhi}\left(\tilde{\eta}\right)$, defined by
\begin{align}\label{eq:superfield_N=4}
\varPhi\left(\eta\right) &= G_{+} + \eta^A  \psi_{A} + \frac{1}{2!} \eta^A \eta^B  \phi_{A B} + \frac{1}{3!} \eta^A \eta^B \eta^C \epsilon_{ABCD} \overline{\psi}^{D}  + \frac{1}{4!} \eta^A \eta^B \eta^C \eta^D \epsilon_{ABCD} G_{-}\,,\\
\overline{\varPhi}\left(\tilde{\eta}\right)& = G_{-} + \tilde{\eta}_A  \overline{\psi}^{A} - \frac{1}{2!} \tilde{\eta}_A \tilde{\eta}_B  \phi^{A B} + \frac{1}{3!} \tilde{\eta}_A \tilde{\eta}_B \tilde{\eta}_C \epsilon^{ABCD} \psi_D  + \frac{1}{4!} \tilde{\eta}_A \tilde{\eta}_B \tilde{\eta}_C \tilde{\eta}_D \epsilon^{ABCD} G_{+}\,.
\end{align}
As a consequence of \cref{eq:helicities} the super fields carry the helicities
\begin{align}
h_i \varPhi_i\left(\eta\right) &= \varPhi_i\left(\eta\right)\,,&&& h_i \overline{\varPhi}_i\left(\tilde{\eta}\right)& = - \overline{\varPhi}_i\left(\tilde{\eta}\right)\,.
\end{align}
The chiral and anti-chiral superfield are related by a Grassmann Fourier transformation
\begin{align}\label{eq:fourier}
\overline{\varPhi}\left(\tilde{\eta}\right) &= \int d^4\eta  \,e^{\eta^A \tilde{\eta}_A} \,\varPhi\left(\eta\right)\,, &&& \varPhi\left(\eta\right) &= \int d^4 \tilde{\eta}\, e^{-\eta^A \tilde{\eta}_A} \,\overline{\varPhi}\left(\tilde{\eta}\right)\,.
\end{align}
Chiral and anti-chiral color ordered superamplitudes $\mathcal{A}_n$ can be defined as functions of the respective superfields
\begin{align}
\mathcal{A}_n&=\mathcal{A}_n (\Phi_1, \Phi_2,\dots,\Phi_n)\,,&&&\overline{\mathcal{A}}_n&=\overline{\mathcal{A}}_n (\overline{\Phi}_1, \overline{\Phi}_2,\dots,\overline{\Phi}_n)\,.
\end{align}
Due to \cref{eq:fourier} both superamplitudes are related by a Grassmann Fourier transformation
\begin{equation}\label{eq:voll_ft}
\mathcal{A}_n (\Phi_1, \Phi_2,\dots,\Phi_n) = \prod_i \int d_i^4 \tilde{\eta}\; e^{-\sum_j \eta_j^A \tilde{\eta}_{jA} } \;\overline{\mathcal{A}}_n (\overline{\Phi}_1, \overline{\Phi}_2,\dots,\overline{\Phi}_n)
\end{equation}
The superamplitudes are inhomogeneous polynomials in the Grassmann odd variables $\eta_i^A$, $\tilde{\eta}_{i\,A}$, whose coefficients are given by the color ordered component amplitudes. A particular component amplitude can be extracted by projecting upon the relevant term
in the $\eta_{i}$ expansion of the super-amplitude via
\begin{align}
G^{+}_{i} &\to \eta^{A}_{i}=0\, ,&
G^{-}_{i} &\to \int d^{4}\eta_{i}\,,&\phi_{i\,AB}&\to \int d\eta^B_{i}d\eta^A_{i}\,,\\
\psi_{i,A} &\to \int d\eta^{A}_i\, , &\bar{\psi}^{A}_{i} 
&\to -\int d^{4}\eta_{i} \, \eta_{i}^{A}\, ,&
\end{align}
and similar in anti-chiral superspace. By construction the chiral  and anti-chiral superamplitudes have a manifest $SU(4)_R$ symmetry. The only $SU(4)_R$ invariants are contractions with the epsilon tensor
\begin{align}\label{eq:RsymmetryInvariants}
\eta_{i}^A\eta_{j}^B\eta_{k}^C\eta_{l}^D\epsilon_{ABCD}\,,&&\text{or}&&\tilde\eta_{i\,A}\tilde\eta_{j\,B}\tilde\eta_{k\,C}\tilde\eta_{l\,D}\epsilon^{ABCD}\,.
\end{align}
Consequently the appearing powers of the Grassmann variables within the superamplitudes need to be multiples of four. As a consequence of supersymmetry the superamplitudes are proportional to the supermomentum conserving delta function
\begin{align}
\delta^{(8)}(q^{\alpha A}):=\prod_{\alpha=1}^2\prod_{A=1}^4q^{\alpha\,A}&&\text{or}&&\delta^{(8)}( \tilde{q}^{\dot{\alpha}}_{A}):=\prod_{\dot\alpha=1}^2\prod_{A=1}^4\tilde{q}^{\dot\alpha}_A\,,
\end{align}
with the chiral $q^{\alpha A}=\sum_i\lambda_{i}^{\a}\eta_i^A$ or anti-chiral conserved supermomentum $\tilde{q}^{\dot{\alpha}}_{A}=\sum_i\tilde{\lambda}_i^{\da}\tilde{\eta}_{i\,A}$.
Since the Grassmann variables carry helicity, \cref{eq:helicities}, their powers keep track of the amount of helicity violation present in the component amplitudes.  Hence, decomposing the superamplitudes into homogeneous polynomials is equivalent to categorizing the component amplitudes according to their degree of helicity violation
\begin{align}\label{eq:MHV_decomposition}
\mathcal{A}_n (\Phi_1, \Phi_2,\dots,\Phi_n) &= \mathcal{A}^{\text{MHV}}_n + \mathcal{A}^{\text{NMHV}}_n + \mathcal{A}^{\text{N}^2\text{MHV}}_n + \dots + \mathcal{A}^{N^{(n-4)}\text{MHV}}_n\,,\\
\overline{\mathcal{A}}_n (\overline{\Phi}_1, \overline{\Phi}_2,\dots,\overline{\Phi}_n) &= \overline{\mathcal{A}}^{\overline{\text{MHV}}}_n + \overline{\mathcal{A}}^{N\overline{\text{MHV}}}_n + \overline{\mathcal{A}}^{N^2\overline{\text{MHV}}}_n + \dots + \overline{\mathcal{A}}^{N^{(n-4)}\overline{\text{MHV}}}_n\,,\end{align}
with
\begin{align}
\mathcal{A}^{\text{N}^p\text{MHV}}_n&=\mathcal{O}(\eta^{4(p+2)})\,,&&&\overline{\mathcal{A}}^{N^p\overline{\text{MHV}}}_n&=\mathcal{O}(\tilde{\eta}^{4(p+2)})\,. 
\end{align}
The highest amount of helicity violation is present in the maximally helicity violating (MHV) superamplitude or in the $\overline{\text{MHV}}$ superamplitude in anti-chiral superspace. In general, $\mathcal{A}^{\text{N}^p\text{MHV}}_n$ and $\overline{\mathcal{A}}^{N^p\overline{\text{MHV}}}_n$ are the (Next to)${}^p$ MHV and the (Next to)${}^p$ $\overline{\text{MHV}}$ superamplitudes . The complexity of the amplitudes is increasing with the degree $p$ of helicity violation, the simplest being the 
MHV superamplitude in chiral superspace \cite{Nair:1988bq}
\begin{equation}\label{MHV_super}
\mathcal{A}^{\text{MHV}}_n =i \frac{\delta^{(4)}(\sum_i p^{\alpha \dot{\alpha}}_i) \delta^{(8)}(\sum_i q^{\alpha A}_i)}{\left<1 2\right> \left<2 3\right> \dots \left<n 1\right>} \,,
\end{equation}
and the $\overline{\text{MHV}}$ superamplitude in anti-chiral superspace
\begin{equation}\label{anti_MHV_super}
\mathcal{A}^{\overline{\text{MHV}}}_n = i(-1)^n\frac{\delta^{(4)}(\sum_i p^{\alpha \dot{\alpha}}_i) \delta^{(8)}(\sum_i \tilde{q}^{\dot{\alpha}}_{iA})}{\left[1 2\right] \left[2 3\right] \dots \left[n 1\right]}\,,
\end{equation}
which are supersymmetric versions of the well known Parke-Taylor formula \cite{Parke:1986gb}. The increasingly complicated formulae for the amplitudes $\mathcal{A}^{\text{N}^p\text{MHV}}_n$ have been obtained in reference \cite{Drummond:2008cr}. Plugging the MHV decomposition, \cref{eq:MHV_decomposition}, into \cref{eq:voll_ft} we obtain the relation
\begin{equation}\label{eq:MHV_MHVbar}
\mathcal{A}_n ^{\text{N}^p\text{MHV}} = \prod_i \int d_i^4 \tilde{\eta} \;e^{-\sum_j \eta_j^A \tilde{\eta}_{jA} } \overline{\mathcal{A}}_n^{N^{n-4-p}\overline{\text{MHV}}}\,,
\end{equation}
simply stating that $\mathcal{A}_n ^{\text{N}^p\text{MHV}}$ and $\overline{\mathcal{A}}_n^{N^{n-4-p}\overline{\text{MHV}}}$ contain the same component amplitudes. Depending on whether $p<n-4-p$ or $p>n-4-p$ it is therefore more convenient to use the chiral or the anti-chiral description of the amplitudes, e.\,g.~the $\text{N}^{n-4}\text{MHV}=\overline{\text{MHV}}$ amplitudes are complicated in chiral superspace whereas they are trivial in anti-chiral superspace. Hence the most complicated amplitudes appearing in an $n$ point chiral or anti-chiral superamplitude are the helicity amplitudes of degree $p=\lfloor\tfrac{n}{2}\rfloor-2$, called minimal helicity violating (minHV) amplitudes .
\subsection{Non-chiral superspace}
\label{ncsuperspace4d}
Besides the well studied chiral and anti-chiral superspaces there is as well the non-chiral superspace 
\begin{equation}
 \{\lambda_i,\tilde\lambda_i,\eta_i^m,\tilde\eta_{i\,m'}\}\,,
\end{equation}
which is more natural from the perspective of the massive amplitudes and the six dimensional
parent theory that we are interested in.
Here the $SU(4)$ indices of the fields get split into two $SU(2)$ indices $m$ and $m'$ according to
\begin{align}
 \psi_A&=\{\psi_m,\psi_{m'}\}\,,&&&\bar\psi^A&=\{\bar\psi^m,\bar\psi^{m'}\}\,,&&&\phi_{AB}&=\{\phi_{mn},\phi_{m'n},\phi_{mn'},\phi_{m'n'}\}\,.
\end{align}
Note that the due antisymmetry the fields $\phi_{mn}=-\phi_{nm}$ and $\phi_{m'n'}=-\phi_{n'm'}$ represent only one scalar field respectively, whereas the $\phi_{mn'}=-\phi_{n'm}$
account for the four remaining scalars. If raising and lowering of the $SU(2)$ indices are defined by left multiplication with $\epsilon=i\sigma_2$ and $\epsilon^{-1}$, the non-chiral superfield reads
\begin{multline}\label{eq:nonChiralSuperfield}
\varUpsilon= \tfrac{1}{2} \phi^{m'}_{\phantom{m}m'} + \eta^{m} \overline{\psi}_{m} +  \tilde{\eta}_{m'}\psi^{m'} +\eta^{m} \tilde{\eta}_{m'}  \phi^{\phantom{m}m'}_{m} + \eta^2 G_{-} + \tilde{\eta}^2 G_{+} \\+ \eta^2 \tilde{\eta}_{m'} \overline{\psi}^{m'} + \tilde{\eta}^2 \eta^{m} \psi_{m} + \tfrac{1}{2} \tilde{\eta}^2 \eta^2 \phi^{m}_{\;\;m}\,,
\end{multline}
with the abbreviations $\eta^2=\tfrac{1}{2}\eta^m\eta_m$, $\tilde\eta^2=\tfrac{1}{2}\tilde\eta_{m'}\tilde\eta^{m'}$. The non-chiral superfield is a scalar and has zero helicity. Obviously, the non-chiral superamplitudes will not have a $SU(4)_R$ symmetry, but rather will be invariant under $SU(2,2)$ transformations.    
With the convention $m\in\{1,4\}$,  $m'\in\{2,3\}$ the non-chiral superfield is related to the chiral and anti-chiral superfield by the half Grassmann Fourier transformations 
\begin{align}\label{eq:relationOfsuperfields}
 \varUpsilon = \int d\eta^{3} d\eta^{2} \;e^{ \eta^{2}\tilde{\eta}_{2} +  \eta^{3}\tilde{\eta}_{3}} \varPhi=\int d\tilde\eta_{1} d\tilde\eta_{4} \;e^{ -\eta^{1}\tilde{\eta}_{1} -  \eta^{4}\tilde{\eta}_{4}} \overline{\varPhi}\,.
\end{align}
As a consequence of supersymmetry, the superamplitudes are proportional to the supermomentum conserving delta functions
\begin{align}
\delta^{(4)}(q^{\alpha m}):=\prod_{\alpha=1}^2\prod_{m=1}^2q^{\alpha\,m}&&\text{and}&&\delta^{(4)}( \tilde{q}^{\dot{\alpha}}_{m'}):=\prod_{\dot\alpha=1}^2\prod_{m'=1}^2\tilde{q}^{\dot\alpha}_{m'}\,,
\end{align}
with the conserved supermomenta  $q_{\alpha}^m=\sum_i\eta^m_i\lambda_{i\,\alpha}$ and $\tilde q_{\dot\alpha}^{m'}=\sum_i\tilde\eta^{m'}_i\tilde\lambda_{i\,\dot\alpha}$. Since we additionally have $h_i \varUpsilon_i=0$, the non-chiral superamplitudes have the general form
\begin{equation}
\label{325}
 \mathcal{A}_n(\varUpsilon_1,\dots,\varUpsilon_n)=\delta^{4}(\sum_i q_{i\,\alpha}^m)\delta^{4}(\sum_i \tilde q_{i\,\dot\alpha}^{m'})f_n(\{p_i,q_i,\tilde q_i\})\,.
\end{equation}
It should be stressed that the dependence of $f_{n}$ only on the momenta 
$\{p_i,q_i,\tilde q_i\}$ is distinct to the situation for the chiral or anti-chiral superamplitudes, where
we have a dependence on the super-spinors $\{\lambda_i,\tilde\lambda_i,\eta_i^A\}$ or
$\{\lambda_i,\tilde\lambda_i,\tilde\eta_i^A\}$.
 Analyzing the half Fourier transform \eqref{eq:relationOfsuperfields} relating the superfields we see that the non-chiral superamplitudes are homogeneous polynomials in the variables $q_i$ and $\tilde{q}_i$ of degree $2n$ and the MHV decomposition \eqref{eq:MHV_decomposition} of the chiral superamplitudes translates to a MHV decomposition of the non-chiral superamplitudes
\begin{equation}
 f_n=f_n^{\text{MHV}}+f_n^{\text{NMHV}}+\dots+f_n^{\overline{\text{MHV}}}\,,
\end{equation}
where the N${}^p$MHV sector corresponds to a fixed degree in the variables $q_i$ and $\tilde{q}_i$ 
\begin{equation}
 f_n^{\text{N${}^p$MHV}}=\mathcal{O}(q^{2p}\tilde{q}^{2n-8-2p})\,.
\end{equation}
This reflects the chiral nature of $\cN=4$ SYM theory.

Each of the three superspaces presented above has an associated dual superspace. In general, dual superspaces naturally arise when studying dual conformal properties of color ordered scattering amplitudes. Part of the spinor variables get replaced by
 the region momenta $x_i$, which are related to the ordinary momenta of the external legs by
\begin{equation}\label{eq:regions}
x_i-x_{i+1}=p_i
\end{equation}
and a new set of dual fermionic variables $\theta_i$ or $\tilde\theta_i$ is introduced, related to the fermionic momenta by
\begin{align}\label{eq:theta}
\theta_i-\theta_{i+1}&=q_i\,,&\tilde\theta_i-\tilde\theta_{i+1}&=\tilde{q}_i\,.
\end{align}
Obviously, the amplitudes will depend on differences of dual variables $x_{ij}=x_i-x_j$,  $\theta_{ij}=\theta_i-\theta_{j}$ and $\tilde\theta_{ij}=\tilde\theta_i-\tilde\theta_{i+1}$, as the dual variables are only defined up to an overall shift. With the identifications $x_1=x_{n+1}$, $\theta_1=\theta_{n+1}$, and $\tilde\theta_1=\tilde\theta_{n+1}$, the dual variables trivialize the momentum and supermomentum conservation.
The dual chiral superspace is given by 
\begin{equation}\label{eq:dual_chiral}
\{\lambda_i^\alpha,x_i^{\dot{\alpha}\alpha},\theta_i^{A\,\alpha}\}
\end{equation}
with the constraints
\begin{align}\label{eq:constraints_dual_chiral}
x_{i\,i+1}^{\dot{\alpha}\alpha}\lambda_{i\,\alpha}&=0\,,&\theta_{i\,i+1}^{A\,\alpha}\lambda_{i\,\alpha}&=0\,.
\end{align}
Analogously, the dual anti-chiral superspace is given by 
\begin{equation}\label{eq:dual_antichiral}
\{\tilde\lambda_i^{\dot\alpha},x_i^{\dot{\alpha}\alpha},\tilde\theta_{i\,A}^{\dot\alpha}\}
\end{equation}
with the constraints
\begin{align}
(x_{i\,i+1})_{\alpha\dot{\alpha}}\tilde\lambda_{i}^{\dot\alpha}&=0\,,&(\tilde\theta_{i\,i+1})_A^{\dot\alpha}\tilde\lambda_{i\,\dot\alpha}&=0\,.
\end{align}
In the case of the dual non-chiral superspace it is possible to completely eliminate all spinor variables and express the superamplitudes solely with  the dual variables 
\begin{equation}\label{eq:dual_nonchiral}
\{x_i^{\dot{\alpha}\alpha},\theta_i^{m\,\alpha},\tilde\theta_i^{m'\,\dot\alpha},y_i^{nn'}\}
\end{equation}
which are subject to the constraints
\begin{align}\label{eq:constraints_dualnonchiral}
x_{i\,i+1}^{\dot{\alpha}\alpha}\theta_{i\,i+1\,\alpha}^m&=0\,,&(x_{i\,i+1})_{\alpha\dot{\alpha}}\tilde\theta_{i\,i+1}^{m'\,\dot\alpha}&=0\,,&x_{i\,i+1}^{\dot{\alpha}\alpha}y_{i\,i+1}^{mm'}&=\theta_{i\,i+1}^{m\,\alpha}\tilde\theta_{i\,i+1}^{m'\,\dot\alpha}\,.
\end{align}
Note that $x_{i\,i+1}^2=0$ is a consequence of \cref{eq:constraints_dualnonchiral}. In fact the Grassmann even dual variables $y_i^{mm'}$ are not independent as they can be expressed by $\{x_i^{\dot{\alpha}\alpha},\theta_i^{m\,\alpha},\tilde\theta_i^{m'\,\dot\alpha}\}$. Hence, the amplitudes will not depend on them. However, the variables $y_i^{mm'}$ are necessary for the construction of the dual non-chiral superconformal symmetry algebra presented in \cref{section:symmetries_N=4,sec:Algebra_Non_Chiral} .

A further possibility is to study superamplitudes using the full superspaces obtained by adding the dual variables to the chiral, anti-chiral and non-chiral superspaces.
The full chiral superspace is given by 
\begin{equation}\label{eq:full_chiral}
\{\lambda_i^\alpha,\tilde\lambda_i^{\dot\alpha},x_i^{\dot{\alpha}\alpha},\eta_i^A,\theta_i^{A\,\alpha}\}
\end{equation}
with the constraints
\begin{align}\label{eq:constraints_full_chiral}
x_{i\,i+1}^{\dot{\alpha}\alpha}&=\lambda_{i}^{\alpha}\tilde\lambda_i^{\dot\alpha}\,,&\theta_{i\,i+1}^{A\,\alpha}=\lambda_i^{\alpha}\eta_i^A\,.
\end{align}
Analogously, the full anti-chiral superspace has the variables
\begin{equation}
\{\lambda_i^\alpha,\tilde\lambda_i^{\dot\alpha},x_i^{\dot{\alpha}\alpha},\tilde\eta_{i\,A},\tilde\theta_{i\,A}^{\dot\alpha}\}
\end{equation}
subject to the constraints
\begin{align}
x_{i\,i+1}^{\dot{\alpha}\alpha}&=\lambda_{i}^\alpha\tilde\lambda_i^{\dot\alpha}\,,&(\tilde\theta_{i\,i+1})_A^{\dot\alpha}=\tilde\lambda_i^{\dot\alpha}\tilde\eta_{i\,A}\,.
\end{align}
Finally, the full non-chiral superspace is given by
\begin{equation}
\{\lambda_i^\alpha,\tilde\lambda_i^{\dot\alpha},x_i^{\dot{\alpha}\alpha},\eta_{i}^m,\tilde\eta_{i}^{m'},\theta_i^{m\,\alpha},\tilde\theta_i^{m'\,\dot\alpha},y_i^{nn'}\}
\end{equation}
with the constraints
\begin{align}\label{eq:constraints_full_nonchiral}
x_{i\,i+1}^{\dot{\alpha}\alpha}&=\lambda_{i}^\alpha\tilde\lambda_i^{\dot\alpha}\,,&\theta_{i\,i+1}^{m\,\alpha}&=\lambda_i^\alpha\eta_{i}^m\,,&\tilde\theta_{i\,i+1}^{m'\,\dot\alpha}&=\tilde\lambda_{i}^{\dot\alpha}\tilde\eta_{i}^{m'}\,&y_{i\,i+1}^{nn'}=\eta_i^m\tilde\eta_i^{m'}\,.
\end{align}
\subsection{Symmetries of non-chiral superamplitudes}\label{section:symmetries_N=4}
We are going to give a complete derivation of the symmetry generators of the non-chiral superamplitudes at tree level, which has not yet been done in full detail in the literature. Part of the results presented here can be found in reference \cite{Huang:2011um}. For recent
textbook treatments of the superconformal and dual superconformal symmetry of the chiral superamplitudes see \cite{Henn:2014yza,Elvang:2013cua}. A detailed presentation of the non-chiral superconformal algebra and its relevant representations is given in \cref{sec:Algebra_Non_Chiral}.
\subsubsection{Superconformal symmetry of non-chiral superamplitudes}
Due to the half Fourier transformation connecting the non-chiral and the chiral superspace, the $SU(4)_R$ symmetry is turned into an $SU(2,2)_R$ symmetry. The conformal symmetry does not involve Grassmann variables, hence the tree-level non-chiral superamplitudes are invariant under the
conformal algebra $su(2,2)$, with generators
\begin{equation}
\{p^{\dot{\alpha} \alpha},m_{\alpha \beta},\overline{m}_{\dot{\alpha} \dot{\beta}},d,k_{\alpha\dot{\alpha}}\}\,.
\end{equation}
As a consequence of the supersymmetry of the chiral and anti-chiral superamplitudes and \cref{eq:relationOfsuperfields} relating the superfields, the non-chiral superamplitudes are invariant under the $(2,2)$-supersymmetry generators
\begin{align}
q^{\alpha n} &= \sum_{i}  \lambda^{\alpha}_{i} \eta^{n}_{i}\,, &\tilde{q}^{\dot{\alpha} n'} &= \sum_{i}  \tilde{\lambda}^{\dot{\alpha}}_{i} \tilde{\eta}_{i}^{n'} 
\end{align}
and their conjugates
\begin{align}
\overline{q}^{\dot{\alpha}}_{n} &= \sum_{i}  \tilde{\lambda}^{\dot{\alpha}}_{i} \partial_{i n} \,, &\overline{\tilde{q}}^{\alpha}_{n'} &= \sum_{i}  \lambda^{\alpha}_{i} \partial_{i n'} \,,&\qquad \mbox{with} \qquad \partial_{i n} &= \frac{\partial}{\partial \eta^{n}_{i}}\,,&\partial_{i n'} &= \frac{\partial}{\partial \tilde{\eta}^{n'}_{i}}\,.
\end{align}
All other symmetry generators now follow from the non-chiral superconformal symmetry algebra listed in \cref{sec:Algebra_Non_Chiral}.
Commuting the supersymmetry generators $q^{\alpha n}$, $\tilde{q}^{\dot{\alpha} n'}$, $\overline{q}^{\dot{\alpha}}_{n}$, $\overline{\tilde{q}}$ with the conformal boost generator $k_{\alpha \dot{\alpha}}$ yields the superconformal generators
\begin{equation}
\begin{aligned}
\begin{aligned}
s_{\alpha n} &= \sum_{i} \partial_{i \alpha} \partial_{i n}\,,& \qquad \overline{s}^{n}_{\dot{\alpha}} &= \sum_{i}  \eta^{n}_{i} \partial_{i \dot{\alpha}} \,,\\
\tilde{s}_{\dot{\alpha} n'} &=\sum_{i}  \partial_{i n'} \partial_{i \dot{\alpha}} \,,& \qquad \overline{\tilde{s}}^{n'}_{\alpha} &=\sum_{i} \tilde{\eta}_{i}^{n'} \partial_{i \alpha} \,,
\end{aligned}&&\text{with}&  &\partial_{i \a} &= \frac{\partial}{\partial \lambda^{\a}_i}\,,&\partial_{i \da} &= \frac{\partial}{\partial \tilde\lambda^{\da}_i}\,,
\end{aligned}
\end{equation}
The central charge $c$ and the hypercharge $b$ are given by:
\begin{equation}
\begin{aligned}
c &=  \tfrac{1}{2} \sum_{i}\left(-\lambda^{\alpha}_{i} \partial_{i \alpha} + \tilde{\lambda}^{\dot{\alpha}}_{i} \partial_{i \dot{\alpha}} + \eta^{n}_{i} \partial_{i n} - \tilde{\eta}^{n'}_{i} \partial_{i n'}\right)\,,& \qquad b &=  \tfrac{1}{2} \sum_{i}\left(\eta^{n}_{i} \partial_{i n} - \tilde{\eta}^{n'}_{i} \partial_{i n'}\right) \,.
\end{aligned} 
\end{equation}
As already stated at the beginning, the non-chiral superamplitudes have a $su(2,2)_R$ symmetry.  Up to the constant in the R-dilatation $\mathpzc{d}$ and some sign ambiguities, its generators $\{\mathpzc{p}^{n n'}$, $\mathpzc{m}_{\,n m}$, $\widetilde{\mathpzc{m}}_{\,n' m'}$, $\mathpzc{d}$, $\mathpzc{k}_{\,\,n n'}\}$ are related to the conformal generators $\{p^{\dot{\alpha} \alpha}$, $m_{\alpha \beta}$, $\overline{m}_{\dot{\alpha} \dot{\beta}}$, $\,d$, $k_{\alpha\dot{\alpha}}\}$ by the replacements  $\lambda\leftrightarrow\eta$ and $\tilde{\lambda}\leftrightarrow\tilde{\eta}$
\begin{equation}
\begin{gathered}
\begin{aligned}
\mathpzc{p}^{n n'} &= \sum_{i} \eta_{i}^{n} \tilde{\eta}_{i}^{n'}\,,& \qquad \mathpzc{k}_{\,\,n n'} &= \sum_{i} \partial_{i n}  \partial_{i n'}\,,\\
\mathpzc{m}_{\,n m}& = \sum_{i} \eta_{i(n} \partial_{i m)} \,,&\qquad \widetilde{\mathpzc{m}}_{\,n' m'} &= \sum_{i} \tilde{\eta}_{i (n'} \partial_{i m')}\,,\\
\end{aligned} \\
\mathpzc{d} = \tfrac{1}{2} \sum_{i} \left(\eta^{n}_{i} \partial_{i n} + \tilde{\eta}^{n'}_{i} \partial_{i n'} - 2\right)\,.
\end{gathered}
\end{equation} 
Whereas the  generators $\mathpzc{m}_{\,n m}$, $\widetilde{\mathpzc{m}}_{\,n' m'}$ and $\mathpzc{d}$ are obvious symmetries of the non-chiral superamplitudes, invariance under $\mathpzc{p}^{n n'}$ and $\mathpzc{k}_{\,\,n n'} $ is unexpected.
\subsubsection{Dual superconformal symmetry of non-chiral superamplitudes}\label{dual-shell-nonchiral}
By analogy to the chiral superamplitudes we expect the non-chiral superamplitudes to have a dual superconformal symmetry as well. Starting point is the dual non-chiral superspace $\{x_i^{\dot{\alpha}\alpha},\theta_i^{m\,\alpha},\tilde\theta_i^{m'\,\dot\alpha},y_i^{mm'}\}$, introduced in \cref{eq:dual_nonchiral}, and the invariance of the non-chiral superamplitudes under the dual super Poincar\'e symmetry
\begin{equation}
\{P_{\alpha\dot\alpha},M_{\alpha\beta},\overline{M}_{\dot\alpha\dot\beta},Q_{\alpha m},\overline{Q}_{\dot\alpha m},\widetilde{Q}_{\dot\alpha m'},\overline{\widetilde{Q}}_{\alpha m'}\}\,
\end{equation}
where
\begin{equation}
\begin{aligned}  
M_{\alpha \beta} &= \sum_{i} \left( \theta^{n}_{i (\alpha} \partial_{i \beta) n} + x_{i (\alpha}^{\dot{\alpha}} \partial_{i \beta) \dot{\alpha}} \right)\,,\\
\overline{M}_{\dot{\alpha} \dot{\beta}} &= \sum_{i} \left( \tilde{\theta}^{n'}_{i (\dot{\alpha}} \partial_{i \dot{\beta}) n'} + x_{i (\dot{\alpha}}^{\alpha} \partial_{i \dot{\beta})\,, \alpha} \right)
\end{aligned} 
\end{equation}
are just the ordinary Lorentz generators $m_{\alpha \beta}$, $\overline{m}_{\dot{\alpha} \dot{\beta}}$ acting in dual non-chiral superspace and we used the abbreviations $\partial_{i \alpha \dot{\alpha}} = \frac{\partial}{\partial x_{i}^{\dot{\alpha}\alpha }}=\tfrac{1}{2}\sigma^{\mu}_{\alpha\dot\alpha}\frac{\partial}{\partial x_i^\mu}$, $\partial_{i \alpha n} = \frac{\partial}{\partial \theta^{\alpha n}_{i}}$, $\partial_{i \dot{\alpha} n'} = \frac{\partial}{\partial \tilde{\theta}^{\dot{\alpha} n'}_{i}}$.
The dual momentum $P_{\alpha \dot{\alpha}}$ and the dual supermomenta $Q_{\alpha m}$, $\widetilde{Q}_{\dot\alpha m'}$ are the generators of translations with respect to the dual variables $x$ and $\theta$, $\tilde\theta$ 
\begin{equation}
\begin{aligned} 
P_{\alpha \dot{\alpha}} &= \sum_{i} \partial_{i \alpha \dot{\alpha}}\,,&\qquad Q_{\alpha n} &= -\sum_{i} \partial_{i \alpha n}\,,& \qquad \widetilde{Q}_{\dot{\alpha} n'} &=- \sum_{i} \partial_{i \dot{\alpha} n'}\,,
\end{aligned} 
\end{equation}
The trivial translation invariance in the dual $y$ variable leads to the dual R-symmetry generator 
\begin{align}
\mathpzc{P}_{m m'} &= -\sum_{i} \partial_{i m m'}&&\text{with}&\partial_{i m m'}=\frac{\partial}{\partial y_i^{mm'}}\,.
\end{align}
The conjugate dual supermomenta $\overline{Q}^{n}_{\dot{\alpha}}$, $\overline{\widetilde{Q}}^{n'}_{\alpha}$ are given by the action of the superconformal generators $\overline{s}^{n}_{\dot{\alpha}}$, $\overline{\widetilde{s}}^{n'}_{\alpha}$ in dual non-chiral superspace. Hence, we have
\begin{equation}
\begin{aligned}  
\overline{Q}^{n}_{\dot{\alpha}} &= \sum_{i} (\theta^{\alpha n}_{i} \partial_{i \alpha \dot{\alpha}} +  y_{i}^{n n'} \partial_{i n' \dot{\alpha}})\,,& \qquad \overline{\widetilde{Q}}^{n'}_{\alpha}& = \sum_{i} (\tilde{\theta}^{\dot{\alpha} n'}_{i} \partial_{i \alpha \dot{\alpha}}  - y_{i}^{n n'} \partial_{i n \alpha}) \,.
\end{aligned} 
\end{equation} 
Similar to the chiral case, the non-chiral dual superconformal symmetry can be obtained by adding the discrete transformation of dual conformal inversion $I$ to the super Poincar\'e group. The conformal generator $K_{\alpha \dot{\alpha}}$ and the superconformal generators $S_{\alpha m}$, $\widetilde{S}_{\dot\alpha m'}$, $\overline{S}_{\dot\alpha m}$, $\overline{\widetilde{S}}_{\alpha m}$ are then given by
\begin{equation}\label{eq:superconformalGenerators_NC}
\begin{gathered}
 K_{\alpha\dot{\beta}}=I P_{\beta\dot{\alpha}} I\\
\begin{aligned}
S_{\alpha m}&=I \overline{Q}_{\dot{\alpha} m} I\,,&\qquad\overline{S}_{\dot\alpha m}&=I Q_{\alpha m} I\,,\\
\widetilde{S}_{\dot\alpha m'}&=I \overline{\widetilde{Q}}_{\alpha m'} I\,,&\qquad\overline{\widetilde{S}}_{\alpha m'}&=I \widetilde{Q}_{\dot\alpha m'} I\,,
 \end{aligned}
 \end{gathered}
\end{equation}
and their commutators and anti-commutators immediately follow from the dual super Poincar\'e algebra and the fact that the inversion is an involution, i.\,e.~$I^2=\mathds{1}$. As we are going to show in \cref{section:BCFWnonChiral}, using the BCFW recursion, the tree-level  non-chiral superamplitudes transform covariantly under inversions
\begin{equation}\label{eq:Inversion_Amp_NC}
I\left[\mathcal{A}_n\right]=x_1^2x_2^2\dots x_n^2 \,\mathcal{A}_n
\end{equation}
if the coordinates of full non-chiral superspace invert as,
\begin{equation}\label{eq:inversion4dNC}
\begin{aligned}
I \left[ x_i^{\dot\alpha\beta} \right]&=-(x_i^{-1})^{\dot\beta\alpha}\,,&I \left[ y_i^{mm'} \right]& = y_i^{mm'} -\langle\theta_i^m|x_i^{-1}|\tilde\theta_i^{m'}]\,, \\
I [ \theta^{\alpha m}_{i} ] &= (x^{-1}_{i})^{\dot{\alpha}\beta} \theta_{i\,\beta}^{m}\,,&I [ \tilde{\theta}^{\dot{\alpha}m'}_{i} ] &= \tilde{\theta}_{i\,\dot{\beta}}^{m'}(x^{-1}_{i})^{\dot\beta\alpha} \,,\\
I \left[ \lambda^{\alpha}_{i} \right] &= (x^{-1}_{i})^{\dot{\alpha}\beta} \lambda_{i\,\beta} \,,& I [ \tilde{\lambda}^{\dot{\alpha}}_{i} ] &= \tilde{\lambda}_{i\,\dot{\beta}}(x^{-1}_{i + 1})^{\dot\beta\alpha} \,, \\
I [ \eta^{m}_{i} ] &=\frac{x_i^2}{x_{i+1}^2}\left(\eta_i^m-\langle\theta_i^m|x_i^{-1}|\tilde\lambda_i]\right)\,,&I [ \tilde\eta^{m'}_{i} ] &=\tilde\eta_i^{m'}-[\tilde\theta_i^{m'}|x_i^{-1}|\lambda_i]\,.\end{aligned}
\end{equation}
The inversion rules of the Levi-Civita tensors
\begin{align}\label{eq:inversion4depsilon}
I[\epsilon_{\alpha\beta}]&=\epsilon_{\dot\alpha\dot\beta}\,,& I[\epsilon_{\dot\alpha\dot\beta}]&=\epsilon_{\alpha\beta}
\end{align}
can be deduced from $I^2[\lambda^{\alpha}_{i}]=\lambda_{i}^\alpha$, and $I^2 [ \tilde{\lambda}^{\dot{\alpha}}_{i} ]= \tilde{\lambda}^{\dot{\alpha}}_{i} $ since the inversion is an involution.
Note that the inversion defined in \cref{eq:inversion4dNC} is compatible with the constraints \cref{eq:constraints_full_nonchiral} in full non-chiral superspace.
The simplest purely bosonic dual conformal covariants are
\begin{align}
I[\,x_{ij}^2\,]&=\frac{x_{ij}^2}{x_{i}^2x_{j}^2}\,,&I[\,\ang{i}{i+1}\,]&=\frac{\ang{i}{i+1}}{x_i^2}\,,&I[\,[i\,i+1]\,]&=\frac{[i\,i+1]}{x_{i+2}^2}\,.
\end{align}
With the help of the inversion rules \eqref{eq:inversion4dNC} and its definition \eqref{eq:superconformalGenerators_NC}, the action of the dual conformal boost generator in dual non-chiral superspace can be calculated by applying the chain rule, 
\begin{align}\label{eq:calculationOfK}
K_{\a\db}&=\sum_i\sum_j\bigg[I\biggl[\frac{\partial I[x_j^{\gamma\dot\delta}]}{\partial x_{i}^{\da\b}}\biggr]\partial_{j\,\gamma\dot{\delta}}+I\biggl[\frac{\partial I[y_j^{mm'}]}{\partial x_{i}^{\da\b}}\biggr]\partial_{j\,mm'}\notag\\
&\=\phantom{\sum_i\sum_j\bigg[}{}+I\biggl[\frac{\partial I[\theta_j^{\gamma\,m}]}{\partial x_{i}^{\da\b}}\biggr]\partial_{j\,\gamma\,m} + I\biggl[\frac{\partial I[\tilde\theta_j^{\dot\gamma\,m'}]}{\partial x_{i}^{\da\b}}\biggr]\partial_{j\,\dot\gamma\,m'}\biggr]\,.
\end{align}
Applying the Schouten identity \eqref{Schouten_4D_1} we obtain
\begin{equation}
\frac{\partial x_{i\delta\dot{\gamma}}^{-1}}{\partial x_{i}^{\da\b}}=\frac{\epsilon_{\b\delta}\epsilon_{\da\dot{\gamma}}}{x_i^2}-\frac{x_{i\,\delta\dot{\gamma}}x_{i\,\b\da}}{x_i^4}{=}-x_{i\b\dot{\gamma}}^{-1}x_{i\delta\da}^{-1}\,,
\end{equation}
immediately leading to e.\,g.
\begin{equation}
I\biggl[\frac{\partial I[x_j^{\gamma\dot\delta}]}{\partial x_{i}^{\da\b}}\biggr]=\delta_{ij}\,x_{i\a}^{\;\;\;\;\dot{\gamma}}\, x_{i\db}^{\;\;\;\;\delta}\,.
\end{equation}
The final result is
\begin{equation}
\begin{gathered}  
K_{\alpha \dot{\alpha}} = \sum_{i}  \left( x_{i \alpha}^{\;\;\; \dot{\beta}} x_{i \dot{\alpha}}^{\;\;\; \beta} \partial_{i \beta \dot{\beta}} + x_{i \dot{\alpha}}^{\;\;\;\beta} \theta_{i \alpha}^{m} \partial_{i m \beta} + x_{i \alpha}^{\;\;\; \dot{\beta}} \tilde{\theta}_{i \dot{\alpha}}^{n'} \partial_{i n' \dot{\beta}} +\theta_{i \alpha}^{n} \tilde{\theta}_{i\dot{\alpha}}^{n'} \partial_{i n n'}\right)\,.
\end{gathered} 
\end{equation}
Note that it would be equally straightforward to obtain the action of $K_{\alpha \dot{\alpha}}$ in full non-chiral superspace from \cref{eq:superconformalGenerators_NC,eq:inversion4dNC}. All other generators of the dual non-chiral superconformal symmetry now follow from the algebra listed in \cref{eq:NCsuperconformal} of appendix \cref{sec:Algebra_Non_Chiral}. Similar to the chiral case, part of the generators of the dual non-chiral superconformal algebra are directly given by the action of chiral generators in dual non-chiral superspace
\begin{equation}
\begin{gathered}
\begin{aligned}
m_{\alpha\beta}&=M_{\alpha\beta}\,,&\overline{m}_{\dot\alpha\dot\beta}&=\overline{M}_{\dot\alpha\dot\beta}\,,\\
\mathpzc{m}_{\,n m}&=\mathpzc{M}_{\,n m}\,,&\widetilde{\mathpzc{m}}_{\,n' m'}&=\widetilde{\mathpzc{M}}_{\,n' m'}\,,\\
\overline{s}^{n}_{\dot{\alpha}}&=\overline{Q}^{n}_{\dot{\alpha}}\,, &\overline{\widetilde{s}}^{n'}_{\alpha}&=\overline{\widetilde{Q}}^{n'}_{\alpha}\,,\\
\overline{q}^{n}_{\dot{\alpha}}&=-\overline{S}^{n}_{\dot{\alpha}}\,, &\overline{\widetilde{q}}^{n'}_{\alpha}&=-\overline{\widetilde{S}}^{n'}_{\alpha}\,,
\end{aligned}\\
\begin{aligned}
d&=-D+n\,,&\mathpzc{d}&=-\mathpzc{D}-n\,,&b=-B\,.
\end{aligned}
\end{gathered}
\end{equation}
Non trivial are the dual superconformal generators
\begin{equation}
\begin{aligned}  
S^{\alpha n} &= \sum_{i}  \left(- \theta_{i}^{\alpha m} \theta_{i}^{\beta n} \partial_{i \beta m} + x_{i}^{\alpha \dot{\beta}} \theta_{i}^{\beta n} \partial_{i \beta \dot{\beta}} - \theta_{i }^{\alpha m} y_{i}^{n m'} \partial_{im m'} + y_{i}^{n m'} x_{i}^{\alpha \dot{\alpha}} \partial_{i\dot{\alpha} m' }\right)\,,\\
\widetilde{S}^{\dot{\alpha} n'} &= \sum_{i}  \left(-\tilde{\theta}_{i}^{\dot{\alpha} m'} \tilde{\theta}_{i}^{\dot{\beta} n'} \partial_{i \dot{\beta} m'} + x_{i}^{\dot{\alpha} \beta} \tilde{\theta}_{i}^{\dot{\beta} n'} \partial_{i \beta \dot{\beta}}  -\tilde{\theta}_{i}^{ \dot{\alpha} m'} y_{i}^{m n'} \partial_{i m m'} - y_{i}^{m n'} x_{i}^{\dot{\alpha} \alpha} \partial_{i \alpha m}\right)\,.
\end{aligned} 
\end{equation}
and the dual R-symmetry boost generator
\begin{equation}
\mathpzc{K}^{n n'} = \sum_{i} \left(- y_{i}^{n m'} y_{i}^{m n'} \partial_{i m m'} - \tilde{\theta}^{\dot{\alpha} n'}_{i} y_{i}^{n m'} \partial_{i \dot{\alpha} m'} - \theta^{\alpha n}_{i} y_{i}^{m n'} \partial_{i \alpha m}  +\theta_{i}^{n \alpha} \tilde{\theta}^{n' \dot{\alpha}} \partial_{\alpha \dot{\alpha}} \right)\,,
\end{equation}
Due to the covariance of the non-chiral superamplitudes under dual conformal inversions, \cref{eq:Inversion_Amp_NC}, some of the generators only act covariantly on the amplitude. From \cref{eq:superconformalGenerators_NC,eq:Inversion_Amp_NC} and the algebra \cref{eq:NCsuperconformal} it follows
\begin{align}
K^{\dot\alpha\alpha}\mathcal{A}_n&=-\sum_i x_i^{\dot\alpha\alpha}\mathcal{A}_n\,,&\mathpzc{K}^{m m'}\mathcal{A}_n&=-\sum_i y_i^{mm'}\mathcal{A}_n\,,\\
S^{\alpha m}\mathcal{A}_n&=-\sum_i \theta_i^{\alpha m}\mathcal{A}_n\,,&\widetilde{S}^{\dot\alpha m'}\mathcal{A}_n&=-\sum_i \tilde\theta_i^{\dot\alpha m'}\mathcal{A}_n\,,\\
D\mathcal{A}_n&=n\,\mathcal{A}_n\,,&\mathpzc{D}\mathcal{A}_n&=-n\,\mathcal{A}_n\,.
\end{align}
For a complete list of the non-chiral superconformal algebra and its dual representation we refer to \cref{sec:Algebra_Non_Chiral}.
\subsubsection{Yangian symmetry of superamplitudes}\label{section:Yangian}
The conventional and dual superconformal algebras present at tree level
close into an infinite dimensional symmetry algebra known as the Yangian $Y[\text{psu}(2,2|4)]$ as was shown for the chiral and anti-chiral
superamplitudes in \cite{Drummond:2009fd}.
This symmetry algebra is a loop-algebra with a positive integer
level structure, whose level zero generators $J_a^{[0]}=\sum_i J_{a\,i}^{[0]}$ with local densities $ J_{a\,i}^{[0]}$ are given by the original superconformal generators 
\begin{equation}
[J_a^{[0]},J_b^{[0]}\}=f_{ab}^{\phantom{ab}c}\,J_c^{[0]}\,,
\end{equation}
where $[\cdot,\cdot\}$ denotes the graded commutator and $f_{ab}^{\phantom{ab}c}$ are the structure constants of the superconformal algebra. Invariance under the level one Yangian generators $J_a^{[1]}$ with the bi-local representation
\begin{equation}\label{eq:level1}
J_a^{[1]}=f_a^{\phantom{a}cb}\sum_{i<j} J_{b\,i}^{[0]} J_{c\,j}^{[0]}
\end{equation}
then follows from the covariance under the non-trivial dual superconformal generators $K_{\a\da}$, $S_{A}^{\a}$.
The level one generators obey the commutation relations
\begin{equation}\label{eq:commutatorsLevel1}
[J_a^{[1]},J_b^{[0]}\}=f_{ab}^{\phantom{ab}c}\,J_c^{[1]}
\end{equation}
as well as the Serre relation, for details we refer to \cite{Drummond:2009fd}.

Similar to the chiral superamplitudes the non-chiral superamplitudes have a Yangian symmetry as well, which has been investigated in \cite{Huang:2011um}. The infinite dimensional Yangian symmetry of the tree-level  superamplitudes is a manifestation of the expected integrability of the planar sector of $\cN=4$ SYM. In principle it should be possible to exploit the algebraic constraints,  that the Yangian invariance puts on the amplitudes, to determine the amplitudes efficiently. The fact that the Yangian symmetry is obscured by the manifest local and unitary Lagrangian formulation of $\cN=4$ SYM theory led to the development of alternative formulations \cite{ArkaniHamed:2009vw,ArkaniHamed:2012nw,Arkani-Hamed:2013jha}, that enjoy a manifest Yangian symmetry but lack manifest locality and manifest unitarity.
\section[Six-Dimensional \texorpdfstring{$\cN=(1,1)$}{N=(1,1)} SYM theory]{Six-Dimensional \texorpdfstring{$\bm{\cN=(1,1)}$}{N=(1,1)} SYM Theory}
\subsection{On-shell superspace and superamplitudes}
In this section we introduce the maximal supersymmetric $\mathcal{N} = (1,1)$ SYM theory in six dimensions based on references 
\cite{Dennen:2009vk, Bern:2010qa, Brandhuber:2010mm, Dennen:2010dh, Huang:2010rn, Huang:2011um, Elvang:2011fx}. The $\mathcal{N} = (1,1)$ SYM theory can be obtained by dimensionally reducing the $\mathcal{N} = 1$ SYM theory in ten dimensions and the dimensional reduction of $\mathcal{N} = (1,1)$ SYM to four dimensions is given by $\mathcal{N} = 4$ SYM theory. Hence, without presenting its Lagrangian we can immediately write down its on-shell degrees of freedom:
\begin{align}
\mbox{gluons:}&\quad g^{a}_{\;\;\dot{a}} &  \mbox{scalars:}&\quad s, s', s'', s''' &  \mbox{gluinos:}&\quad \chi^{a}, \lambda^{a}& \mbox{anti-gluinos:}&\quad \tilde{\chi}^{\dot{a}}, \tilde{\lambda}_{\dot{a}}
\end{align} 
The amplitudes of $\mathcal{N} = (1,1)$ SYM theory are most conveniently studied using the six dimensional spinor helicity formalism introduced in \cref{section:spinor6d} and the non-chiral  on-shell superspace introduced in \cite{Dennen:2009vk}
\begin{equation}
\{\, 
\lambda^{A\, a}_{i}\, , \, \tilde\lambda_{i\, A\, \dot a}\, ,\, 
\xi_{i\, a}\, ,\,  \tilde\xi^{\dot a}_{i}\, \}\, ,
\end{equation}
whose Grassmann variables $\xi_{a}$, $\tilde\xi^{\dot a}$ carry little group indices and can be used to
encode all the on-shell degrees of freedom into the scalar superfield
\begin{multline}\label{eq:superfield6d}
\Omega=s+\chi^a\,\xi_a+s'\,\xi^2+\tilde\chi_{\dot a}\,\tilde\xi^{\dot a} +g^a_{\phantom{a}\dot b}\,\xi_a\tilde\xi^{\dot b}+\tilde\lambda_{\dot b} \, \tilde\xi^{\dot b}\xi^2+s''\,\tilde\xi^2+\lambda^a\,\xi_a \tilde\xi^2+s'''\, \xi^2\tilde\xi^2\,,
\end{multline}
with the abbreviations $\tilde\xi^2=\tfrac{1}{2} \tilde\xi_{\dot a}\tilde\xi^{\dot a}$, $\xi^2=\tfrac{1}{2}\xi^a\xi_a$. Superamplitudes can now be defined as functions of the superfields
\begin{equation}
\begin{gathered}
\mathcal{A}_n=\mathcal{A}_n(\Omega_1, \Omega_2,\dots,\Omega_n)\,.
\end{gathered}
\end{equation}   
By construction these superamplitudes are invariant under the $SU(2)\times SU(2)$ little group but, as explained in \cite{Dennen:2009vk}, do not have the $SU(2)_{R}\times SU(2)_{R}$ symmetry of $\mathcal{N} = (1,1)$ SYM theory. As a consequence of the missing $R$-symmetry, the superamplitudes can not be decomposed according to the degree of helicity
 violation as in four dimensions \eqref{eq:MHV_decomposition}. 

The non-chiral superamplitudes are homogeneous polynomials of degree $n+n$ in the Grassmann variables
\begin{equation}
\mathcal{A}_n(\{\, 
\lambda^{A\, a}_{i}\, , \, \tilde\lambda_{i\, A\, \dot a}\, ,\, 
\alpha\xi_{i\, a}\, ,\,  \tilde\alpha\tilde\xi^{\dot a}_{i}\, \}) = \alpha^n \tilde\alpha^n\mathcal{A}_n(\{\, 
\lambda^{A\, a}_{i}\, , \, \tilde\lambda_{i\, A\, \dot a}\, ,\, 
\xi_{i\, a}\, ,\,  \tilde\xi^{\dot a}_{i}\, \})
\end{equation}
The tree-level superamplitudes of $\mathcal{N} = (1,1)$ are known only up to five external legs \cite{Dennen:2009vk}. 

We now review the known amplitudes starting with $n=3$.	 The special three point kinematics require the introduction \cite{Cheung:2009dc} of the bosonic spinor variables $u_i^{a}$, $w_i^{a}$, $\tilde{u}_{i \dot{a}}$ and $\tilde{w}_{i \dot{a}}$, defined in appendix \cref{appendix:threePoint}. With the definition
\begin{align}
{\bf u}_i &= u_i^{a} \xi_{i a}\,,& \qquad \tilde{{\bf u}}_i &= \tilde{u}_{i \dot{a}} \tilde{\xi}_i^{\dot{a}}\,,& \qquad {\bf w}_i &= w_i^{a} \xi_{i a}\,,& \qquad \tilde{{\bf w}}_i &= \tilde{w}_{i \dot{a}} \tilde{\xi}_i^{\dot{a}}
\end{align}
the three point amplitude reads \cite{Cheung:2009dc}
\begin{equation}\label{eq:A3_6D}
 \mathcal{A}_3 = -i \delta^{6}( \sum_{i} p^{AB} ) \left({\bf u}_1 {\bf u}_2 + {\bf u}_2 {\bf u}_3  + {\bf u}_3 {\bf u}_1 \right) \left( \sum_{i = 1}^3 {\bf w}_i \right) \left(\tilde{{\bf u}}_1 \tilde{{\bf u}}_2 + \tilde{{\bf u}}_2 \tilde{{\bf u}}_3  + \tilde{{\bf u}}_3 \tilde{{\bf u}}_1 \right) \left( \sum_{i = 1}^3 \tilde{{\bf w}}_i \right)\,,
\end{equation}
and has a manifest cyclic symmetry, and symmetry under chiral conjugation.
The four point amplitude has the nice and simple form
\begin{equation}\label{eq:A4_6D}
 \mathcal{A}_4 = - \delta^{6}\left(p \right) \delta^{4}\left(q^A \right) \delta^{4}\left( \tilde{q}_{A} \right) \frac{i}{x_{1 3}^2 x_{2 4}^2} \,.
\end{equation}
with the conserved supermomenta being given by
\begin{align}
q^A&=\sum_{i}\lambda_i^{Aa}\xi_{ia}\,,&\tilde{q}_{A}&=\sum_{i}\tilde\lambda_{iA\dot{a}}\tilde\xi_{i}^{\dot{a}}\,,
\end{align}
and the Grassmann delta functions
\begin{align}
 \delta^{4}\left(q^A \right)&=\tfrac{1}{4!}\epsilon_{ABCD}q^Aq^Bq^Cq^D\,,&\delta^{4}\left( \tilde{q}_{A} \right)&=\tfrac{1}{4!}\epsilon^{ABCD}\tilde{q}_{A}\tilde{q}_{B}\tilde{q}_{C}\tilde{q}_{D}\,.
\end{align}
The five point amplitude can be computed using the BCFW recursion, presented in \cref{section:BCFW6d}. The result, obtained in \cite{Bern:2010qa}, has the form
\begin{equation}\label{eq:5pkt_6d}
\begin{gathered}
 \mathcal{A}_5 = - \delta^{6}\left( p \right) \delta^{4}\left( q \right) \delta^{4}\left( \tilde{q} \right) \frac{i}{x_{1 3}^2 x^2_{2 4} x_{3 5}^2 x_{4 1}^2 x_{5 2}^2} \bigl(\langle q_1 |p_2 p_3 p_4 p_5 |\tilde{q}_1]  + \mbox{cyclic permutations}\\
 \begin{aligned}
&+ \tfrac{1}{2} \langle q_1 | p_2 p_3 p_4 p_5 - p_2 p_5 p_4 p_3 | \tilde{q}_2] + \tfrac{1}{2} \langle q_3 | p_4 p_5 p_1 p_2 - p_4 p_2 p_1 p_5 | \tilde{q}_4] +\text{c.c.}\\
&+ \tfrac{1}{2} \langle q_4 | p_5 p_1 p_2 p_3 - p_5 p_3 p_2 p_1 | \tilde{q}_5] + \tfrac{1}{2} \langle q_3 | p_5 p_1 p_2 p_3 - p_5 p_3 p_2 p_1 | \tilde{q}_5] +\text{c.c.}\;\bigr)\,.
%\\
%&+ \tfrac{1}{2} [\tilde{q}_1 | p_2 p_3 p_4 p_5 - p_2 p_5 p_4 p_3 | q_2\rangle + \tfrac{1}{2} [\tilde{q}_3 | p_4 p_5 p_1 p_2 - p_4 p_2 p_1 p_5 | q_4\rangle + \\
%&+ \tfrac{1}{2} [\tilde{q}_4 | p_5 p_1 p_2 p_3 - p_5 p_3 p_2 p_1 | q_5\rangle + \tfrac{1}{2} [\tilde{q}_3 | p_5 p_1 p_2 p_3 - p_5 p_3 p_2 p_1 |q_5\rangle
\end{aligned}
\end{gathered} 
\end{equation}
This representation of the five-point superamplitude lacks any manifest non-trivial symmetry apart from supersymmetry and is much more complicated than the four point amplitude \cref{eq:A4_6D}. As the  five point amplitude indicates, superamplitudes  with more than three partons have the general form
\begin{equation}
 \mathcal{A}_n=\delta^{(6)}\left(p\right)\delta^{(4)}\left(q\right)\delta^{(4)}\left(\tilde q\right)f_n(\{p_i,q_i,\tilde q_i\})\,.
\end{equation}
Judging from the increase in complexity going from $n=4$ to $n=5$, any straightforward application of the BCFW recursion, using \cref{eq:5pkt_6d} as initial data, cannot be expected to yield reasonable results for amplitudes with more than five external legs. Obviously new strategies are necessary to investigate higher point tree amplitudes of $\mathcal{N} = (1,1)$ SYM theory.
\subsection{Symmetries of superamplitudes}\label{generators_max_6d}
\subsubsection{Superpoincar\'e symmetry}

Although a part of the symmetries of tree-level  $\mathcal{N} = (1,1)$ SYM theory amplitudes appear in the literature, e.\,g. in \cite{Elvang:2011fx, Dennen:2010dh}, a complete list of all generators and their algebra is missing. This section aims to close this gap. 

We start with the symmetries of the tree level superamplitudes in on-shell superspace $\{\, 
\lambda^{A\, a}_{i}\, , \, \tilde\lambda_{i\, A\, \dot a}\, ,\, 
\xi_{i\, a}\, ,\,  \tilde\xi^{\dot a}_{i}\, \}$.
In contrast to its four-dimensional daughter theory, $\cN = 4$ SYM theory, the  six-dimensional $\cN = (1,1)$ SYM theory has no conformal symmetry since the gauge coupling constant in six dimensions is not dimensionless.  However, we have a super Poincar\'e symmetry
\begin{equation}
\{p_{AB},m^A_{\;\;B},q^A,\overline{q}_A,\widetilde{q}_A,\overline{\widetilde{q}}^A\}\,.
\end{equation}
The super Poincar\'e algebra is given by the supersymmetry algebra
\begin{equation}\label{komutator_susy}
\begin{aligned}
\left\{q^A, \overline{q}^{B}\right\} &= {p}^{A B} \,,&\qquad \left\{\widetilde{{q}}_A, \widetilde{\overline{{q}}}_{B}\right\} = {p}_{A B} 
\end{aligned}
\end{equation}
and the commutators involving the $m^{A}_{\;\;B}$ of the $SO(1,5)$ Lorentz symmetry 
with covering group $SU^{\ast}(4)$ read
\begin{equation}
\begin{aligned}
{}[m^{A}_{\;\;B}, m^{C}_{\;\;D}] &= \delta^{C}_{B} m^{A}_{\;\;D}-\delta^{A}_{D} m^{C}_{\;\;B}\,,&\qquad [m^{A}_{\;\;B}, p_{CD}] &= \delta^{A}_{[C} p_{D]B}+\tfrac{1}{2}\delta^{A}_{B} p_{CD}\,,\\
[m^{A}_{\;\;B}, {q}^C] &= \delta^{C}_{B} {q}^A - \tfrac{1}{4} \delta^{A}_{B} {q}^C \,,&\qquad [m^{A}_{\;\;B}, \overline{q}^C]& = \delta^{C}_{B} \overline{{q}}^A - \tfrac{1}{4} \delta^{A}_{B} \overline{{q}}^C\,, \\
[m^{A}_{\;\;B}, \widetilde{{q}}_C] &= - \delta^{A}_{C} \widetilde{{q}}_B + \tfrac{1}{4} \delta^{A}_{B} \widetilde{{q}}_C \,,&\qquad [m^{A}_{\;\;B}, \overline{\widetilde{{q}}}_C] &= - \delta^{A}_{C} \overline{\widetilde{{q}}}_B + \tfrac{1}{4} \delta^{A}_{B}  \overline{\widetilde{{q}}}_C \, .
\end{aligned}
\end{equation}
The translation symmetry is trivially given by momentum conservation
\begin{align}
p_{AB}=\sum_i \tilde\lambda_{iA\dot a}\tilde\lambda_{iB}^{\dot a}\,,
\end{align}
and the representation of the $(1,1)$ supersymmetry generators and their conjugates is
\begin{align}
&\begin{aligned}
q^A &= \sum_{i} \lambda_{i}^{A a} \xi_{i a}\,,& \qquad \widetilde{q}_A &= \sum_{i} \tilde{\lambda}_{i A \dot{a}} \tilde{\xi}_{i}^{\dot{a}}\,, \\
\overline{q}^A &= \sum_{i} \lambda_{i}^{A a} \partial_{i a}\,,& \qquad \overline{\widetilde{q}}_A &= \sum_{i} \tilde{\lambda}_{i A \dot{a}}\partial_{i}^{\dot{a}} \,,
\end{aligned}&& \mbox{ with } &  \partial_{i a} &= \frac{\partial}{\partial \xi_{i}^{a}}\,, & \partial_{i}^{\dot{a}} &= \frac{\partial}{\partial \tilde{\xi}_{i \dot{a}}} \,.
\end{align}
The correct form of the $su(4)$ Lorentz generators 
\begin{equation}
m^{A}_{\;\;B}=\sum_i \lambda_i^{Aa}\partial_{iBa}-\tilde\lambda_{iB\dot a}\partial_i^{A\dot a}-\tfrac{1}{4}\delta^A_B\lambda_i^{Ca}\partial_{iCa}+\tfrac{1}{4}\delta^A_B\tilde\lambda_{iC\dot a}\partial_i^{C\dot a}\,.
\end{equation}
is a bit more involved since the chiral and anti-chiral spinors are subject to the constraints
\begin{align}
\lambda_i^{A\, a}\lambda^{B}_{i\,a}&=\tfrac{1}{2}\epsilon^{ABCD}\tilde\lambda_{iC\, \dot a}\tilde\lambda_{iD}^{\dot a}\,,&\lambda_i^{A\, a}\tilde\lambda_{iA\, \dot a}&=0\,.
\end{align}
However, it is straightforward to show that the generators $m^{A}_{\;\;B}$ given above commute with these constraints.

Besides the super Poincar\'e symmetry there are a few additional trivial symmetries. First of all, we have the dilatation symmetry whose generator
\begin{equation}
d=\tfrac{1}{2}\sum_i\bigl[ \lambda_i^{Aa}\partial_{iAa} +\tilde\lambda_{iA\dot a}\partial_i^{A\dot a}\bigr]+n+2
\end{equation}
simply measures the dimension of a generator $\mathds{G}$
\begin{equation}
[{d}, \mathds{G}] = \dim(\mathds{G})\mathds{G}\,.
\end{equation}
The non-zero dimensions are
\begin{align}
\dim(\,p\,)&=1\,&\dim\left(\,q\,\right)&=\dim\left(\,\overline{q}\,\right)=\dim\left(\,\widetilde q\,\right)=\dim(\,\overline{\widetilde q}\,)=\tfrac{1}{2}\,.
\end{align}
As already mentioned before, the on-shell superfield and consequently the superamplitudes are manifest symmetric under the $SO(4)\simeq SU(2)\times SU(2)$ little group, whose generators are given by
\begin{align}
h_{a b} &= \sum_{i} \lambda^{A}_{i (a}\partial_{iA b)} - \xi_{i (a} \partial_{i b)} \,,& \tilde{h}_{\dot{a} \dot{b}} &= \sum_{i} \tilde{\lambda}_{iA(\dot{a}}\partial_{i\dot{b})}^A  - \tilde{\xi}_{i (\dot{a}} \partial_{i \dot{b})}\,.
\end{align}
Finally there are two hyper charges 
\begin{align}
b &= \sum_{i} \left(\xi_{i a} \partial_{i}^{a} - 1\right)\,,&  \qquad \tilde{b} &=  \sum_{i} \left(\tilde{\xi}_{i}^{\dot{a}} \partial_{i \dot{a}} - 1\right) 
\end{align}
 that correspond to a $U(1) \times U(1)$ subgroup of the $SU(2) \times SU(2)$ $R$-symmetry that we sacrificed for the manifest little group invariance. The action of the hyper charges on some generator $\mathds{G}$ are given by
\begin{align}
[b, \mathds{G}]& = \text{hyper}(\mathds{G})\mathds{G}\,,& [\tilde{b}, \mathds{G}] &=  \widetilde{\text{hyper}}(\mathds{G})\mathds{G}\,,
\end{align}
and the non-zero values are
\begin{align}\label{eq:Rcharges}
\text{hyper}(\,q\,)&=\widetilde{\text{hyper}}\left(\,\widetilde q\,\right)=1\,,&\text{hyper}(\,\overline{q}\,)&=\widetilde{\text{hyper}}\left(\,\overline{\widetilde q}\,\right)=-1\,.
\end{align}
Note that the constants in $d$, $b$, $\tilde{b}$ are not fixed by the algebra and have been chosen such that they annihilate the superamplitude.

\subsubsection{Enhanced dual conformal symmetry}

All the symmetries presented up to this point exactly match the expectations. Beautifully
 there is an additional non-trivial symmetry of the superamplitudes  \cite{Dennen:2010dh}.  Similar to $\mathcal{N} = 4$ SYM theory in four dimensions, the $\mathcal{N} = (1,1)$ SYM theory in six dimensions has a tree-level dual conformal symmetry. Due to the lack of a superconformal symmetry, the dual conformal symmetry does get not promoted to a full dual superconformal symmetry. 
 
In analogy to four dimensions we extend the on-shell superspace by dual variables to the full non-chiral superspace
\begin{equation}\label{eq:fullSuper6d}
\{\, 
\lambda^{A\, a}_{i}\, , \, \tilde\lambda_{i\, A\, \dot a}\, ,\, 
\xi_{i\, a}\, ,\,  \tilde\xi^{\dot a}_{i}\,,\,x_i^{AB},\theta_i^A,\tilde\theta_{i\,A} \}\,.
\end{equation}
The variables are subject to the constraints
\begin{equation}\label{eq:constraints6d}
\begin{aligned}
x_{i i+1}^{A B} & = \lambda^{A a}_{i} \lambda^{B}_{i a}\,,&  x_{i i+1 \; A B} &=\tilde{\lambda}_{i A \dot{a}} \tilde{\lambda}_{i B}^{\dot{a}} \\
\theta_{i i+1}^{A} & =\lambda^{A a}_{i} \xi_{i a} \,,& \tilde{\theta}_{i i+1 \; A} &= \tilde{\lambda}_{i A \dot{a}} \tilde{\xi}_{i}^{\dot{a}} 
\end{aligned}
\end{equation}
Similar to the non-chiral superamplitudes of $\mathcal{N} = 4$ SYM theory, it is possible to express the superamplitudes of $\mathcal{N} = (1,1)$ SYM solely using the dual superspace variables $\{x,\theta,\tilde\theta\}$. The amplitudes only depend on differences of dual variables, resulting in translation symmetries with respect to each of the dual variables. Hence, we define the dual translation generator to be
\begin{equation}
\begin{aligned}
P_{A B} &= \sum_i \partial_{i A B} \,,& \mbox{ with } && \partial_{i A B} &= \frac{\partial}{\partial x_{i}^{A B}}=\tfrac{1}{2}\widetilde{\Sigma}^{\mu\,BA}\frac{\partial}{\partial x_i^{\mu}}\,,
\end{aligned}
\end{equation} 
and the dual supermomenta are
\begin{align}
Q_{A} &= \sum_i \partial_{i A}\,, & \widetilde{Q}^{A} &= \sum_i \partial_{i}^{A}\,, &\mbox{ with }&&  \partial_{i A}& = \frac{\partial}{\partial \theta_{i}^{A}}\,& \partial_{i}^{A} &= \frac{\partial}{\partial \tilde{\theta}_{i A}}\,.
\end{align}
Although it is easy to algebraically construct conjugates $\overline{Q}_{A}$, $\overline{\widetilde{Q}}^{A}$ to the dual supermomenta, these conjugates would imply the invariance under the superconformal generators  $\overline{s}_{A}=\sum\xi_i^a\partial_{iAa}$ and $\overline{\widetilde{s}}^{A}=\sum\widetilde{\xi}_{i\dot{a}}\partial^{iA\dot a}$, which is not the case. We conclude that the amplitudes have an supersymmetry enhanced dual Poincar\'e symmetry
\begin{equation}
\{P_{AB},M^A_{\;\;B},Q_A,\widetilde{Q}^A\} 
\end{equation}
Though we do not have a full dual super Poincar\'e symmetry we have a dual conformal symmetry, which we are going to derive in what follows. First we recall that for $n>3$ the superamplitudes have the form
\begin{equation}
 \mathcal{A}_n=\delta^{(6)}\left(p\right)\delta^{(4)}\left(q\right)\delta^{(4)}\left(\tilde q\right)f_n\,.
\end{equation}
It is possible to define a dual conformal inversion $I$ of the variables of the full superspace \cref{eq:fullSuper6d} such that the function $f_n$ inverts covariantly
\begin{equation}\label{eq:inversionA6d}
I[f_n]=\left(\prod_i x_i^2\right)\,f_n\,.
\end{equation}
In contrast to four dimensions the product of momentum and supermomentum conserving delta functions is not dual conformal invariant due to the mismatch of the degrees of momentum and supermomentum conserving delta functions
\begin{equation}
I[\delta^{(6)}(x_{1\,n+1})\delta^{(4)}(\theta_{1\,n+1})\delta^{(4)}(\tilde{\theta}_{1\,n+1})]=(x_1^2)^2\delta^{(6)}(x_{1\,n+1})\delta^{(4)}(\theta_{1\,n+1})\delta^{(4)}(\tilde{\theta}_{1\,n+1})\,.
\end{equation}
The inversion leading to \cref{eq:inversionA6d} is defined as
\begin{align}
I[x_{i}^{\mu}]&=-(x^{-1}_{i})_{\mu}=-\frac{x_{i\,\mu}}{x_i^2}\,,&I[x_{i}^{AB}]&=(x^{-1}_{i})_{AB}\,,\label{eq:inversion6d_first}\\
I[\theta_{i}^{A}]&=\theta_{i}^{B}(x^{-1}_{i})_{BA}\,,&I[\tilde\theta_{i\,A}]&=(x^{-1}_{i})^{AB}\tilde\theta_{i\,B}\,\\
I[\lambda_{i}^{Aa}]&=\frac{x_{i\,AB}\lambda_{i\,a}^{B}}{\sqrt{x_i^2x_{i+1}^2}}\,,&I[\tilde\lambda_{iA\dot a}]&=\frac{x_{i}^{AB}\tilde\lambda_{iB}^{\dot a}}{\sqrt{x_i^2x_{i+1}^2}}\,,\\
I[\xi_{i\,a}]&=\sqrt{\frac{x_i^2}{x_{i+1}^2}}\left(\xi_{i}^a+\langle\theta_i|x_i^{-1}|i^a\rangle\right)\,,&I[\tilde\xi_{i}^{\dot{a}}]&=-\sqrt{\frac{x_i^2}{x_{i+1}^2}}\left(\tilde\xi_{i\,\dot{a}}+[\tilde\theta_i|x_i^{-1}|i_{\dot a}]\right)\,,\\
I[u_{i\,a}]&=\frac{\beta \,u_i^a}{\sqrt{x_{i+2}^2}}\,,&I[\tilde{u}_{i\,\dot a}]&=\frac{\tilde{u}_i^{\dot a}}{\beta\sqrt{x_{i+2}^2}}\,,
\end{align}
where $\beta$ is some arbitrary constant.
Equations \eqref{eq:inversion6d_first} and the fact that the inversion needs to be an involution on the dual variables, i.\,e.~$I^2=\mathds{1}$, imply the inversion rules of the sigma matrices 
\begin{align}
I[\Sigma_{AB}^\mu]&=\widetilde{\Sigma}_{\mu}^{BA}\,,&I[\widetilde{\Sigma}_{\mu}^{AB}]&=\Sigma_{BA}^\mu\,.
\end{align}
Consistency between the inversions of $x$ and the chiral and anti-chiral spinors requires the following inversion of the epsilon tensors of the little group
\begin{align}\label{eq:inversion6d_last}
I[\epsilon_{ab}]&=\epsilon^{ba}\,,&I[\epsilon_{\dot{a}\dot{b}}]&=\epsilon^{\dot{b}\dot{a}}\,.
\end{align}
Consequently, we have $I^2=-\mathds{1}$ on all variables carrying a little group index. Since the superamplitude is little group invariant this is no obstacle.
We note that the inversion defined in eqs.~\eqref{eq:inversion6d_first} to \eqref{eq:inversion6d_last} differs from the one presented in \cite{Dennen:2010dh} by some signs which are necessary in order to yield the desired inversion of the amplitudes. The proof of \cref{eq:inversionA6d} is straightforward using the BCFW recursion and will be presented in \cref{section:ProofDual}.

Similar to the four dimensional case we now define the generators
\begin{align}\label{eq:dualconformal6d}
K^{AB}&=IP_{AB}I\,,& \overline{S}^A&=IQ_AI\,,&\overline{\widetilde{S}}_A&=I\widetilde{Q}^AI\,.
\end{align}
From \cref{eq:inversionA6d} it immediately follows, that $f_n$ is annihilated by the dual superconformal generators $ \overline{S}^A$, $\overline{\widetilde{S}}_A$, but is covariant under dual conformal boosts
\begin{equation}\label{eq:actionK6d}
K^{AB}\,f_n=-\left(\sum_i x_i^{AB}\right)f_n\,.
\end{equation}
From the inversion rules eqs.~\eqref{eq:inversion6d_first} to \eqref{eq:inversion6d_last} and the defining equation \eqref{eq:dualconformal6d} we can obtain the action of the dual conformal boost generator by applying the chain rule, c.f.~four-dimensional case \cref{eq:calculationOfK}. Since for the action of the inversion operator on all variables carrying little group indices we have $I^2=-\mathds{1}$, the action of $K^{AB}$ on a little group invariant object is given by
\vspace*{0.5cm}
\begin{align}
K^{AB} &=\sum_i\sum_j\bigg[I\biggl[\frac{\partial I[x_j^{CD}]}{\partial x_{i\,AB}}\biggr]\partial_{j\,CD}+I\biggl[\frac{\partial I[\theta_j^{C}]}{\partial x_{i\,AB}}\biggr]\partial_{j\,C}+I\biggl[\frac{\partial I[\tilde\theta_{j\,D}]}{\partial x_{i\,AB}}\biggr]\partial_{j}^{D}\notag\\[+0.5cm]
&\=\phantom{\sum_i\sum_j\bigg[}{}-I\left[\frac{\partial I[\lambda_j^{Ca}]}{\partial x_{i\,AB}}\right]\partial_{j\,C a}-I\left[\frac{\partial I[\tilde{\lambda}_{j\,E\dot{a}}]}{\partial x_{i\,AB}}\right]\partial_{j}^{E\dot{a}}\notag\\[+0.5cm]
&\=\phantom{\sum_i\sum_j\bigg[}{}-I\left[\frac{\partial I[\xi_{j \,a}]}{\partial x_{i\,AB}}\right]\partial_{j}^{a}-I\left[\frac{\partial I[\tilde\xi_{j }^{\dot a}]}{\partial x_{i\,AB}}\right]\partial_{j\,\dot{ a}}\biggr]\,.
\end{align}
The coefficients of the derivatives are straightforward to obtain leading to
\begin{equation}\label{eq:K6d}
\begin{aligned}
K^{AB} &= \sum_{i} \biggl[ x_{i}^{AC} x_{i}^{BD} \partial_{i\, CD} - \theta^{[A}_{i} x_{i}^{B] C} \partial_{i C} -  \epsilon^{ABCD} \tilde{\theta}_{i C} x_{i DE} \partial_{i}^{E}  \\
&\=\phantom{\sum_{i} \biggl[}- \tfrac{1}{2}\lambda_{i}^{[A a} \left(x_i + x_{i+1}\right)^{B] C} \partial_{i C a} - \tfrac{1}{2}\epsilon^{ABCD} \tilde{\lambda}_{i C \dot{a}} \left(x_{i} + x_{i+1}\right)_{DE} \partial_{i}^{E \dot{a}}  \\
&\=\phantom{\sum_{i} \biggl[}+ \tfrac{1}{2}\left(\theta_{i} + \theta_{i+1}\right)^{[A}\lambda^{B]}_{i a} \partial_{i}^{a}+\tfrac{1}{2} \epsilon^{ABCD} (\tilde{\theta}_{i} + \tilde{\theta}_{i+1})_{C} \tilde{\lambda}_{i D}^{\dot{a}} \partial_{i \dot{a}}\biggr]\,,
\end{aligned}
\end{equation}
In an analogue calculation or by calculating the commutators of  $K^{AB}$ with the dual supermomenta $Q^A$, $\tilde{Q}_A$ we obtain
\begin{align}\label{eq:S_dual}
\overline{S}^{A} &= \sum_i x_{i}^{A B} \partial_{i B} - \lambda^{A}_{ia} \partial^{a}\,,&  \overline{\widetilde{S}}_A &=\sum_i x_{i A B} \partial_{i}^{B} - \tilde{\lambda}_{i A \dot{a}} \partial^{\dot{a}}\,.
\end{align}
Obviously the dual superconformal  generators $\overline{S}^{A}$, $\overline{\widetilde{S}}_A$ are related to the conformal generators $\overline{q}^{A}$,  $\overline{\widetilde{q}}_A$ by $\overline{S}^{A}=-\overline{q}^{A}$ and $\overline{\widetilde{S}}_A=-\overline{\widetilde{q}}_A$. 

Adding dual conformal inversions promotes the enhanced Poincar\'e symmetry to an enhanced dual conformal symmetry
\begin{equation}
\{P_{AB},M^A_{\;\;B},D,K_{AB},Q_A,\widetilde{Q}^A,\overline{S}^{A},\overline{\widetilde{S}}_A\}\,.
\end{equation}
The generators $M^A_{\;\;B}$ of the $SU(4)$ Lorentz symmetry\footnote{We drop the star
 of $SU^{\ast}(4)$ from now on.} act canonically on all generators carrying $SU(4)$ indices 
\begin{equation}
\begin{gathered}
{}[M^{A}_{\;\;B}, M^{C}_{\;\;D}] = \delta^{C}_{B} M^{A}_{\;\;D}-\delta^{A}_{D} M^{C}_{\;\;B}\,,\\[+.2cm]
\begin{aligned}
{} [M^{A}_{\;\;B}, P_{CD}] &= \delta^{A}_{[C} P_{D]B}+\tfrac{1}{2}\delta^{A}_{B} P_{CD}\,,&[M^{A}_{\;\;B}, K_{CD}] &= \delta^{A}_{[C} K_{D]B}+\tfrac{1}{2}\delta^{A}_{B} K_{CD}\,,\\
[M^{A}_{\;\;B}, Q_C] &= - \delta^{A}_{C}Q_B + \tfrac{1}{4} \delta^{A}_{B} Q_C \,,& [M^{A}_{\;\;B}, \overline{\widetilde{{S}}}_C] &= - \delta^{A}_{C} \overline{\widetilde{{S}}}_B + \tfrac{1}{4} \delta^{A}_{B}  \overline{\widetilde{{S}}}_C\,,\\
[m^{A}_{\;\;B}, \widetilde{{Q}}^C] &= \delta^{C}_{B} \widetilde{{Q}}^A - \tfrac{1}{4} \delta^{A}_{B} \widetilde{{Q}}^C \,,&\qquad [M^{A}_{\;\;B}, \overline{S}^C]& = \delta^{C}_{B} \overline{{S}}^A - \tfrac{1}{4} \delta^{A}_{B} \overline{S}^C\,.
\end{aligned}
\end{gathered}
\end{equation}
The remaining non-zero commutation relations are
\begin{equation}\label{eq:Dualconformal6d}
\begin{gathered}
\begin{aligned}[t]
[{K}^{AB}, {Q}_{C}] &= \delta^{[A}_{C} \overline{{S}}^{B]} \,,&&\hspace{3cm}& [{K}_{AB}, \widetilde{{Q}}^{C}] &=  \delta_{[A}^{C} \overline{\widetilde{S}}_{B]}\,,\\
[{P}^{AB}, \overline{\widetilde{S}}_{C}] &= \delta^{[A}_{C} \widetilde{{Q}}^{B]} \,,&&\hspace{3cm}&[{P}_{AB}, \overline{{S}}^{C}] &=  \delta_{[A}^{C} Q_{B]}\,,\\
\end{aligned}\\[+.2cm]
[K_{AB},P^{CD}]=\delta_{[A}^{[C}M^{D]}_{\;\;B]}+\delta_{[A}^{C}\delta_{B]}^{D}D\,.
\end{gathered}
\end{equation}
The dual dilatation generator is given by
\begin{equation}
\begin{gathered}
D = -\tfrac{1}{2}\sum_{i}\left(\lambda_{i a}^{A} \partial_{i A}^{a} + \tilde{\lambda}_{i A \dot{a}} \partial_{i}^{A \dot{a}} + \theta_{i}^{A} \partial_{i A} + \tilde{\theta}_{i A} \partial_{i}^{A} + x_{i}^{A B} \partial_{i\,AB} \right)
\end{gathered}
\end{equation}
and, as a consequence of \cref{eq:actionK6d,eq:Dualconformal6d}, acts covariantly
\begin{equation}
D\,f_n=n\,f_n\,.
\end{equation}
The dual Lorentz generators $M^{A}_{\;\;B}$ are equal to the action of the on-shell Lorentz generators $m^{A}_{\;\;B}$ in the full superspace. Their representation can be obtained from the dual conformal algebra \cref{eq:Dualconformal6d} and is given by
\begin{equation}
\begin{aligned}
M^{A}_{\;\;B} = \sum_i\bigl[ x_{i}^{AC} \partial_{i\,B C} - \tfrac{1}{4} \delta^{A}_{B} x_{i}^{C D} \partial_{i\,CD}&+\lambda_{i}^{A a} \partial_{i B a} - \tfrac{1}{4} \delta^{A}_{B} \lambda_{i}^{C a} \partial_{i C a} + \theta_{i}^{A} \partial_{i B} - \tfrac{1}{4} \delta^{A}_{B} \theta_{i}^{C} \partial_{i C} \\
&- \tilde{\lambda}_{i B}^{\dot{a}} \partial_{i}^{A \dot{a}\phantom{i}} + \tfrac{1}{4} \delta^{A}_{B} \tilde{\lambda}_{i C}^{\dot{a}} \partial_{i}^{C \dot{a}\phantom{i}} - \tilde{\theta}_{i B} \partial_{i}^{A} + \tfrac{1}{4} \delta^{A}_{B} \tilde{\theta}_{i C} \partial_{i}^{C} \bigr]
\end{aligned}
\end{equation}
Finally, we define the dual $R$-symmetry hyper charges  to be
\begin{align}
B &= \sum_{i} \left(\xi_{i a} \partial_{i}^{a} + \theta_{i}^{A a} \partial_{i A}^{a}\right) -n+4\,,& \widetilde{B} &= \sum_{i} \left(\tilde{\xi}_{i}^{\dot{a}} \partial_{i \dot{a}} + \tilde{\theta}_{i A \dot{a}} \partial_{i}^{A \dot{a}} \right) -n+4\,.
\end{align}
The non-zero charges are $\text{hyper}(Q)=\widetilde{\text{hyper}}(\widetilde{Q})=-\text{hyper}(\overline{S})=-\widetilde{\text{hyper}}(\overline{\widetilde{S}})=1$, and the constants in the definitions of $B$ and $\widetilde{B}$ have been fixed such that $f_n$ gets annihilated.
\subsection{Dimensional reduction to massless ${\mathcal{N} = 4}$ SYM}\label{section:dimensional_reduction}
In this section we explain how the six dimensional tree-level superamplitudes can be mapped to non-chiral superamplitudes of massless $\mathcal{N} = 4$ SYM. Similar mappings can be found in references \cite{Elvang:2011fx, Bern:2010qa, Dennen:2009vk,Huang:2011um}.

In order to perform the dimensional reduction we restrict the six dimensional momenta to the preferred four dimensional subspace $p_4=p_5=0$. Because of our special choice of six dimensional Pauli matrices, compare \cref{eq:SigmaGamma}, one can express the six dimensional spinors in terms of four dimensional ones
\begin{align}\label{eq:MapSpinors6d4d}
\lambda^{Aa} &= \left(\begin{array}{cc} 0 & \lambda_{\alpha} \\  \tilde{\lambda}^{\dot{\alpha}} & 0 \end{array}\right)\,,&  \tilde{\lambda}_{A\dot{a}} &= \left(\begin{array}{cc} 0 & \lambda^{\alpha} \\  -\tilde{\lambda}_{\dot{\alpha}} & 0 \end{array}\right)\,.
\end{align}
In the four dimensional subspace the contractions with the six-dimensional Pauli matrices read
\begin{align}\label{eq:MapMomenta6d4d}
p_{AB} &= \left(\begin{array}{cc} 0 & -p^{\alpha}_{\;\;\dot \beta} \\  p^{\;\;\beta}_{\dot \alpha} & 0 \end{array}\right)\,,& p^{AB} &= \left(\begin{array}{cc} 0 & -p_{\alpha}^{\;\;\dot \beta} \\   p^{\dot{\alpha}}_{\;\;\beta} & 0 \end{array}\right)\,,
\end{align}
and the supermomenta are
\begin{align}\label{eq:456}
q^{A} &= \lambda^{A a} \xi_{a} = \left(\begin{array}{c} \lambda_{\alpha} \xi_{2} \\  \tilde{\lambda}^{\dot{\alpha}} \xi_{1} \end{array}\right) \,,&
\tilde{q}_{A} &= \tilde{\lambda}_{A \dot{a}} \tilde{\xi}^{\dot{a}} = \left(\begin{array}{cc} \lambda^{\alpha} \tilde\xi^{\dot{2}}, & -\tilde{\lambda}_{\dot{\alpha}} \tilde{\xi}_{\dot{1}} \end{array}\right)\,.
\end{align}
Obviously, both, $\xi_{a}$ and $\xi^{\dot{a}}$ have to be mapped to  $\eta^{m}$ and $\tilde{\eta}_{m'}$. Here we make the choice
\begin{align}\label{eq:grassmann_map}
\xi_{a} &= \left(\tilde{\eta}_{3},\eta^{1} \right)\,,&  \tilde{\xi}^{\dot{a}}& = \left(\tilde{\eta}_{2}, -\eta^{4}\right)\,,
\end{align}
recall that we are using the convention $m\in\{1,4\}$ and $m'\in\{2,3\}$ for the 
non-chiral 4d superspace of section \ref{ncsuperspace4d}.
This implies the maps of the supermomenta
\begin{align}\label{eq:MapSupermomenta6d4d}
q^{A} &= \left(\begin{array}{c} q_\alpha^1 \\  \tilde{q}^{\dot{\alpha}}_{3} \end{array}\right) \,,&
\tilde{q}_{A} & = \left(\begin{array}{cc} -q^{\alpha\,4} ,  & -\tilde{q}_{\dot{\alpha}\,2}  \end{array}\right)\,,
\end{align}
and supermomentum conserving delta functions
\begin{equation}\label{eq:projektion_delta}
\delta^{4} \left(\sum_{i} q_{i}^{A} \right) \delta^{4} \left(\sum_{i} \tilde{q}_{i A} \right)= \delta^{4} \left(\sum_{i}  q_{i \alpha}^{m} \right) \delta^{4} \left(\sum_{i} \tilde{q}^{\dot{\alpha}}_{i m'} \right)\,.
\end{equation}
Applying the map of the Grassmann variables \cref{eq:grassmann_map} to the six dimensional superfield \cref{eq:superfield6d} and comparing it with the four dimensional non-chiral superfield \cref{eq:nonChiralSuperfield} yields the following map of the six and four dimensional on-shell states
\begin{equation}\label{eq:Map4d6d}
\begin{aligned}
\mbox{scalars:}&&\hspace{1cm}& \begin{aligned} s &= \phi_{2 3}\,, & s' &= \phi_{ 2 1}\,, & s''& = \phi_{4 3}\,,& s''' &= \phi_{4 1} \,,\end{aligned}\\
\mbox{gluinos:}&&\hspace{1cm}& \begin{aligned}\chi^{a} &= \left(\overline{\psi}^{4} ,-\psi_{2}\right)\,,& \lambda^{a} &= \left(\psi_{4},-\overline{\psi}^{2} \right)\,, \\
\tilde{\chi}_{\dot a} &=\left(-\psi_{3}, -\overline{\psi}^{1} \right)\,, & \tilde{\lambda}_{\dot{a}} &= \left(-\psi_{1}, -\overline{\psi}^{3} \right)\,,\end{aligned}\\
\mbox{gluons:}& &\hspace{1cm}& g^a_{\phantom{a}\dot{a}} = \left(\begin{array}{cc}G_{+}  &\phi_{4 2}  \\\phi_{3 1}  & -G_{-}\end{array}\right)\,.
\end{aligned}
\end{equation}
With the help of \cref{eq:MapSpinors6d4d,eq:MapMomenta6d4d,eq:grassmann_map,eq:MapSupermomenta6d4d} it is possible to perform the dimensional reduction of any six dimensional superamplitude.

For a detailed analysis of the connection between the massless amplitudes in six and four dimensions and an investigation of a potential uplift from four to six dimensions we refer to \cref{section:uplift_huang}.

 \section{From massless 6d to massive 4d superamplitudes}\label{section:DimRedmassive}
\subsection{On-shell massive superspace in 4d from dimensional reduction}
In \cref{section:dimensional_reduction} we dimensionally reduced the massless six-dimensional amplitudes to massless four-dimensional ones. In analogy, we now want to perform the dimensional reduction of the superamplitudes of ${\mathcal N}=(1,1)$ SYM to the massive Coulomb branch amplitudes of $\mathcal{N}=4$ SYM. 
When performing the dimensional reduction we need to choose an appropriate set of massive four-dimensional on-shell variables. For the bosonic part of the on-shell variables we choose \emph{two} sets of helicity spinors $\{\lambda_{\alpha},\tilde{\lambda}_{\dot \alpha}\}$
and $\{\mu_{\alpha},\tilde{\mu}_{\dot \alpha}\}$ to write the bispinor representation of a four dimensional massive momentum as
\begin{align}
p_\mu\sigma^\mu_{\a\da}&=p_{\a\da}= \lambda_{\a}\tilde{\lambda}_{\da} + \mu_{\a}\tilde{\mu}_{\da}\,.
\end{align}
We introduce abbreviations for the spinor contractions
\begin{equation}
\label{massshell}
\ang{\lambda}{\mu} = m \, ,\qquad 
[\tilde\mu \,\tilde\lambda] = \bar m \, ,
\end{equation}
where the mass parameters $m$ and $\bar m$ are in general complex numbers, related to the physical mass by $p^2=m \bar m$. 

For the particular representation of the six-dimensional Pauli matrices listed in \cref{appendix:Spinors}, the six-dimensional spinors can be expressed using the two sets of four dimensional spinors introduced above
\begin{align}\label{eq:DimRedSpinors}
\lambda^{A\,a}&=\begin{pmatrix}
               -\mu_\alpha&\lambda_\alpha\\
		\tilde\lambda^{\dot\alpha}&\tilde\mu^{\dot\alpha}
              \end{pmatrix}&\text{and}&&
\tilde\lambda_{A\,\dot a}&=\begin{pmatrix}
               \bar\rho\mu^\alpha&\lambda^\alpha\\
		-\tilde\lambda_{\dot\alpha}&\rho\tilde\mu_{\dot\alpha}
              \end{pmatrix}&\text{with}&&\rho&=\bar\rho^{-1}=\frac{m}{\bar m}\,,
\end{align}
and the six-dimensional momenta and dual momenta are given by
\begin{align}\label{eq:DimRedMomenta}
p_{AB}&=\begin{pmatrix}
      -\bar m\,\epsilon^{\alpha\beta}&-p^\alpha_{\phantom{\alpha}\dot\beta}\\
        p_{\dot \alpha}^{\phantom{\dot \alpha}\beta}&m\,\epsilon_{\dot\alpha\dot\beta}
     \end{pmatrix}&p^{AB}&=\begin{pmatrix}m \,\epsilon_{\alpha\beta}&-p_{\alpha}^{\phantom{\alpha}\dot\beta}\\
p^{\dot\alpha}_{\phantom{\dot\alpha}\beta}&-\bar m\,\epsilon^{\dot\alpha\dot\beta}\end{pmatrix}
\end{align}
and
\begin{align}\label{eq:DimRedRegions}
x_{AB}&=\begin{pmatrix}
      -\bar n\,\epsilon^{\alpha\beta}&-x^\alpha_{\phantom{\alpha}\dot\beta}\\
        x_{\dot \alpha}^{\phantom{\dot \alpha}\beta}&n\,\epsilon_{\dot\alpha\dot\beta}
     \end{pmatrix}&x^{AB}&=\begin{pmatrix}n\,\epsilon_{\alpha\beta}&-x_{\alpha}^{\phantom{\alpha}\dot\beta}\\
x^{\dot\alpha}_{\phantom{\dot\alpha}\beta}&-\bar n\,\epsilon^{\dot\alpha\dot\beta}\end{pmatrix}\,.
\end{align}
Here $p_{\a\da}=p_\mu\sigma^\mu_{\a\da}$, $x_{\a\da}=x_\mu \sigma^\mu_{\a\da}$ are the contractions of the first four components of the six-dimensional vectors with the four-dimensional Pauli matrices and $m=p_5-ip_4$, $n=x_5-ix_4$. Our conventions for four dimensional spinors can be found in \cref{appendix:Spinors}.

Since we are interested in massive four dimensional amplitudes in the following, we from now on set the fourth spatial component of all six-dimensional vectors to zero, thereby effectively performing the dimensional reduction from a massless five-dimensional to a massive four dimensional theory. This is equivalent to setting $n=\bar{n}=x_5$ and imposing the constraint $m=\bar m$ on the spinor variables, which together with the reality condition for the momenta $\lambda^{\ast}=\pm \tilde\lambda$, $\mu^{\ast}=\pm \tilde\mu$ results in the 5 real degrees of freedom of a massive four dimensional momentum and a spin quantization axis\footnote{Each helicity spinor starts out with 4 real degrees of freedom, the reality condition
$\lambda^{\ast}=\pm \tilde\lambda$ and the $U(1)$ helicity scaling $\lambda\to\exp[i\alpha]\lambda$
cuts this down to 3 real degrees of freedom. The further condition 
$\ang{\lambda}{\mu}=\bracket{\tilde\mu}{\tilde\lambda}$ brings us to 5=3+3-1 degrees of freedom.}.

Inserting the dimensional reduction of the spinors into the definition of the supermomenta we obtain
\begin{align}
 q^{A} &= \lambda^{A a} \xi_{a} = \left(\begin{array}{c}-\mu_{\alpha} \xi_{1}+ \lambda_{\alpha} \xi_{2} \\  \tilde{\lambda}^{\dot{\alpha}} \xi_{1}+\tilde{\mu}^{\dot{\alpha}} \xi_{2} \end{array}\right) \,,&
\tilde{q}_{A} &= \tilde{\lambda}_{A \dot{a}} \tilde{\xi}^{\dot{a}} = \left(\begin{array}{cc} \mu^{\alpha} \tilde\xi^{\dot{1}}+\lambda^{\alpha} \tilde\xi^{\dot{2}}, & -\tilde{\lambda}_{\dot{\alpha}} \tilde{\xi}_{\dot{1}} +\tilde{\mu}_{\dot{\alpha}} \tilde{\xi}_{\dot{2}} \end{array}\right)\,,
\end{align}
generalizing the four-dimensional massless case of \cref{eq:456}. It is then
convenient to define the Grassmann part of our four-dimensional massive on-shell variables to be
\begin{align}\label{eq:MapGrassmann}
\zeta^a&=\begin{pmatrix}\xi_{1}\\-\tilde\xi^{\dot 1}\end{pmatrix}\,,&\bar\zeta^a&=\begin{pmatrix}\xi_{2}\\\tilde\xi^{\dot 2}\end{pmatrix}\,,
\end{align}
leading to the four-dimensional supermomenta
\begin{align}
 q_{\alpha}^a&=\lambda_{\alpha}\bar\zeta^a-\mu_{\alpha}\zeta^a&
\tilde{q}_{\dot\alpha}^a&=\tilde\lambda_{\dot\alpha}\zeta^a+\tilde\mu_{\dot\alpha}\bar\zeta^a
\end{align}
related to the six-dimensional ones by
\begin{align}\label{eq:DimRedSuper}
 q^{A} &= \left(\begin{array}{c}q_\alpha^1 \\  \tilde{q}^{\dot\alpha\,1} \end{array}\right) \,,&
\tilde{q}_{A} &=  \left(\begin{array}{cc} q^{\alpha\,2}, & \tilde{q}_{\dot\alpha}^2 \end{array}\right)\,.
\end{align}
The dual fermionic momenta $\theta^{a}_{i\,\alpha}$, $\tilde\theta^{a}_{i\,\dot\alpha}$ are defined by
\begin{align}
 (\theta_i-\theta_{i+1})^{a}_\alpha&= q^{a}_{i\,\alpha}\\
(\tilde\theta_i-\tilde\theta_{i+1})^{a}_{\dot\alpha}&= \tilde{q}^{a}_{i\,\dot\alpha}\,,
\end{align}
and are related to the six-dimensional dual fermionic momenta by
\begin{align}\label{eq:DimRedDual}
  \theta^{A} &= \left(\begin{array}{c}\theta_\alpha^1 \\  \tilde{\theta}^{\dot\alpha\,1} \end{array}\right) \,,&
\tilde{\theta}_{A} &=  \left(\begin{array}{cc} \theta^{\alpha\,2}, & \tilde{\theta}_{\dot\alpha}^2 \end{array}\right)\,.
\end{align}
In conclusion the massive Coulomb branch amplitudes of $\mathcal{N}=4$ SYM may be expressed either by the on-shell variables
\begin{equation}\label{eq:onshellMassive}
 \{\lambda_i^\alpha,\mu_i^\alpha,\tilde\lambda_i^{\dot\alpha},\tilde\mu_i^{\dot\alpha};\zeta_i^a,\bar{\zeta}_i^a\}
\end{equation}
or the dual variables
\begin{equation}\label{eq:dualMassive}
\{x_i^{\dot{\alpha}\beta},n_i,\theta^{a}_{i\,\alpha},\tilde\theta^{a}_{i\,\dot\alpha}\}\,.%,\bar{n}_i
\end{equation}
In the associated full superspace the constraints on the variables read
\begin{align}
(x_i-x_{i+1})_{\alpha\dot\alpha}&=p_{i\,\alpha\dot\alpha}\label{eq:constraint1}\,,\\
n_i-n_{i+1}&=m_i\,,\\
%\bar n_i-\bar n_{i+1}&=\bar m_i\,,\\
m_i&=\bar m_i\label{eq:constraint3}\,,\\
(\theta_i-\theta_{i+1})^{a}_\alpha&= q^{a}_{i\,\alpha}\label{eq:constraint4}\,\\
(\tilde\theta_i-\tilde\theta_{i+1})^{a}_{\dot\alpha}&= \tilde{q}^{a}_{i\,\dot\alpha}\label{eq:constraint7}\,.
\end{align}
With the help of the maps \cref{eq:DimRedSpinors,eq:DimRedMomenta,eq:MapGrassmann,eq:DimRedRegions,eq:DimRedDual,eq:DimRedSuper} it is straightforward to translate any representation of a six-dimensional superamplitude into our four-dimensional variables. From the general form or the six-dimensional superamplitudes we can deduce the general form of the massive amplitudes to be
\begin{equation}\label{eq:massiveAMPS}
 \mathcal{A}_n=\delta^{(1)}(n_{1\,n+1})\delta^{(4)}(x_{1\,n+1})\delta^{(4)}(\theta_{1\,n+1}^{\a\,a})\delta^{(4)}(\tilde\theta_{1\,n+1}^{\da\,a})f_n(\{x_{ij},n_{ij},\theta_{ij},\tilde\theta_{ij}\})\,.
\end{equation}
\subsection{Symmetries of massive $\mathcal{N}=4$ superamplitudes on the Coulomb branch}\label{section:symmetriesMassive}
\subsubsection{Super-Poincar\'e symmetry}

We now want to investigate the symmetries of the massive amplitudes using the on-shell variables \cref{eq:onshellMassive} introduced in the last section. To be more precisely we are interested in the symmetries of $f_n$, defined in \cref{eq:massiveAMPS}, on the support of the delta functions. Similar to the massless four-dimensional case we define shorthand notations for derivatives with respect to spinors 
\begin{align}
\partial_{i\,\alpha} &= \frac{\partial}{\partial \lambda^{\alpha}_{i}}\,,&
\partial_{i\,\dot{\alpha}} &= \frac{\partial}{\partial
   \tilde{\lambda}^{\dot{\alpha}}_{i}}\,, &
   \delta_{i\, \alpha} &= \frac{\partial}{\partial \mu_{i}^{\alpha}}\,, &
\delta_{i\,\dot{\alpha}} &= \frac{\partial}{\partial
   \tilde{\mu}_{i}^{\dot{\alpha}}}\,, 
\end{align}
Judging from the symmetries of the six-dimensional superamplitudes, presented in \cref{generators_max_6d}, and the imposed constraint $m=\bar{m}$, we expect a five-dimensional super Poincar\'e symmetry. It remains to show how this symmetry is realized on the on-shell variables \cref{eq:onshellMassive}.

Obviously we have translation invariance 
\begin{align}
 p^{\alpha\dot\alpha}&=\sum_i \lambda_i^{\a}\tilde{\lambda}_i^{\da} + \mu_i^{\a}\tilde{\mu}_i^{\da}\,,&m&= \sum_{i} \ang{\lambda_{i}}{\mu_{i}}=\sum_{i} \sqb{\tilde\mu_{i}}{\tilde\lambda_{i}}\,.
\end{align}
as well as the Lorentz generators
\begin{align}
 l_{\alpha\beta}&=\sum_i\lambda_{i\,(\alpha}\partial_{i\,\beta)}+\mu_{i\,(\alpha}\delta_{i\,\beta)}\,,&
\bar l_{\dot\alpha\dot\beta}&=\sum_i\tilde\lambda_{i\,(\dot\alpha}\partial_{i\,\dot\beta)}+\tilde\mu_{i\,(\dot\alpha}\delta_{i\,\dot\beta)}\,,
\end{align}
associated to rotations in the first four spatial directions. Lorentz rotations $l^{\mu 5}$ involving the fifth spatial dimension correspond to the generator
\begin{align}
 w_{\alpha\dot\alpha}&=\sum_i\tilde\mu_{i\,\dot\alpha}\partial_{i\,\alpha}-\tilde\lambda_{i\,\dot\alpha}\delta_{i\,\alpha}+\mu_{i\,\alpha}\partial_{i\,\dot\alpha}-\lambda_{i\,\alpha}\delta_{i\,\dot\alpha}\,.
\end{align}
Supersymmetry is realized as
\begin{align}
q_{\alpha}^a&=\sum_i\lambda_{i\,\alpha}\bar\zeta_i^a-\mu_{i\,\alpha}\zeta_i^a\,,&
\tilde{q}_{\dot\alpha}^a&=\sum_i\tilde\lambda_{i\,\dot\alpha}\zeta_i^a+\tilde\mu_{i\,\dot\alpha}\bar\zeta_i^a\,,\\
 \bar q_{\dot\alpha\,a}&=\sum_i\tilde\lambda_{i\,\dot\alpha}\frac{\partial}{\partial\bar\zeta_i^a}-
\tilde\mu_{i\,\dot\alpha}\frac{\partial}{\partial\zeta_i^a}\,,&\bar{\tilde{q}}_{\alpha\,a}&=\sum_i\lambda_{i\,\alpha}\frac{\partial}{\partial\zeta_i^a}+
\mu_{i\,\alpha}\frac{\partial}{\partial\bar{\zeta}_i^a}\,.
\end{align}
Trivially we have a dilatation symmetry with the generator
\begin{align}\label{eq:dilatation}
d&=\tfrac{1}{2}\sum_i(\lambda^{\alpha}_i\partial_{i\,\alpha}+\tilde\lambda^{\dot\alpha}_i\partial_{i\,\dot\alpha}+\mu^{\alpha}_i\delta_{i\,\alpha}+\tilde\mu_i^{\dot\alpha}\delta_{i\,\dot\alpha}+2)\,.
\end{align}
Performing the dimensional reduction of the spinors, \cref{eq:DimRedSpinors}, the independence of $\lambda_A$ and $\tilde\lambda^A$ gets lost. As a consequence only one $SU(2)$ factor of the  $SU(2)\times SU(2)$ little group symmetry survives the dimensional reduction. Indeed we have the $SU(2)$ helicity generators
\begin{align}
h_+&=\frac{1}{\sqrt{2}}\sum_i\left(\lambda_i^\a\delta_{i\,\a}-\tilde\mu_i^{\da}\partial_{i\,\da}+\zeta_i^a\frac{\partial}{\partial\bar \zeta_i^a}\right),&h_-&=\frac{1}{\sqrt{2}}\sum_i\left(\mu_i^\a\partial_{i\,\a}-\tilde\lambda_i^{\da}\delta_{i\,\da}+\bar\zeta_i^a\frac{\partial}{\partial \zeta_i^a}\right),\notag\\
h&=\frac{1}{2}\sum_i\left(\lambda_i^\a\partial_{i\,\a}+\tilde\mu_i^{\da}\delta_{i\,\da}-\mu_i^{\a}\delta_{i\,\a}-\tilde\lambda_i^{\da}\partial_{i\,\da}+\zeta_i^a\frac{\partial}{\partial \zeta_i^a}-\bar\zeta_i^a\frac{\partial}{\partial \bar\zeta_i^a}\right).\hspace{-3.7cm}&& 
\end{align}
They fulfill the following closing algebra
\begin{equation}\label{eq:OnShellAlgebra}
\begin{aligned}
{}[h_+,h_-]&=h&\hspace{2cm}[h,h_{\pm}]&=\pm h_{\pm}\\
[l_{\a\b},l_{\gamma\delta}]&=2\epsilon_{\gamma(\a}l_{\b)\delta}+2\epsilon_{\delta(\a}l_{\b)\gamma}&[\bar l_{\da\db},\bar l_{\dot\gamma\dot\delta}]&=2\epsilon_{\dot\gamma(\da}\bar l_{\db)\dot\delta}+2\epsilon_{\dot\delta(\da}\bar l_{\db)\dot\gamma}\\
[w_{\alpha\dot\alpha}, w_{\beta\dot\beta}]&=2\epsilon_{\alpha\beta}\bar l_{\dot\alpha\dot\beta}+2\epsilon_{\dot\alpha\dot\beta} l_{\alpha\beta}&&\\
[\, l_{\b\g}, w_{\a\da}\, ] &= \epsilon_{\a(\b}\, w_{\g)\da}&[\, \bar l_{\db\dg}, w_{\a\da}\, ] &= - w_{\a(\db}\,\epsilon_{\dg)\da}\\
[\, l_{\b\g}, p_{\a\da}\, ] &= \epsilon_{\a(\b}\, p_{\g)\da}&[\, \bar l_{\db\dg}, p_{\a\da}\, ] &= - p_{\a(\db}\,\epsilon_{\dg)\da}\\
[w_{\alpha\dot\alpha},m]&=p_{\alpha\dot\alpha}&[w_{\alpha\dot\alpha},p_{\beta\dot\beta}]&=2\epsilon_{\alpha\beta}\epsilon_{\dot\alpha\dot\beta}m\\
[\, l_{\b\g}, q_{\a}^a\, ] &= \epsilon_{\a(\b}\, q_{\g)}^a&[\, l_{\b\g}, \bar{\tilde{q}}_{\a\,a}\, ] &= \epsilon_{\a(\b}\, \bar{\tilde{q}}_{\g)\,a}\\
[\, \bar l_{\db\dg}, \tilde{q}_{\da}^a\, ] &= \epsilon_{\da(\db}\, \tilde{q}_{\dg)}^a&[\, \bar l_{\db\dg}, \bar q_{\da\,a}\, ] &= \epsilon_{\da(\db}\, \bar q_{\dg)\,a}\\
[w_{\alpha\dot\alpha},q_{\beta}^a]&=-\epsilon_{\alpha\beta}\tilde{q}_{\dot\alpha}^a&[w_{\alpha\dot\alpha},\tilde{q}_{\dot\beta}^a]&=\epsilon_{\dot\alpha\dot\beta}q_{\alpha}^a\\
[w_{\alpha\dot\alpha},\bar{\tilde{q}}_{\beta\,a}]&=\epsilon_{\alpha\beta}\bar q_{\dot\alpha\,a}&[w_{\alpha\dot\alpha},\bar q_{\dot\beta\, a}]&=-\epsilon_{\dot\alpha\dot\beta}\bar{\tilde{q}}_{\alpha\, a}\\
\{q_{\alpha}^a,\bar q_{\dot\alpha\,b}\}&=p_{\alpha\dot\alpha}\,\delta^a_b&\{\tilde{q}_{\dot\alpha}^a,\bar{\tilde{q}}_{\alpha\,b}\}&=p_{\alpha\dot\alpha}\,\delta^a_b\\
\{q_{\alpha}^a,\bar{\tilde{q}}_{\beta\,b}\}&=m\,\epsilon_{\alpha\beta}\,\delta_b^a&\{\tilde{q}_{\dot\alpha}^a,\bar q_{\dot\beta\,b}\}&=-m\,\epsilon_{\dot\alpha\dot\beta}\,\delta_b^a
\end{aligned}
\end{equation}
along with the generic $[d, j] =\dim(j)\, j$ for any generator $j$, all other commutators vanishing. A necessary condition for the generators to be well defined on the massive amplitudes under consideration is that they commute with the constraint $m=\bar{m}$. One indeed shows that this is the case, e.g.
\begin{equation}
[\, w_{\a\da}, \ang{\lambda_{i}}{\mu_{i}} - [\tilde\mu_{i}\,\tilde\lambda_{i}]\, ] = 0\,.
\end{equation}

Clearly the nice form of the algebra is suggesting the existence of a $SU(2)$ symmetry with respect to the Grassmann label $a$, introduced in \cref{eq:MapGrassmann}. However,  at this point we see no indication that such a symmetry is realized on the massive superamplitudes \eqref{eq:massiveAMPS} for multiplicities larger than four and the introduction of the Grassmann variables $\zeta^a$, $\bar\zeta^a$ and their dual partners $\theta^a$, $\tilde\theta^a$ should be regarded as a very convenient way to compactly write down the algebra. Indeed, the $SU(2)$ symmetry of the algebra will be explicitly broken if we include the generators $r_1$, $r_2$ of $U(1)\times U(1)$  $R$-symmetry realized on the massive superamplitudes \eqref{eq:massiveAMPS}
\begin{align}
 r_1&=\sum_i \left(\zeta_i^1\frac{\partial}{\partial \zeta_i^1}+\bar\zeta_i^1\frac{\partial}{\partial \bar\zeta_i^1}\right) -n+4\,&r_2=\sum_i\left( \zeta_i^2\frac{\partial}{\partial \zeta_i^2}+\bar\zeta_i^2\frac{\partial}{\partial \bar\zeta_i^2}\right)-n+4\,.
\end{align}
Invariance under $r_a$ follows from the hyper charges $b$, $\tilde{b}$ of  \eqref{eq:Rcharges} of the six-dimensional superamplitudes. We have
\begin{align}
[r_a,q_{\alpha}^b]&=\delta_a^b q_{\alpha}^b\,,&[r_a,\tilde{q}_{\da}^b]&=\delta_a^b \tilde{q}_{\da}^b\,,&[r_a,\bar{q}_{\da\,b}]&=-\delta_a^b \bar{q}_{\da\,b}\,,&[r_a,\bar{\tilde q}_{\a\,b}]&=-\delta_a^b \bar{\tilde q}_{\a\,b}\,.
\end{align}

\subsubsection{Enhanced dual conformal symmetry}

We now want to investigate the symmetries in the dual superspace \eqref{eq:dualMassive}. Similar to the on-shell case we already know from the the six-dimensional amplitudes that we will have an extended dual conformal symmetry. Obviously the massive amplitudes have an extended 
dual Poincar\'e symmetry with generators
\begin{equation}
\{P_{\alpha\dot{\alpha}},\,M,\,L_{\a\b},\,\bar L_{\da\db},\,W_{\a\da}\} \,.
\end{equation}
Translation invariance in the dual variables implies the symmetries
\begin{align}
 P_{\alpha\dot{\alpha}}&=\sum_i \frac{\partial}{\partial x_{i}^{\dot{\alpha}\alpha}}\,,&  M&=\sum_i \frac{\partial}{\partial n_{i}}\,.
\end{align}
and
\begin{align}
 Q_{\a\,a}&=\sum_i \frac{\partial}{\partial \theta_{i}^{\a\,a}}\,,& \tilde{Q}_{\da\,a}&=\sum_i \frac{\partial}{\partial \theta_{i}^{\da\,a}}\,.
\end{align}
The Lorentz generators $L_{\a\b}$, $\bar L_{\da\db}$, $W_{\a\da}$ are simply given by the action of the on-shell Lorentz generators $l_{\a\b}$, $\bar l_{\da\db}$, $w_{\a\da}$ in dual superspace
\begin{align}
L_{\a\b}&=\sum_i\left(x_{i(\alpha}^{\da}\partial_{i\beta)\da}+\theta_{i(\alpha}^a\frac{\partial}{\partial\theta_i^{\beta)a}}\right)\,,&\bar{L}_{\da\db}&=\sum_i\left(x_{i(\da}^{\a}\partial_{i\db)\a}+\tilde\theta_{i(\da}^a\frac{\partial}{\partial\tilde\theta_i^{\db)a}}\right)\,,
\end{align}
and
\begin{equation}
W_{\a\da}=\sum_i\left(x_{\a\da}\frac{\partial}{\partial n}+2\,n\frac{\partial}{\partial x_{i}^{\dot{\alpha}\alpha}}+\tilde{\theta}^a_{i\,\da}\frac{\partial}{\partial \theta_i^{\a a}}-\theta^a_{i\,\a}\frac{\partial}{\partial \tilde\theta_i^{\da a}}\right)
\end{equation}
making the relation of $W_{\a\da}$ to the Lorentz rotations $l_{\mu 5}$ more obvious than in on-shell superspace. The dual dilatation is given by
\begin{equation}
D=-\tfrac{1}{2}\sum_i\bigl[2x_i^{\da\a}\partial_{i\,\a\da}+2n\frac{\partial}{\partial n}+{\theta}^{\a\,a}_{i}\frac{\partial}{\partial \theta_i^{\a a}}+\tilde{\theta}^{\da\,a}_{i}\frac{\partial}{\partial \tilde\theta_i^{\da a}}\bigr] 
\end{equation}
and acts covariantly on the amplitude
\begin{equation}
 D f_n=n\,f_n\,.
\end{equation}
From the six-dimensional superamplitude we know that the massive tree amplitudes are covariant under dual conformal inversion
\begin{equation}\label{eq:inversionMassive}
I[f_n]=\left(\prod_i (x_i^2-n_i^2)\right) f_n\,,
\end{equation}
and we only need to find the representation of the dual conformal boost generator in the dual variables \cref{eq:dualMassive}. We emphasize that in order to obtain the correct expression for the $\mu=0,1,2,3$ components of the dual conformal boost generator we cannot simply plug the 4$d$ variables into the expression for $K^{AB}$ given in \cref{eq:K6d} since this leads to the wrong result. The four-dimensional spinor variables solve the constraint \eqref{eq:6dspinorConstraint} on the six-dimensional spinors and thus spoil the assumed independence of chiral and anti-chiral spinors $\frac{\partial \tilde\lambda_A}{\partial \lambda^B}=0$ in the six-dimensional representation of the dual conformal boost generator $K^{AB}$. 

Since there is no obstacle in translating the inversion rules of the six-dimensional dual momenta \eqref{eq:inversion6d_first}, one possibility to obtain the action of the dual conformal boost generator $K_{\a\db}=IP_{\b\da}I$ in the full superspace is to start with the inversion rules for the bosonic dual variables
\begin{align}
I[x_{\a\db}]&=-\frac{x_{\b\da}}{x^2-n^2}\,,& I[n]&=\frac{n}{x^2-n^2}\,,
\end{align}
and extend the corresponding part of the dual conformal boost generator $K_{\a\da}$ acting only on the bosonic dual variables
\begin{equation}\label{eq:Kx}
K_{\alpha\dot\alpha}\bigr\rvert_{x,n}=\sum_{i}\left( x_{i\, \alpha\dot\gamma}\,x_{i\, \dot\alpha\gamma}\,\frac{\partial}{\partial
x_{i\, \gamma\dot\gamma}} + x_{i\, \alpha\dot\alpha}\, n_{i}\frac{\partial}{\partial n_{i}} + n_{i}^2\,
\, \frac{\partial}
{\partial x_i^{\dot\alpha\alpha}}\right)
\end{equation}
 such that it commutes with the constraints \eqref{eq:constraint1} to \eqref{eq:constraint7}. Note that the additional minus sign in the inversion rules for $n$ originates from the six-dimensional mostly minus metric $\eta_{55}=-1$.

Requiring that the dual conformal generator $K_{\alpha\dot\alpha}\bigr\rvert_{x,n}$ commutes with the bosonic constraints \eqref{eq:constraint1} to \eqref{eq:constraint3} leads to
\begin{align}\label{eq:Kboson}
K_{\alpha\dot\alpha}^{\text{boson}}&=K_{\alpha\dot\alpha}\bigr\rvert_{x,n}+\tfrac{1}{2}\sum_i\biggl[(x_i+x_{i+1})^\beta_{\phantom{\beta}\dot\alpha}\;l_{i\,\alpha\beta}+(x_i+x_{i+1})^{\phantom{\alpha}\dot\beta}_\alpha\; \bar l_{i\,\dot\alpha\dot\beta} \notag\\
&\=\phantom{K_{\alpha\dot\alpha}^{\text{$x$ space}}+\sum_i\biggl[}+(x_i+x_{i+1})_{\alpha\dot\alpha}\,(d_i-1)+(n_i+n_{i+1})\,w_{i\,\alpha\dot\alpha}\biggr]
\end{align}
Since $K_{\alpha\dot\alpha}^{\text{boson}}$ has a non-vanishing commutator with the right hand side of the fermionic constraints \eqref{eq:constraint4} and \eqref{eq:constraint7}, we have to introduce the following fermionic terms:
\begin{align}\label{eq:Kfermion}
K_{\alpha\dot\alpha}^{\text{fermion}}&=\sum_i\biggl[\,\theta^a_{i\,\alpha}\,x_{i\,\beta\dot\alpha}\,\frac{\partial}{\partial\theta^a_{i\,\beta}}+\,\tilde\theta^a_{i\,\dot\alpha}\,x_{i\,\alpha\dot\beta}\,\frac{\partial}{\partial\tilde\theta^a_{i\,\dot\beta}}+\,n_i\,\tilde\theta^a_{i\,\dot\alpha}\,\frac{\partial}{\partial\theta_{i}^{\alpha\,a}}+\, n_i\,\theta^a_{i\,\alpha}\,\frac{\partial}{\partial\tilde\theta_{i}^{\dot\alpha\,a}}\notag\\
&\=\phantom{\sum_i\biggl[}+\tfrac{1}{2}(\theta_{i}+\theta_{i+1})_\alpha^a\,\bar q_{i\,\dot\alpha\,a}+\tfrac{1}{2}(\theta_{i}+\theta_{i+1})_{\dot\alpha}^a\,\bar{\tilde{q}}_{i\,\alpha\,a}\biggr]
\end{align}
Their sum $K_{\alpha\dot\alpha}=K_{\alpha\dot\alpha}^{\text{boson}}+K_{\alpha\dot\alpha}^{\text{fermion}}$ commutes with all constraints. The part of $K_{\alpha\dot\alpha}$ acting on the on-shell variables $\{\lambda_{i}, \tilde\lambda_{i}, \mu_{i},\tilde\mu_{i};
\zeta_{i},\bar\zeta_{i}\}$ is given by 
\begin{align}\label{eq:K_onshell}
 %\begin{aligned}
  \hspace{-.3cm}K_{\alpha\dot\alpha}\bigr\rvert_{\text{on-shell}}&=\tfrac{1}{2}\sum_i\Bigl[ (x_i+x_{i+1})^\beta_{\phantom{\beta}\dot\alpha}\;l_{i\,\alpha\beta}+(x_i+x_{i+1})^{\phantom{\alpha}\dot\beta}_\alpha\; \bar l_{i\,\dot\alpha\dot\beta}+(n_i+n_{i+1})\,w_{i\,\alpha\dot\alpha}\hspace{.6cm}\\*
&\=\phantom{\tfrac{1}{2}\sum_i}{}+(x_i+x_{i+1})_{\alpha\dot\alpha}\,(d_i-1)+(\theta_{i}+\theta_{i+1})_\alpha^a\,\bar q_{i\,\dot\alpha\,a}+(\theta_{i}+\theta_{i+1})_{\dot\alpha}^a\,\bar{\tilde{q}}_{i\,\alpha\,a}\Bigr]\,.\notag
 %\end{aligned}
\end{align}
The representation of $K_5=IMI$ in four-dimensional variables may be obtained in a similar way or by Lorentz rotation $[W_{\a\da},K_{\b\db}]=\epsilon_{\a\b}\epsilon_{\da\db}K_{5}$ of $K_{\alpha\dot\alpha}$. The representations of $K_{\alpha\dot\alpha}$ and $K_5$ in dual superspace are
\begin{align}
K_{\a\da}&=\sum_{i}\begin{aligned}[t]\biggl[&  x_{i\, \alpha\dot\gamma}\,x_{i\, \dot\alpha\gamma}\,\frac{\partial}{\partial
x_{i\, \gamma\dot\gamma}} + x_{i\, \alpha\dot\alpha}\, n_{i}\frac{\partial}{\partial n_{i}}
 + n_{i}^2\,
\, \frac{\partial}
{\partial x_i^{\dot\alpha\alpha}}\label{eq:K}\\
&+\theta^a_{i\,\alpha}\,x_{i\,\beta\dot\alpha}\,\frac{\partial}{\partial\theta^a_{i\,\beta}}+\,\tilde\theta^a_{i\,\dot\alpha}\,x_{i\,\alpha\dot\beta}\,\frac{\partial}{\partial\tilde\theta^a_{i\,\dot\beta}}+\,n_i\,\tilde\theta^a_{i\,\dot\alpha}\,\frac{\partial}{\partial\theta_{i}^{\alpha\,a}}-\, n_i\,\theta^a_{i\,\alpha}\,\frac{\partial}{\partial\tilde\theta_{i}^{\dot\alpha\,a}}\biggr]\,,\end{aligned}\\
K_5&=\sum_{i}\begin{aligned}[t]\biggl[& n_i^2\,\frac{\partial}{\partial
n_{i}} + 2\,n_{i}\,x_i^{\dot\alpha\alpha} \frac{\partial}
{\partial x_i^{\dot\alpha\alpha}}+x_i^{2}\,\frac{\partial}{\partial n_i}
\\&+\theta^{\a \,a}_{i}\,x_{i\,\a\db}\,\frac{\partial}{\partial\tilde\theta^a_{i\,\db}}+\,\tilde\theta^a_{i\,\dot\alpha}\,x_{i}^{\da\beta}\,\frac{\partial}{\partial\theta^{\b a}_{i}}+\,n_i\,\theta^{\a\,a}_{i}\,\frac{\partial}{\partial\theta_{i}^{\alpha\,a}}+\, n_i\,\tilde\theta^{a\,\da}_{i}\,\frac{\partial}{\partial\tilde\theta_{i}^{\dot\alpha\,a}}\biggr]\label{eq:K5}\,.\end{aligned}
\end{align}
and the action of $K_5$ on the on-shell variables is given by
\begin{multline}\label{eq:K5onshell}
 K_5\bigr\rvert_{\text{on-shell}}=\tfrac{1}{2}\sum_i\biggl[w_{i\,\a\da}(x_i+x_{i+1})^{\da\a}+2(d_i-1)(n_i+n_{i+1})\\-(\tilde{\theta}_i-\tilde{\theta}_{i+1})^{\da\,a}\bar{q}_{\da\,a}+({\theta}_i-{\theta}_{i+1})^{\a\,a}\bar{\tilde q}_{\a\,a}\biggr]
\end{multline}
The dual superconformal generators
\begin{align}
\bar{S}_{\da\,a}&=\sum_i x_{i\,\a\da}\frac{\partial}{\partial \theta_{i\,\a}^a}-n_i\frac{\partial}{\partial \theta_{i}^{\da\,a}}\,,&\bar{\tilde S}_{\a\,a}&=\sum_i x_{i\,\a\da}\frac{\partial}{\partial \tilde\theta_{i\,\da}^a}+n_i\frac{\partial}{\partial \theta_{i}^{\a\,a}}\,.
\end{align}
can be obtained from the commutators of  $K_{\a\da}$ with the dual supermomenta $Q_a^{\b}$ and $\tilde{Q}_a^{\db}$. In full superspace they coincide with the supersymmetry generators $\bar{q}_{\da\,a}$, $\bar{\tilde q}_{\a\,a}$
\begin{align}
 \bar{S}_{\da\,a}&=\bar{q}_{\da\,a}\,,& \bar{\tilde{S}}_{\da\,a}&=\bar{\tilde{q}}_{\a\,a}\,,
\end{align}
similar to the massless case. The dual conformal algebra reads 
\begin{equation}\label{eq:algebraDualMassive}
\begin{gathered}
 \begin{aligned}[t]
[M,K_{\a\da}]&=W_{\a\da}&[M,K_5]&=-2\,D \\
[W_{\a\da},K_{\b\db}]&=\epsilon_{\a\b}\epsilon_{\da\db}K_{5}&[W_{\a\da},K_5]&=2K_{\a\da}\\
[K_{\a\da},Q_a^{\b}]&=\delta_{\a}^{\b}\bar{S}_{\da\,a}&[K_{\a\da},\tilde{Q}_a^{\db}]&=\delta_{\da}^{\db}\bar{\tilde{S}}_{\a\,a}\\
[K_{5},Q_a^{\a}]&=-\bar{\tilde{S}}_{\a\,a}&[K_{5},\tilde{Q}_a^{\da}]&=\bar{S}_{\da\,a}\\
 \end{aligned}\\
[K_{\a\da},P^{\db\b}]=\delta_{\a}^{\b}\delta_{\da}^{\db}D+\delta_{\a}^{\b}\bar{L}_{\da}^{\;\;\db}+\delta_{\da}^{\db}{L}_{\a}^{\;\;\b}
\end{gathered}
\end{equation}
along with the generic $[D,J]=\dim(J)J$ for all generators $J$.
We omitted all commutators that are either vanishing or equal to the corresponding commutators in the on-shell algebra \cref{eq:OnShellAlgebra}.
The action of the $R$-symmetry charges $r_a$ in dual superspace are given by
\begin{align}
 R_1&=\sum_i \left(\theta_{i\a}^1\frac{\partial}{\partial \theta_i^{\a\,1}}+\tilde\theta_{i\da}^1\frac{\partial}{\partial \tilde\theta_i^{\da\,1}}\right) -n+4\,&R_2&=\sum_i \left(\theta_{i\a}^2\frac{\partial}{\partial \theta_i^{\a\,2}}+\tilde\theta_{i\da}^2\frac{\partial}{\partial \tilde\theta_i^{\da\,2}}\right) -n+4\,,
\end{align}
with the non-vanishing commutators
\begin{align}
 [R_a,Q_{b}]&=-\delta_a^b Q_{b}\,,&[R_a,\tilde{Q}_{b}]&=-\delta_a^b \tilde{Q}_{b}\,,&[R_a,\bar{S}_{b}]&=-\delta_a^b \bar{S}_{b}\,,&[R_a,\bar{\tilde S}_{b}]&=-\delta_a^b \bar{\tilde S}_{b}\,.
\end{align}

Some further remarks are in order here. as we already mentioned, the generator $w_{\a\da}$
arises from the Lorentz-generators $l^{\mu 5}$, just as $m$ is related to the momentum in the extra dimensional direction $p^{5}$. 
As has been shown in \cite{Dennen:2010dh}, if the loop momentum is restricted to be four-dimensional, which is equivalent to the Higgs regularization described in \cite{Alday:2009zm}, the cut constructible parts of the loop amplitudes invert as
\begin{equation}\label{eq:inversionLoop}
 I\left[\int\left(\prod_i^Ld^4x_{l_i}\right)\mathcal{I}_n^L\right]=\left(\prod_i^nx_i^2\right) \int\left(\prod_i^Ld^4x_{l_i}\right)\mathcal{I}_n^L\,.
\end{equation}
Due to the four dimensional loop momenta, the five dimensional Lorentz invariance as well as the dual translation invariance in the $x^{5}$ direction are lost. Hence, $w_{\a\da}$ is a manifest symmetry of the tree-superamplitudes but no symmetry of the Higgs regularized loop amplitudes. Since the dual conformal boost generator is given by $K^\mu=IP_\mu I$, the inversion properties \eqref{eq:inversionLoop}  only imply that $(K^\mu+2\sum_i x_i^\mu)$ is a symmetry of the regularized loop amplitudes for $\mu=0,1,2,3$, whereas the tree-amplitudes have the full five-dimensional dual conformal symmetry.
\subsubsection{Yangian symmetry}

The obvious question now arises: Can one reinterpret the dual conformal operator in six dimensions
as a level-one Yangian generator in a four dimensional massive theory? To answer this we proceed in great analogy to the work \cite{Drummond:2009fd} where a Yangian symmetry 
of tree superamplitudes was established for ${\mathcal N}=4$ SYM as reviewed in\cref{section:Yangian}.
We continue by translating the expression for $K_{\alpha\dot\alpha}+\sum_i x_{i\,\alpha\dot\alpha}$ to four dimensional on-shell variables. Inserting
\begin{align}
 x_i^{\da\a}&=x_1^{\da\a}-\sum_{j=1}^{i-1}p_j^{\da\a}&n_i&=n_1-\sum_{j=1}^{i-1}m_j\\
\theta^a_{i\,\a}&=\theta^a_{1\,\a}-\sum_{j=1}^{i-1}q^a_{i\,\a}&\theta^a_{i\,\da}&=\theta^a_{1\,\da}-\sum_{j=1}^{i-1}\tilde{q}^a_{i\,\da}
\end{align}
into the part  of the dual conformal boost generator acting on the on-shell variables \cref{eq:K_onshell}, one finds the non-local result
\begin{align}
K_{\alpha\dot\alpha}\+\sum_i x_{i\,\alpha\dot\alpha}&=-\sum_{j<i}\biggr[p_{j\phantom{\beta}\dot\alpha}^{\phantom{j}\beta}l_{i\,\alpha\beta}+p_{j\,\alpha}^{\phantom{j\,\alpha}\dot\beta}\bar l_{i\,\dot\alpha\dot\beta}+p_{j\,\alpha\dot\alpha}d_i+m_j w_{i\,\alpha\dot\alpha}+q_{j\,\dot\alpha}^a\bar{\tilde{q}}_{i\,\alpha\, a}+q_{j\,\alpha}^a\bar q_{i\,\dot\alpha\, a}\biggr]\notag\\
&\=-\tfrac{1}{2}\sum_{i=1}^n\biggr[p_{i\phantom{\beta}\dot\alpha}^{\phantom{i}\beta}l_{i\,\alpha\beta}+p_{i\,\alpha}^{\phantom{i\,\alpha}\dot\beta}\bar l_{i\,\dot\alpha\dot\beta}+p_{i\,\alpha\dot\alpha}d_i+m_i w_{i\,\alpha\dot\alpha}+q_{i\,\dot\alpha}^a\bar{\tilde{q}}_{i\,\alpha\, a}+q_{i\,\alpha}^a\bar q_{i\,\dot\alpha\, a}\biggr]
\label{Kadaonshell}
\end{align}
Here we dropped the terms
\begin{align}\label{eq:drop}
+\,(x_1)_{\dot\alpha}^{\phantom{\dot\alpha}\beta}l_{\alpha\beta}+\,(x_1)_{\alpha}^{\phantom{\alpha}\dot\beta}\bar l_{\dot\alpha\dot\beta}+\,(x_{1})_{\alpha\dot\alpha}\,d+\, n_1\, w_{\a\dot\alpha}
+(\theta_{1})_\alpha^a\,\bar q_{\dot\alpha\,a}+(\theta_{1})_{\dot\alpha}^a\,\bar{\tilde{q}}_{\alpha\,a}+\tfrac{1}{2}p_{\a\da}
\end{align}
which annihilate the tree amplitudes on their own because they are each proportional to symmetry generators. Since the tree superamplitude is independent of $x_1,\theta_1$, $n_1$ and $K_{\alpha\dot\alpha}+\sum_i x_{i\,\alpha\dot\alpha}$ annihilates it, one could also apply the reverse logic by concluding from \eqref{eq:drop} that $d, l_{\alpha\beta}, \bar l_{\dot\alpha\dot\beta}, w_{\a\dot\alpha}, \bar q_{\dot\alpha\,a}, \bar{\tilde{q}}_{\alpha\,a}$ are symmetries of the tree amplitudes. The Higgs regularized loop amplitudes explicitly depend on $n_1$ and are not invariant under 
$w_{\a\dot\alpha}$. Consequently, the term  $n_1\, w_{\a\dot\alpha}$ cannot be dropped at loop level.

Let us proceed by investigating the structure of the dual conformal boost generator in on-shell variables a bit further. Upon adding to $(K_{\a\da} +\sum_{i}x_{i\,\a\da})$ of \cref{Kadaonshell} the quantity
\begin{equation}
\Delta K_{\a\da} = \,
\tfrac{1}{2}\biggr[p_{\dot\alpha}^{\beta}\, l_{\alpha\beta}+p_{\alpha}^{\dot\beta}
\,\bar l_{\dot\alpha\dot\beta}+p_{\alpha\dot\alpha}\,
d+ m\, \bar w_{\alpha\dot\alpha}
+ q^{a}{}_{\da}\, {\bar q}_{\a a} + q^{a}{}_{\a}\,
\bar q_{\da a} 
\biggr] \, ,
\end{equation}
which is a manifest symmetry of the super-amplitudes, as $\{p_{\a\da},m,l_{\a\b},\bar l_{\da\db}
\}$ annihilate it, we find the bi-local representation of the level-one $p^{(1)}_{\a\da}$ generator,
\begin{align}
p^{(1)}_{\a\da} &=  K_{\a\da}+ \Delta K_{\a\da} +\sum_{i}x_{i\,\a\da} \nn\\
&=- \tfrac{1}{2}\sum_{j<i}\, \Biggl [ p_{j}^{\b\db}\, (\epsilon_{\da\db}\, l_{i\, \a\b}
+ \epsilon_{\a\b}\, \bar l_{i\, \da\db} + \epsilon_{\a\b}\, \epsilon_{\da\db}\, d_{i}\, )
+ m_{j}\, w_{i\, \a\da}\nn\\ &\=\phantom{- \sum_{j<i}\, \Biggl [}
+ q_{j}^{a}{}_{\da}\, {\bar q}_{i\, \a a} + q_{j}^{a}{}_{\a}\,
\bar q_{i\, \da a} - (i\leftrightarrow j) \Biggr ]\, .
\end{align}
which indeed obeys a level-one Yangian like relation, \cref{eq:commutatorsLevel1},
\begin{equation}
[\, w_{\a\da}, p^{(1)}_{\b\db}\, ] = 2\, \epsilon_{\a\b}\, \epsilon_{\da\db}\, m^{(1)}\, ,
\end{equation}
giving rise to the novel level one generator
\begin{equation}
m^{(1)} = - \tfrac{1}{4}\sum_{j<i} \Biggl [ p_{j}^{\g\dg}\, w_{i\, \g\dg} + 2m_{j}\, d_{i}
+ q^{a\g}_{j}\, {\bar q}_{i\, \g a}+ q_{j}^{a}{}_{\dg}\, {\bar q}_{i}^{\dg}{}_{a} 
- (i\leftrightarrow j)\Biggr]\, .
\end{equation}
One checks that it indeed obeys the commutation relation
\begin{equation}
[\, w_{\a\da}, m^{(1)}\, ] = p^{(1)}_{\a\da}\, .
\end{equation}
We note that $m^{(1)}$ can also be obtained from the action  of $K_5$ on the on-shell variables \eqref{eq:K5onshell} in the same way as $ p^{(1)}$ has been obtained from $K_{\a\da}$ in \eqref{eq:Kx}.

A natural question to be addressed in future work is whether or not there exist the level-one fermionic generators $q_{a\,\a}^{(1)}$, $q_{a\,\da}^{(1)}$. However, already at 
this point it is clear that the non-local symmetry generators found will not lift to
the complete super Poincar\'e algebra but rather stay confined to the super-translational
piece. In particular there will be no level-one $w^{(1)}_{\a\da}$ symmetry generator.

\section{BCFW on-shell recursion relations for tree-level amplitudes}
\subsection{General remarks}
The BCFW on-shell recursion \cite{Britto:2004ap, Britto:2005fq} is a valuable tool in calculating color ordered tree-level amplitudes in gauge theories, as it allows to recursively calculate an $n$ point amplitude from lower point amplitudes. As a direct consequence, the knowledge of the three point amplitudes and the BCFW recursion relation are sufficient to obtain all color ordered tree amplitudes of a particular gauge theory. In what follows we will briefly outline the general form of the BCFW recursion, for some more details we refer to to the excellent review \cite{Bern:2007dw}. 

The basic idea is to analytically continue two external momenta by introducing light-like shifts proportional to the complex parameter $z$ that neither spoil the on-shell condition of the two shifted momenta nor the overall momentum conservation. If the shift vector $r$ has the properties
\begin{align} \label{eq:shiftvector}
r^2 &= 0\,, &r \cdot p_1 &= 0\,,& r \cdot p_n &= 0\,,  
\end{align}
then the shift
\begin{align}\label{eq:shifts}
p_1\rightarrow p_{\hat{1}}(z) &= p_1 + z r\,,& p_n\rightarrow p_{\hat{n}}(z) &= p_n - z r\,,&A_n&\rightarrow \widehat{A}_n(z)\,,
\end{align}
has the desired properties
\begin{align}
p_{\hat{1}}^2 &= p_{\hat{n}}^2 = 0&  p_{\hat{1}} + p_{\hat{n}} &= p_1 + p_n\,.
\end{align}
Using region momenta instead, the shifts in \cref{eq:shifts} can be reproduced by the single shift
\begin{equation}
 x_1\rightarrow x_{\hat{1}}=x_1+z\,r\,.
\end{equation}
Color ordered tree amplitudes have simple analytic structure since they only have poles where sums of consecutive momenta go on-shell, i.\,e.~$x_{i\,j}^2=0$. As a consequence $\widehat{A}_n(z)$ is an analytical function  that has only the simple poles $z_j$ solving the on-shell condition 
\begin{align}
x_{\hat{1}\,j+1}^2=(x_{1\,j+1}+z\,r )^2&=x_{1\,j+1}^2+2z \,r\cdot x_{1\,j+1}=0\,,
\end{align}
i.\,e. the poles are given by
\begin{equation}\label{eq:poles}
z_{j} = -\frac{x^2_{1\,j+1}}{2r\cdot x_{1\,j+1}}\,.
\end{equation}
If the analytically continued amplitude $\widehat{A}_n$ is vanishing as $|z|\rightarrow \infty$ it is a simple fact that the contour integral of $\frac{\widehat{A}_n}{z}$ over a circle at infinity is vanishing. By virtue of the residue theorem this allows to relate the physical amplitude to the residues of $\frac{\widehat{A}_n}{z}$ at the poles $z_j$
\begin{equation}\label{eq:residues}
 \frac{1}{2\pi i}\oint d z\frac{\widehat{A}_n}{z}=A_n+\sum_{j=2}^{n-2}\mathop{\mathrm{Res}}_{z=z_j}\frac{\widehat{A}_n}{z}=0\,.
\end{equation}
Due to the general factorization properties of tree amplitudes, these residues are given by products of lower-point on-shell amplitudes multiplied by the residue
\begin{equation}
 -\mathop{\mathrm{Res}}_{z=z_j}\left(\frac{1}{z}\frac{i}{x_{\hat{1}\,j+1}^2}\right)=\frac{i}{x_{1\,j+1}^2}\,.
\end{equation}
Introducing the abbreviations
\begin{align}\label{eq:Def_Pj}
 \hat{P}_j&=x_{\hat{1}\,j+1}\,,&P_j=x_{1\,j+1}\,,
\end{align}
the final form of the BCFW on-shell recursion is 
\begin{equation} \label{eq:BCFW}
A_n= \sum_{j = 2}^{n-2} \sum_{h} A_{j+1}(p_{\hat{1}}, p_2,\dots,p_j, -\hat{P}^{(-h)}_{j}) \frac{i}{P^2_{j}}  A_{n-j+1}(\hat{P}^{(h)}_{j}, p_{j+1},\dots,p_{\hat{n}})\Biggr\rvert_{z=z_j}
\end{equation}
where the sum goes over all poles $z_j$ and over all helicities of the intermediate states. Note that we assumed the vanishing of $\widehat{A}_n(z)$ for large $z$ to derive the recursion relation which is not a general feature for all gauge theories and all possible shifts.  For details we refer to \cite{Cheung:2008dn} and \cite{ArkaniHamed:2008yf}.

In the following sections we will derive supersymmetric versions of the BCFW recursion \cref{eq:BCFW} for the four-dimensional $\cN=4$ SYM theory and the six-dimensional $\cN=(1,1)$ SYM theory.
\subsection{Supersymmetric BCFW for $\mathcal{N}=4$ SYM in non-chiral superspace}\label{section:BCFWnonChiral}
As it has not been done in the literature before, we are going to present the BCFW recursion in the non-chiral super space $\{\lambda_i^\alpha,\tilde\lambda_i^{\dot\alpha},\eta_i^m,\tilde\eta_i^{m'}\}$, introduced in \cref{section:superamps4d}. Additionally we will use the BCFW recursion to prove the postulated covariance in eq.~\eqref{eq:Inversion_Amp_NC} of the non-chiral superamplitudes under the dual conformal inversions \eqref{eq:inversion4dNC}, as well as to calculate the four-, five- and six-point superamplitudes.

 Based on the previous section it is straightforward to write down a set of shifts preserving both bosonic and fermionic momentum conservation
\begin{align}
\lambda_1\rightarrow \lambda_{\hat{1}}(z) &=  \lambda_1 + z  \lambda_n \,,& \tilde{\lambda}_n\rightarrow\tilde{\lambda}_{\hat{n}}(z) &= \tilde{\lambda}_n - z \tilde{\lambda}_1\,,\\
 \eta_n\rightarrow\eta_{\hat{n}}(z) &= \eta_n - z \eta_1\,,&\tilde{\eta}_{1}\rightarrow \tilde{\eta}_{\hat{1}}(z) &= \tilde{\eta}_{1} + z \tilde{\eta}_{n}\,,
\end{align}
leading to the poles, \cref{eq:poles}, 
\begin{equation}
 z_{j} = - \frac{x_{1\, j+1}^2}{\langle n| x_{1\, j+1}| 1]}\,.
\end{equation}
of the shifted superamplitude. The corresponding dual shifts are
\begin{align}
x_{\hat{1}}^{\dot\alpha\alpha}&=x_1 ^{\dot\alpha\alpha}+z\, \tilde\lambda^{\dot\alpha}_1\lambda^\alpha_n\,,&
\tilde{\theta}_{\hat{1}}^{\dot{\alpha}\, m'} &=  \tilde{\theta}_{1}^{\dot{\alpha}\,m'} + z \tilde{\lambda}^{\dot{\alpha}}_1 \tilde{\eta}_{n}^{m'}\,,&
\theta_{\hat{1}\,m}^{\alpha}&=\theta_{1\,m}^{\alpha}+ z\,  \lambda^{\alpha}_n  \eta_1^m\,.
\end{align}
According to the same arguments as in chiral and anti-chiral superspace, the BCFW recursion in non-chiral superspace is given by
\begin{equation}\label{eq:BCFW_non-chiral}
\mathcal{A}_n = \sum_{j = 2}^{n-2} \int d^2 \eta_{\hat{P}_j}\int d^2 \tilde\eta_{\hat{P}_j}  \mathcal{A}_{j+1}(p_{\hat{1}},\dots,p_j,-\hat{P}_j)\frac{-i}{P^2_{j}}\mathcal{A}_{n-j+1}(\hat{P}_j, p_{j+1},\dots,p_{\hat{n}})\Biggr\rvert_{z=z_j}\,,
\end{equation}
with the explicit minus sign originating from the definition of the Grassmann integration measures $d^2\eta=\tfrac{1}{2}d\eta^m d\eta_{m}$, $d^2 \tilde\eta=\tfrac{1}{2}d\tilde\eta_{m'}d \tilde\eta^{m'}$.
The starting point for this recursion can be obtained by a half Fourier transform of the MHV and $\overline{\text{MHV}}$ three point amplitudes in chiral or anti-chiral superspace yielding
\begin{align}
\mathcal{A}_3^{\text{MHV}} &=-i \frac{\delta^4(p) \delta^4(q) \delta^2(\tilde{\eta}_{1} \langle2 3\rangle + \tilde{\eta}_{2} \langle3 1\rangle + \tilde{\eta}_{3} \langle1 2\rangle)}{\langle1 2\rangle \langle2 3\rangle \langle3 1\rangle}\,,\\
\mathcal{A}_3^{\overline{\text{MHV}}} &=i \frac{\delta^4(p)\delta^4(\tilde q) \delta^2(\eta_1 [2 3] + \eta_2 [3 1] + \eta_3 [1 2] ) }{[1 2] [2 3] [3 1]}\,,
\end{align} 
where the two dimensional delta functions of objects $\chi_m$, $\tilde\chi_{m'}$ carrying Grassmann indices have the definition $\delta^2(\chi_m)=\tfrac{1}{2}\chi_m\chi^m$, $\delta^2(\tilde\chi_{m'})=\tfrac{1}{2}\tilde\chi^{m'}\chi_{m'}$ such that $\int d^2 \eta\, \delta^2(\eta)=\int d^2 \tilde\eta \,\delta^2(\tilde\eta)=1$.
We recall from eq.~(\ref{325}) that the superamplitudes with $n>3$ partons in non-chiral superspace have the form
\begin{equation}
\mathcal{A}_n=\delta^4(q)\delta^4(\tilde q)f_n(\{x_{ij},\theta_{ij},\tilde\theta_{ij}\}) 
\end{equation}
i.\,e.~the only $\eta_{\hat{P}_j}$, $\tilde\eta_{\hat{P}_j}$ dependence of the integrand in the BCFW recursion \cref{eq:BCFW_non-chiral} originates from delta functions of the three point amplitudes and the delta functions of the fermionic momenta, making the Grassmann integrations straightforward. For the four point amplitude we obtain
\begin{equation}\label{eq:A4nonchiral}
\mathcal{A}_4=-i\delta^4(q)\delta^4(\tilde q)\frac{1}{x_{13}^2x_{24}^2}\,,
\end{equation}
In agreement with \cite{Huang:2011um}.
Introducing the definitions
\begin{align}
|B_{ijk}\rangle&=x_{ij}x_{jk}|\theta_{ki}\rangle+x_{ik}x_{kj}|\theta_{ji}\rangle\,,& |\tilde B_{ijk}]&=x_{ij}x_{jk}|\tilde\theta_{ki}] +x_{ik}x_{kj}|\tilde\theta_{ji}]
\end{align}
we present the results of the Grassmann integrations in \cref{eq:BCFW_non-chiral} for the three different cases $j=2$, $2<j<n-2$ and $j=n-2$. In the case $j=2$ the left superamplitude has to be $\mathcal{A}_3^{\overline{\text{MHV}}}$ since $\mathcal{A}_3^{\text{MHV}}$ 
does not exist for the three point kinematics of this case. We obtain 
\begin{equation}\label{eq:B1}
\begin{aligned}
\mathcal{B}_2&=\int d^2 \eta_{\hat{P}_2}\int d^2 \tilde\eta_{\hat{P}_2}  \mathcal{A}_3^{\overline{\text{MHV}}}(p_{\hat{1}},p_2,-\hat{P}_2)\frac{i}{P^2_{2}}\mathcal{A}_{n-1}(\hat{P}_2, p_{3},\dots,p_{\hat{n}})\Biggr\rvert_{z=z_2}\\
&=\frac{\delta^4(q)\delta^4(\tilde{q})[1\,2]\delta^2([P_2|\tilde{\theta}_{\hat{1}3}])f_{n-1}(x_{\hat{1}},x_3,\dots,x_n)}{x_{2n}^2 x_{13}^2[1\,P_2] [2\,P_2]}\biggr\rvert_{z=z_2}\,.
\end{aligned}
\end{equation}
 For practical applications it is convenient to rewrite $\mathcal{B}_2$ as
\begin{equation}
\mathcal{B}_2=\frac{\delta^4(q)\delta^4(\tilde{q})\delta^2([1|\tilde{B}_{13n}])f_{n-1}(x_{\hat{1}},x_3,\dots,x_n)\bigr\rvert_{z=z_2}}{x_{2n}^2 x_{13}^2[1|x_{13}|n\rangle [1\,n]}\,.
\end{equation}
For $2<j<n-2$ we have 
\begin{multline}\label{eq:B2}
\mathcal{B}_j= \int d^2 \eta_{\hat{P}_j}\int d^2 \tilde\eta_{\hat{P}_j}  \mathcal{A}_{j+1}(p_{\hat{1}},\dots,p_j,-\hat{P}_j)\frac{i}{P^2_{j}}\mathcal{A}_{n-j+1}(\hat{P}_j, p_{j+1},\dots,p_{\hat{n}})\Biggr\rvert_{z=z_j}=\\
 \frac{i\,\delta^4(q)\delta^4(\tilde{q})\delta^2([P_j|\tilde{\theta}_{\hat{1}j+1}])\delta^2(\langle P_j|\theta_{\hat{1} j+1}])f_{j+1}(x_{\hat{1}},\dots,x_{j+1})f_{n-j+1}(x_{\hat{1}},x_{j+1},\dots,x_n)\bigr\rvert_{z=z_j}}{x_{1j+1}^2}\,.
\end{multline}
For practical applications it is more convenient to use the following expression for $\mathcal{B}_j$
\begin{equation}
\frac{i\,\delta^4(q)\delta^4(\tilde{q})\delta^2([1|\tilde{B}_{1j+1n}])\delta^2(\langle 2|B_{21j+1}])f_{j+1}(x_{\hat{1}},\dots,x_{j+1})f_{n-j+1}(x_{\hat{1}},x_{j+1},\dots,x_n)\bigr\rvert_{z=z_j}}{x_{1j+1}^2\langle n|x_{1j+1}|1]^2[n\,1]^2\ang{1}{2}^2}\,.\!\!
\end{equation}
In the case $j=n-2$ the right superamplitude has to be $\mathcal{A}_3^{\text{MHV}}$ due to the special three point kinematics and the integration gives
\begin{equation}\label{eq:B3}
\begin{aligned}
\mathcal{B}_{n-2} &=\int\!\! d^2 \eta_{\hat{P}_{n-2}}\int\!\! d^2 \tilde\eta_{\hat{P}_{n-2}}  \mathcal{A}_{n-1}(p_{\hat{1}},\dots,p_{n-2},-\hat{P}_{n-2})\frac{i}{P^2_{n-2}}\mathcal{A}^{\text{MHV}}_{3}(\hat{P}_{n-2}, p_{n-1},p_{\hat{n}})\Biggr\rvert_{z=z_{n-2}}\\
&=\frac{\delta^4(q)\delta^4(\tilde{q})\ang{n}{n-1}\delta^2(\langle P_{n-2}|\theta_{\hat{1}n-1}])f_{n-1}(x_{\hat{1}},x_2,\dots,x_{n-1})\bigr\rvert_{z=z_{n-2}}}{x_{1n-1}^2\ang{P_{n-2}}{n-1}\ang{P_{n-2}}{n}}\,,
\end{aligned}
\end{equation}
which may be rewritten as
\begin{equation}
\mathcal{B}_{n-2} =\frac{\delta^4(q)\delta^4(\tilde{q})\delta^2(\langle 2|B_{21n-1}])f_{n-1}(x_{\hat{1}},x_2,\dots,x_{n-1})\bigr\rvert_{z=z_{n-2}}}{x_{1n-1}^2\langle n|x_{1n-1}|1]\ang{1}{2}^2[1\,n]}\,.
\end{equation}
Now the integrated non-chiral BCFW recursion relation reads
\begin{equation}\label{eq:BCFWnonChiralIntegrated}
 \mathcal{A}_n=\sum_{j=2}^{n-2}\mathcal{B}_{j}\,.
\end{equation}
In this form it is straightforward to prove the dual conformal symmetry of the non-chiral superamplitudes. Applying the inversion rules \cref{eq:inversion4dNC}, we find
\begin{equation}
\begin{aligned}
I\left(\,[j\,P_j]\,\right)&=\frac{[j\,P_j]}{x_{j+1}^2}\,,&I\left(\,[1\,P_j]\,\right)&=\frac{x_{\hat{1}}^2}{x_2^2x_{j+1}^2}[1\,P_j]\,,\\
I\left(\,\langle j+1\,P_j\rangle\,\right)&=\frac{\langle j+1\,P_j\rangle}{x_{\hat{1}}^2}\,,&I\left(\,\langle n\,P_j\rangle\,\right)&=\frac{\langle n\,P_j\rangle}{x_n^2}\,,\\
I\left(\,[P_{j}|\tilde{\theta}_{\hat{1}j+1}]\,\right)&=\frac{[P_{j}|\tilde{\theta}_{\hat{1}j+1}]}{x_{j+1}^2}\,,&I\left(\,\langle P_{j}|\theta_{\hat{1}j+1}\rangle\,\right)&=\frac{\langle P_{j}|\theta_{\hat{1}j+1}\rangle}{x_{\hat{1}}^2}\,.\end{aligned}
\end{equation}
Hence, it follows from \cref{eq:B1,eq:B2,eq:B3} together with the inductive assumption
on $I[f_{k<n}]$ of eqn.~(\ref{eq:inversionA6d}) that 
\begin{equation}
I[\mathcal{B}_j]=\left(\prod_i x_i^2\right)\mathcal{B}_j
\end{equation}
which proves the covariance \eqref{eq:Inversion_Amp_NC} of the non-chiral superamplitude  under the dual conformal inversions \eqref{eq:inversion4dNC}.

In order to obtain useful representations of the non-chiral superamplitudes from the integrated BCFW recursion \cref{eq:BCFWnonChiralIntegrated} it remains to remove the hats from the the shifted dual point $\hat{1}$ by using identities like e.\,g.
\begin{equation}
 x_{\hat{1}k\,}^2\bigr\rvert_{z=z_j}=-\frac{\langle n|x_{nj+1}x_{j+1k}x_{k1}|1]}{\langle n| x_{1\, j+1}| 1]}\,,
\end{equation}
or
\begin{equation}
 \langle B_{kk+1\hat{1}}|\,\bigr\rvert_{z=z_j}=\frac{1}{\langle 1|x_{1k}|k]\langle n|x_{nj+1}|1]}\Bigl(\begin{aligned}[t]
                                                                                                        &\langle n|x_{nj+1}x_{j+12}x_{2k}|k]\langle B_{kk+11}|\\
													 & \qquad\qquad+x_{1j+1}^2\langle n|x_{nk}|k]\langle B_{kk+12}|\Bigr)\,.
                                                                                                       \end{aligned}
\end{equation}
After removing all hats the obtained expression may still contain spinors. However, these spinors can be removed by multiplying and dividing with the chiral conjugate spinor brackets. The final expression will only depend on $\{x_i,\,\theta_i,\,\tilde\theta_i\}$ and besides $x_{ij}^2$ it can be expressed by the dual conformal covariant objects
\begin{equation}\label{eq:defBB}\begin{gathered}
\Tr(i_1 \dots i_{2k}):=\Tr(x_{i_1\,i_2}x_{i_2\,i_3}\,\dots\, x_{i_{2k-1}\,i_{2k}}x_{i_{2k}\,i_1})\,,\qquad\qquad\qquad x_{ij}^2 =-\tfrac{1}{2}\Tr(i\,j)\,,\\
\langle B_{ijk}|i_1\,\dots \,i_{2k+1}|B_{ijk}\rangle:=\tfrac{1}{2}\langle (B_{ijk})^m | x_{i\,i_1}x_{i_1\,i_2}\dots x_{i_{2k+1}\,i} | (B_{ijk})_m\rangle\\
[ \tilde B_{ijk}|i_1\,\dots \,i_{2k+1}|\tilde B_{ijk}]:=\tfrac{1}{2}[ (\tilde B_{ijk})_{m'} |x_{i\,i_1}x_{i_1\,i_2}\dots x_{i_{2k+1}\,i}| (\tilde B_{ijk})^{m'}]\,,\end{gathered}                                                                                                                                    
\end{equation}
where the prefactor of $\frac{1}{2}$ has been introduced for convenience. Carrying out the recursion step from four to five points we obtain
\begin{align}\label{eq:f_5_4d}
\mathcal{A}_5=
 i\delta^4(q)\delta^4(\tilde q)\frac{ \langle B_{5 4 2}|\, 1\, 2\, 3\,| B_{5 4 2}\rangle + 
   [ \tilde B_{5 4 2}|\, 1\, 2\, 3\,| \tilde B_{5 4 2}]}{x_{1 3}^2 x_{2 4}^4 x_{2 5}^4 x_{3 5}^2 x_{4 1}^2}
\end{align}
and for the six-point amplitude we get
\begin{align}\label{eq:f_6_4d}
\mathcal{A}_6&=
 i\delta^4(q)\delta^4(\tilde q)\Bigl(\frac{\langle B_{625}|\,4\,3 \,2 \,| B_{625}\rangle\langle B_{235}|\,6 \,5 \,1 \,| B_{235}\rangle}{x_{1 3}^2x_{24}^2x_{35}^4x_{46}^2x_{51}^2x_{52}^4x_{62}^4\Tr(6235)}+\text{chiral conjugate}\notag\\
&\=\phantom{i\delta^4(q)\delta^4(\tilde q)\Bigl(}+\frac{[ \tilde B_{136}|\, 2\,3 \,5 \,| \tilde B_{136}]\langle B_{436}|\,5 \,6 \,1 \,| B_{436}\rangle}{x_{1 3}^4x_{26}^2x_{35}^2x_{36}^2x_{46}^4x_{51}^2\Tr(2356)\Tr(3461)}\notag\\
&\=\phantom{i\delta^4(q)\delta^4(\tilde q)\Bigl(}-\frac{[ \tilde B_{325}|\, 4\,5 \,1 \,| \tilde B_{325}]\langle B_{215}|\,6 \,5 \,3 \,| B_{215}\rangle}{x_{1 3}^2x_{15}^2x_{24}^2x_{25}^2x_{35}^4x_{51}^2x_{62}^2\Tr(1245)\Tr(2356)}\notag\\
&\=\phantom{i\delta^4(q)\delta^4(\tilde q)\Bigl(}-\frac{[ \tilde B_{146}|\,2\,4 \,5 \,| \tilde B_{146}]\langle B_{214}|\,6 \,4 \,3 \,| B_{214}\rangle}{x_{1 3}^2x_{14}^2x_{24}^4x_{46}^4x_{51}^2x_{62}^2\Tr(1245)\Tr(3461)}\Bigr)
\end{align}
Dual conformal invariance of these expressions is easy to verify by simply counting the inversion weights on each dual point.
  
In principle all non-chiral amplitudes could be obtained by a half Fourier transform of the known chiral or anti-chiral superamplitudes. However, it is in general nontrivial to carry out these integrations in a way that leads to a useful representation of the amplitude. One exception are the MHV and $\overline{\text{MHV}}$ part of the non-chiral superamplitude, which can be obtained by either solving the BCFW recursion or by performing the half Fourier transform in the way described in \cite{Huang:2011um}. The result we found and also checked numerically is
%\hspace{-0.6cm}
\begin{multline}
\hspace{-0.4cm}\mathcal{A}_n^{\overline{\text{MHV}}}=\\\hspace{0.4cm}\frac{i\delta^4(q)\delta^4(\tilde q)}{\prod_{k=1}^nx_{kk+2}^2} \frac{\langle B_{n\,2\,n-1}|\,n\!-\!2\,n\!\!-\!\!3 \,n\!-\!4 \,| B_{n\,2\,n-1}\rangle}{x_{n-3n-1}^2x_{n2}^2}\prod_{k=1}^{n-5}\frac{\langle B_{k + 1\, k + 2\, n - 1}|\,n\, n\! -\!1\, k \,| B_{k + 1\, k + 2\, n - 1}\rangle}{x_{n-1k+1}^2\Tr(n\,k\!+\!1\,k\!+\!2\,n\!-\!1)},\hspace{-0.4cm}
\end{multline}
and similar for the MHV part. Note that our result differs from the one presented in \cite{Huang:2011um}.
\subsection[Supersymmetric BCFW for \texorpdfstring{$\cN=(1,1)$}{N=(1,1)} SYM]{Supersymmetric BCFW for \texorpdfstring{$\bm{\cN=(1,1)}$}{N=(1,1)} SYM}\label{section:BCFW6d}
The supersymmetric BCFW recursion of $\mathcal{N} = (1,1)$ SYM theory in six dimensions will play a central role when investigating massive amplitudes in \cref{section:6damps,section:uplift_huang}. It has been introduced in reference \cite{Dennen:2009vk}. In what follows we will closely follow the detailed review presented in reference \cite{Bern:2010qa}. At the end of this section we will use the BCFW recursion relation to prove the dual conformal covariance, \cref{eq:dualconformal6d}, of the superamplitudes. 

As a first step we introduce the shift vector
\begin{equation}\label{eq:shift6d}
 r^\mu=\frac{1}{2 s_{1n}}X_{a\dot{a}}\langle 1^a|\Sigma^\mu p_n|1^{\dot a}]\,,
\end{equation}
that obviously has the desired properties $r\cdot p_1=0=r\cdot p_n$. The requirement $r^2=0$, implies $0=\epsilon^{ab}\epsilon^{\dot{a}\dot{b}}X_{a\dot{a}}X_{b\dot{b}}=2\det(X)$. Hence $X_{a\dot{a}}$ is some arbitrary rank one matrix and has a spinor helicity representation $X_{a\dot{a}}=x_a\tilde{x}_{\dot a}$. \Cref{eq:shift6d} implies
\begin{align}\label{eq:shift6d_matrix}
r^{AB}&=X_{a\dot a}\frac{{}^{[A}| p_n|1^{\dot{a}}]\langle 1^a|^{B]}}{s_{1n}}\,,&r_{AB}&=-X_{a\dot a}\frac{{}_{[A}|1^{\dot{a}}]\langle 1^a|p_n|_{B]}}{s_{1n}}\,,
\end{align}
and the shifts of the momenta $p_1$ and $p_n$ \eqref{eq:shifts} can be reinterpreted as shifts of the chiral and anti-chiral spinors. The equations
\begin{equation}
\begin{aligned}
p_{\hat{1}}^{AB}&=\lambda_{\hat{1}}^{A\,a}\lambda_{\hat{1}\,a}^{B}\,,& p_{\hat{n}}^{AB}&=\lambda_{\hat{n}}^{A\,a}\lambda_{\hat{n}\,a}^{B}\,,\\[+0.3cm]
p_{\hat{1}\,AB}&=\tilde\lambda_{\hat{1}\,A\,\dot{a}}\tilde\lambda_{\hat{1}\,B}^{\dot{a}}\,,\qquad&p_{\hat{n}\,AB}&=\tilde\lambda_{\hat{n}\,A\,\dot{a}}\tilde\lambda_{\hat{n}\,B}^{\dot{a}}                 
\end{aligned}
\end{equation}
have the simple solutions
\begin{equation}\label{eq:shiftSpinors6d}
 \begin{aligned}
\lambda_{\hat{1}}^{A\,a}&= s_{1n}^{-1}\langle 1^a|p_n\,p_{\hat{1}}|^A=\lambda_{1}^{A\,a} + \frac{z}{s_{1n}} \langle 1^a|p_n\,r|^A\,,\\[+0.3cm]
\lambda_{\hat{n}}^{A\,b}&= s_{1n}^{-1}\langle n^a|p_1\,p_{\hat{n}}|^A=\lambda_{n}^{A\,a} - \frac{z}{s_{1n}} \langle n^a|p_1\,r|^A\,,\\[+0.3cm]
\tilde{\lambda}_{\hat{1}A\dot{a}} &= s_{1n}^{-1}[1_{\dot{a}}|p_n\,p_{\hat{1}}|_A=\tilde\lambda_{1\,A\,\dot{a}} + \frac{z}{s_{1n}} [1_{\dot{a}}|p_n\,r|_A\,,\\[+0.3cm]
\tilde{\lambda}_{\hat{n}A\dot{b}} &= s_{1n}^{-1}[n_{\dot{a}}|p_1\,p_{\hat{n}}|_A=\tilde\lambda_{n\,A\,\dot{a}} - \frac{z}{s_{1n}} [n_{\dot{a}}|p_1\,r|_A\,.
 \end{aligned}
\end{equation}
Or after inserting the definition \eqref{eq:shift6d_matrix} of the shift vector 
\begin{equation}\label{eq:shiftSpinors6d_inserted}
 \begin{aligned}
\lambda_{\hat{1}}^{Aa}&= \lambda^{Aa}_{1} + \frac{z}{s_{1n}}X^{a\dot{a}} [1_{\dot a}| n_b\rangle \lambda^{A\,b}_{n}\,,&
\lambda_{\hat{n}}^{Ab}& = \lambda^{Ab}_{n} - \frac{z}{s_{1n}}X_{a\dot{a}} [1^{\dot a}| n^b\rangle \lambda^{A\,a}_{1} \,,\\
\tilde{\lambda}_{\hat{1}A\dot{a}} &= \tilde{\lambda}_{1A\dot{a}} + \frac{z}{s_{1n}}X_{a\dot{a}} [n_{\dot b}| 1^a\rangle \tilde{\lambda}_{nA}^{\dot b}\,,&
\tilde{\lambda}_{\hat{n}A\dot{b}} &= \tilde{\lambda}_{nA\dot{b}} + \frac{z}{s_{1n}}X_{a\dot{a}} [n_{\dot b}| 1^a\rangle \tilde{\lambda}_{1A}^{\dot a}\,.
 \end{aligned}
\end{equation}
Supermomentum conservation can only be maintained if the Grassmann variables of legs $1$ and $n$ are shifted as well
\begin{equation}\label{eq:shiftGrassmann6d}
\begin{aligned}
\xi_{\hat{1} a} &= \xi_{1 a}+ z X_{a \dot{a}} [1^{\dot{a}}|q_n\rangle/s_{1n}\,,&  \xi_{\hat{n} b} &= \xi_{n b} + z X_{a\dot{a}} [1^{\dot{a}}|n_{b}\rangle \xi_{1}^a/s_{1n}\,,\\
\tilde{\xi}^{\dot{a}}_{1} &= \tilde{\xi}^{\dot{a}}_{1}-z X^{a \dot{a}} [\tilde{q}_n|1_{a}\rangle /s_{1n}\,,&\tilde{\xi}^{\dot{b}}_{\hat{n}} &= \tilde{\xi}^{\dot{b}}_{n} - z X_{a\dot{a}} [n^{\dot{b}}|1^{a}\rangle \tilde{\xi}^{\dot{a}}_{1}/s_{1n}\,,
\end{aligned}
\end{equation}
resulting in the following shifts of the supermomenta
\begin{equation}
 \begin{aligned}\label{eq:shiftsSuperMomenta6d}
q^A_{\hat{1}} &=[\tilde{\chi}|p_{\hat{1}}|^A =q^A_1 + z\, s^A \,,& q^A_{\hat{n}} &= [\tilde\chi|p_{\hat{n}}|^A=q^A_n -z\,s^A\,, \\
\tilde{q}_{\hat{1} A}&=\langle\chi|p_{\hat{1}}|_A= \tilde{q}_{1 A} + z\,\tilde{s}_A\,,&\tilde{q}_{\hat{n} A} &=\langle\chi|p_{\hat{n}}|_A= \tilde{q}_{n A} -z \,\tilde{s}_A \,,
\end{aligned}
\end{equation}
with
\begin{equation}
 \begin{aligned}
\chi&=s_{1n}^{-1}([ \tilde{q}_1|p_n|+[\tilde{q}_n|p_1| )\,,&\tilde\chi&=s_{1n}^{-1}(\langle q_1|p_n|+\langle q_n|p_1|)\\
s&=  [\tilde\chi|r|\,,&\tilde{ s}&=\langle \chi|r|\,,
 \end{aligned}
\end{equation}
or with the definition of $r$ being inserted
\begin{equation}
 \begin{aligned}
s^A&=  \frac{X_{a\dot{a}}}{s^2_{1n}}\left(  \langle q_{1}  | p_n| 1^a\rangle[1^{\dot a}| p_n|^A  + s_{1n}\langle q_{n}|1^{\dot{a}}]\lambda_{1}^{A \,a} \right)\,,\\
\tilde s_A&=\frac{X_{a\dot{a}}}{s^2_{1n}}\left( - [\tilde{q}_1| p_n|1^{\dot{a}}]\langle 1^a | p_n|_A - s_{1n} [\tilde{q}_n|1^{\dot{a}}\rangle\tilde{\lambda}_{1\,A}^{\dot{a}}\right)\,.
 \end{aligned}
\end{equation}
The dual shifts are given by
\begin{align}\label{eq:shiftDual6d}
x_{\hat{1}}&=x_1+z\,r &\theta_{\hat{1}}&=\theta_1+z\,s&\tilde{\theta}_{\hat{1}}&=\tilde{\theta}_1+z\,\tilde{s}\,.
\end{align}
Note that the Grassmann shift variables $s^A$ and $\tilde{s}_A$ can alternatively be obtained by solving the equations
\begin{align}
\langle \theta_{\hat{1}2}|x_{\hat{1}2}|&=0\,,&\langle \theta_{n\hat{1}}|x_{n\hat{1}}|&=0\,,&[\tilde\theta_{\hat{1}2}|x_{\hat{1}2}|&=0\,,&[\tilde\theta_{n\hat{1}}|x_{n\hat{1}}|&=0\,. 
\end{align}
The above set of supersymmetry preserving shifts leads to a shifted superamplitude whose residues at the poles \cref{eq:poles} are given by a product of two lower point superamplitudes. Similar to the supersymmetric BCFW recursions of $\mathcal{N} = 4$ SYM, the sum over intermediate states is realized by an integration with respect to the Grassmann variables of the intermediate leg.
Using the abbreviations introduced in \cref{eq:Def_Pj} the BCFW recursion of $\cN=(1,1)$ SYM theory in six dimensions reads
\begin{equation}\label{eq:BCFW_6D}
\!\mathcal{A}_n\!\left(p_1,\dots,p_n\right) = \sum_{j = 2}^{n-2} \int\!\! d^2 \tilde{\xi}_{\hat{P}_{j}} \int\!\! d^2 \xi_{\hat{P}_{j}}  \mathcal{A}_{j+1}(\hat{p}_1,\dots,p_j, -\hat{P}_{j}) \frac{-i}{P^2_{j}}  \mathcal{A}_{n-j+1}(\hat{P}_{j}, p_{j+1},\dots,p_{\hat{n}})\Bigr\rvert_{z=z_j}\!
\end{equation}
Similar to the non-chiral BCFW recursion in four dimensions, \cref{eq:BCFW_non-chiral}, the explicit minus sign originates from the choice $d^2\xi=\tfrac{1}{2}d\xi^a d\xi_{a}$, $d^2 \tilde\xi=\tfrac{1}{2}d\tilde\xi_{\dot{a}}d \tilde\xi^{\dot{a}}$ for the integration measure and can be fixed by projecting the four point function resulting from the six-dimensional BCFW recursion \cref{eq:BCFW_6D} to four dimensions and comparing it with \cref{eq:A4nonchiral}.
Starting point for the recursion is the three-point superamplitude of \cref{eq:A3_6D} 
\cite{Cheung:2009dc}. For applications of the BCFW recursion it is more convenient to use the following alternative representation of the three point amplitude
\begin{equation}\label{eq:A3_6d_alternative}
 \mathcal{A}_3 =i \delta^{6}(p) ({\bf u}_1 -{\bf u}_2)({\bf\tilde u}_1- {\bf \tilde u}_2)\left({\bf u}_3 -\tfrac{1}{2}({\bf u}_1+{\bf u}_2) \right)\left({\bf \tilde u}_3 -\tfrac{1}{2}({\bf\tilde u}_1+{\bf\tilde u}_2) \right) \delta\left( {\bf w} \right)  \delta\left( \tilde{{\bf w}} \right)\,.
\end{equation}
As has been shown in \cite{Dennen:2009vk}, the BCFW recursion yields the four point function
\begin{equation}\label{eq:A4_6d}
 \mathcal{A}_4 = - \delta^{6}\left( p \right) \delta^{4}\left( q\right) \delta^{4}\left(\tilde{q} \right) \frac{i}{x_{1 3}^2 x_{2 4}^2}\,.  
\end{equation}
Note that the four-point amplitude is fixed up to a numerical factor by supersymmetry and dual conformal symmetry.

In the remainder of this section we will explicitly carry out the Grassmann integrations in the BCFW recursion \cref{eq:BCFW_6D}. First of all we recall that for $n \geq 4$ an $n$-point superamplitude has the form
\begin{equation}\label{eq:superamps6d}
\mathcal{A}_{n} = \delta^6(p)\delta^{4} \left(q\right) \delta^{4} \left(\tilde{q} \right) f_n(\{x_i,\theta_i,\tilde\theta_i\})
\end{equation}
In order to consistently treat ingoing and outgoing particles we adopt the prescription
\begin{equation}
\begin{aligned}
\lambda_{(-p)} &= i\, \lambda_{p}\,,&  \tilde{\lambda}_{(-p)} &= i\, \tilde{\lambda}_{p}\,,&\xi_{(-p)} &= i\, \xi_{p}\,,&\tilde{\xi}_{(-p)}& = i\, \tilde{\xi}_{p}\,.
\end{aligned}
\end{equation}
Structurally there are the three different cases $j=2$, $2<j<n-2$ and $j=n-2$ to be analyzed. Starting with the contribution $j=2$ in \cref{eq:BCFW_6D}, we want to evaluate
\begin{equation}
\mathcal{B}_{2}=\frac{-i}{x_{13}^2} \int d^2 \xi_{\hat{P}} d^2\tilde{\xi}_{\hat{P}_2}\mathcal{A}_3\left(p_{\hat{1}},p_2,-\hat{P}_2\right)\delta^4 \left(q_{\hat{P}_2}+\theta_{3\hat{1}}\right) \delta^4 \left(\tilde{q}_{\hat{P}_2}+\tilde\theta_{3\hat{1}} \right) f_{n-1} \left(x_{\hat 1},x_3,\dots,x_n\right)
\end{equation}
Taking the representation \cref{eq:A3_6d_alternative} of $\mathcal{A}_3$, the only dependence on $\xi_{\hat{P}_2}$, $\tilde{\xi}_{\hat{P}_2}$ is contained in Grassmann delta functions, and the integration boils down to solving the linear equations 
\begin{align}
 {\bf u}_{K}&=\tfrac{1}{2}({\bf u}_{\hat 1}+{\bf u}_{2})\,,&{\bf w}_{K}&=-{\bf w}_{\hat 1}-{\bf w}_{2}\,,\\
{\bf \tilde u}_{K}&=\tfrac{1}{2}({\bf \tilde u}_{\hat 1}+{\bf \tilde u}_{2})\,,&{\bf \tilde w}_{K}&=-{\bf \tilde w}_{\hat 1}-{\bf \tilde w}_{2}\,,
\end{align}
for $\xi_{\hat{P}_2}$, $\tilde{\xi}_{\hat{P}_2}$, with the abbreviation $K=-\hat{P}_2$. The solution is
\begin{align}
\xi_{\hat{P}_2}^a &=- \tfrac{i}{2} \left({\bf u}_{\hat{1}} + {\bf u}_{2}\right) w_{K}^a -  i\left({\bf w}_{\hat{1}} + {\bf w}_{2}\right)u_{K}^a\,,& \tilde{\xi}_{\hat{P}_2}^{\dot{a}} &=  \tfrac{i}{2}  \left(\tilde{{\bf u}}_{\hat{1}} + \tilde{{\bf u}}_{2}\right)\tilde{w}_{K}^{\dot{a}} +i  \left(\tilde{{\bf w}}_{\hat{1}} + \tilde{{\bf w}}_{2}\right)\tilde{u}_{K}^{\dot{a}}\,.
\end{align}
Using \cref{eq:P1,eq:P2,eq:P3} it is straightforward to show that on the support of $({\bf u}_{\hat 1} -{\bf u}_2)({\bf\tilde u}_{\hat 1}- {\bf \tilde u}_2)$ this implies
\begin{align}\label{deltas_ersetzung}
q_{\hat{P}_2} &= q_{\hat{1}} + q_{2}\,,& \tilde{q}_{\hat{P}_2}& = \tilde{q}_{\hat{1}} + \tilde{q}_{2} \,,
\end{align}
and therefore
\begin{equation}
\mathcal{B}_{2}=\delta^4 \left(q\right) \delta^4 \left(\tilde{q}\right) \frac{-i}{x_{13}^2} f_{n-1}  \left(x_{\hat 1},x_3,\dots,x_n\right) \int d^2 \xi_{\hat{P}_2} d^2\tilde{\xi}_{\hat{P}_2}\, \mathcal{A}_3\left(p_{\hat{1}},p_2,-\hat{P}_2\right)\,.
\end{equation}
The integral of the three-point amplitude has the solution
\begin{equation}\label{eq:B26d}
i\left({\bf u}_{\hat{1}} - {\bf u}_{2}\right)\left({{\bf\tilde u}}_{\hat{1}} - {{\bf\tilde u}}_{2}\right) = i \left(\frac{\langle q_{2}|k_2 p_{\hat{1}}|\tilde{q}_{2}]}{2\,p_{2}\cdot k_2}- \langle q_{\hat{1}} |\tilde{q}_{2}] + \langle q_{2}| \tilde{q}_{\hat{1}}]  - \frac{\langle q_{\hat{1}}| k_1 p_2|\tilde{q}_{\hat{1}}]}{2\,p_{\hat{1}}\cdot k_1}  \right)\,,
\end{equation}
where $k_1$ and $k_2$ are some arbitrary reference vectors and $u^a w_a=1=\tilde{u}^{\dot{a}}\tilde{w}_{\dot{a}}$ has been used. The final result is
\begin{equation}\label{p_Ref}
\mathcal{B}_{2} = \delta^4 \left(q \right) \delta^4 \left(\tilde{q}\right) \frac{f_{n-1} \left(x_{\hat 1},x_3,\dots,x_n\right)}{x_{13}^2}  \left(\frac{\langle q_{2}|k_2 p_{\hat{1}}|\tilde{q}_{2}]}{2\,p_{2}\cdot k_2} - \langle q_{\hat{1}} |\tilde{q}_{2}] + \langle q_{2}| \tilde{q}_{\hat{1}}]-\frac{\langle q_{\hat{1}}| k_1 p_2|\tilde{q}_{\hat{1}}]}{2\,p_{\hat{1}}\cdot k_1}   \right),
\end{equation}
evaluated at $z=z_2$. In the case $j=n-2$ we need to evaluate
\begin{equation}\label{eq:B_2}
 \mathcal{B}_{n-2}=\frac{-i}{x_{1\,n-1}^2}\delta^4 \left(q\right) \delta^4 \left(\tilde{q} \right) f_{n-1} \left(x_{\hat 1},\dots,x_{n-1}\right) \int d^2 \xi_{\hat{P}_{n-2}} d^2\tilde{\xi}_{\hat{P}_{n-2}}\mathcal{A}_3\left(p_{n-1},p_{\hat{n}},\hat{P}_{n-2}\right)\,.
\end{equation}
Here we already exploited that on the support of the three-point amplitude we have
\begin{align}
 \xi_{\hat{P}_{n-2}}^a &= \tfrac{1}{2} \left({\bf u}_{\hat{n}} + {\bf u}_{n-1}\right) w_{\hat{P}_{n-2}}^a +  \left({\bf w}_{\hat{n}} + {\bf w}_{n-1}\right)u_{\hat{P}_{n-2}}^a\,,\\ \tilde{\xi}_{\hat{P}_{n-2}}^{\dot{a}} &=  -\tfrac{1}{2}  \left(\tilde{{\bf u}}_{\hat{n}} + \tilde{{\bf u}}_{n-1}\right)\tilde{w}_{\hat{P}_{n-2}}^{\dot{a}} -  \left(\tilde{{\bf w}}_{\hat{n}} + \tilde{{\bf w}}_{n-1}\right)\tilde{u}_{\hat{P}_{n-1}}^{\dot{a}}\,
\end{align}
or more conveniently
\begin{align}
  q_{\hat{P}_{n-2}}&=-q_{\hat{n}}-q_{n-1}\,,&\tilde{q}_{\hat{P}_{n-2}}&=-\tilde{q}_{\hat{n}}-q_{n-1}\,.
\end{align}
The remaining integral of the three-point amplitude in \cref{eq:B_2} is given by
\begin{multline}
 i\left( {\bf u}_{n-1}-{\bf u}_{\hat{n}}\right)\left({{\bf\tilde u}}_{n-1}-{{\bf\tilde u}}_{\hat{n}}\right) =\\ i \left(\frac{\langle q_{\hat{n}}| k_n p_{n-1}|\tilde{q}_{\hat{n}}]}{2\,p_{\hat{n}}\cdot k_n}- \langle q_{n-1}| \tilde{q}_{\hat{n}}]+ \langle q_{\hat{n}} |\tilde{q}_{n-1}]  - \frac{\langle q_{n-1}|k_{n-1} p_{\hat{n}}|\tilde{q}_{n-1}]}{2\,p_{n-1}\cdot k_{n-1}}  \right)\,,
\end{multline}
leading to
\begin{multline}
 \mathcal{B}_{n-2}=\delta^4 \left(q\right) \delta^4 \left(\tilde{q} \right) \frac{f_{n-1} \left(x_{\hat 1},\dots,x_{n-1}\right)}{x_{1\,n-1}^2} \times\\
\left(\frac{\langle q_{\hat{n}}| k_n p_{n-1}|\tilde{q}_{\hat{n}}]}{2\,p_{\hat{n}}\cdot k_n}- \langle q_{n-1}| \tilde{q}_{\hat{n}}]+ \langle q_{\hat{n}} |\tilde{q}_{n-1}]  - \frac{\langle q_{n-1}|k_{n-1} p_{\hat{n}}|\tilde{q}_{n-1}]}{2\,p_{n-1}\cdot k_{n-1}}  \right)\,,
\end{multline}
evaluated at $z=z_{n-2}$. Similar to the case $j=2$, arbitrary reference momenta $k_{n}$, $k_{n-1}$ have been introduces in order to get rid of the $u$, $\tilde u$ variables. Finally there is the general case $2<j<n-2$ with no three-point amplitudes involved
\begin{multline}
\mathcal{B}_{j} = \frac{-i}{x_{1\,j+1}^2}\delta^4 \left(q \right) \delta^4 \left(\tilde{q} \right) f_{j + 1}\left(x_{\hat{1}},\dots,x_{j+1}\right)f_{n - j + 1}\left(x_{\hat{1}},x_{j+1},\dots,x_n\right)\times\\\int d^2 \xi_{\hat{P}_j} d^2\tilde{\xi}_{\hat{P}_j} \delta^4 \left(q_{\hat{P}_j}+\theta_{j+1\,\hat{1}} \right) \delta^4 \left(\tilde{q}_{\hat{P}_j} +\theta_{j+1\,\hat{1}}\right)
\end{multline}
To carry out the integration we want to rewrite the fermionic delta functions. Due to the algebra \cref{eq:Sigma} of the six-dimensional Pauli matrices, we have the identity
\begin{align}
 \delta^A_C=s_{ij}^{-1}(p_i^{AB}p_{j\,BC}+p_i^{AB}p_{j\,BC})\,,
\end{align}
which implies
\begin{equation}
\begin{aligned}
   q^A_i+ q^A_j+ Q^A&=(\xi_{i\,a}+s_{ij}^{-1} \langle i_a|p_j|Q\rangle) \lambda_i^{A\,a}+(\xi_{j\,a}+s_{ij}^{-1} \langle j_a|p_i|Q\rangle)\lambda_j^{A\,a}\,,\\
 \tilde{q}_{i\,A}+\tilde{q}_{j\,A}+\tilde{Q}_A&=(\tilde\xi_{i}^{\dot a}+s_{ij}^{-1}[i^{\dot a}|p_j|\tilde{Q}])\tilde{\lambda}_{i\,A\,\dot a}+(\tilde\xi_{j}^{\dot a}+s_{ij}^{-1}[j^{\dot a}|p_i|\tilde{Q}])\tilde{\lambda}_{j\,A\,\dot a}\,.
\end{aligned}
 \end{equation}
Consequently the fermionic delta functions can be rewritten as follows
\begin{equation}
\begin{aligned}
  \delta^4( q_i+ q_j+ Q)&=-s_{ij}\delta^2(\xi_{i\,a}+s_{ij}^{-1}\langle i_a|p_j|Q\rangle)\delta^2(\xi_{j\,a}+s_{ij}^{-1} \langle j_a|p_i|Q\rangle)\,,\\
 \delta^4(\tilde{q}_i+\tilde{q}_j+\tilde{Q})&=-s_{ij}\delta^2(\tilde\xi_{i}^{\dot a}+s_{ij}^{-1}[i^{\dot a}|p_j|\tilde{Q}])\delta^2(\tilde\xi_{j}^{\dot a}+s_{ij}^{-1}[j^{\dot a}|p_i|\tilde{Q}])\,.
\end{aligned}
 \end{equation}
The two-dimensional Grassmann delta functions are defined as $\delta^2(\chi_a)=\tfrac{1}{2}\chi_a\chi^a$ and $\delta^2(\tilde\chi^{\dot{a}})=\tfrac{1}{2}\tilde\chi^{\dot{a}}\tilde\chi_{\dot{a}}$ such that $\int d^2\xi \,\delta^2(\xi_a)=1=\int d^2\tilde\xi \,\delta^2(\tilde\xi^{\dot a})$.
This allows us to easily carry out the Grassmann integrations
\begin{align}
 \int d^2 \xi_{\hat{P}_j} \,\delta^4 \left(q_{\hat{P}_j}+\theta_{j+1\,\hat{1}} \right)&=-s_{\hat{P}_j\,\hat{n}}^{-1}\delta^2(\langle \hat{n}_a|\hat{P}_j|\theta_{j+1\,\hat{1}}\rangle)\notag\\
&=-\tfrac{1}{2}s_{\hat{P}_j\,\hat{n}}^{-1}\langle \hat{n}_a|\hat{P}_j|\theta_{j+1\,\hat{1}}\rangle \langle\hat{n}^a|\hat{P}_j|\theta_{j+1\,\hat{1}}\rangle\notag\\
&=-\tfrac{1}{2}s_{\hat{P}_j\,\hat{n}}^{-1}\langle \theta_{j+1\,\hat{1}}|\hat{P}_jp_{\hat{n}}\hat{P}_j|\theta_{j+1\,\hat{1}}\rangle\notag\\
&=-\tfrac{1}{2}\langle \theta_{j+1\,\hat{1}}|x_{\hat{1}\,j+1}|\theta_{j+1\,\hat{1}}\rangle\,,
\end{align}
and similarly for the anti-chiral integration
\begin{equation}
 \int d^2\tilde{\xi}_{\hat{P}_j} \delta^4 \left(\tilde{q}_{\hat{P}_j} +\theta_{j+1\,\hat{1}}\right)=-\tfrac{1}{2}[\tilde\theta_{j+1\,\hat{1}}|x_{\hat{1}\,j+1}|\tilde\theta_{j+1\,\hat{1}}]\,.
\end{equation}
The full contribution is
\begin{multline}\label{eq:Bj}
\mathcal{B}_{j} =- i\,\delta^4 \left(q \right) \delta^4 \left(\tilde{q} \right) f_{j + 1}\left(x_{\hat{1}},\dots,x_{j+1}\right)f_{n - j + 1}\left(x_{\hat{1}},x_{j+1},\dots,x_n\right)\frac{1}{x_{1\,j+1}^2}\times\\
\tfrac{1}{4}\langle \theta_{j+1\,\hat{1}}|x_{\hat{1}\,j+1}|\theta_{j+1\,\hat{1}}\rangle[\tilde\theta_{j+1\,\hat{1}}|x_{\hat{1}\,j+1}|\tilde\theta_{j+1\,\hat{1}}]\,,
\end{multline}
evaluated at $z=z_j$. Hence, given all lower point amplitudes, the $n$-point super amplitude is simply given by
\begin{equation}\label{eq:BCFW_6D_integrated}
\mathcal{A}_n = \sum_{j = 2}^{n-2} \mathcal{B}_{j}\,.
\end{equation}
This expression is straightforward to implement numerically. Unfortunately, it is ill suited to directly obtain reasonable analytical expressions for higher point amplitudes because of the auxiliary variable $X_{a\dot{a}}=x_a\tilde{x}_{\dot a}$ contained in the shift \cref{eq:shift6d}. In contrast to four dimensions the shift vector is not fixed by requiring $r^2=0$, $r\cdot p_1=0=r\cdot p_n$. This ambiguity is reflected by the presence of $X_{a\dot{a}}$ in the definition of the shift vector. Obviously the amplitudes are independent of the shift vector, i.\,e.~independent of $X_{a\dot{a}}$. In principle it should be possible to remove the shift vector from the right hand side of \cref{eq:BCFW_6D_integrated} without inserting its definition \cref{eq:shift6d_matrix}, only using its general properties \cref{eq:shiftvector}. Unfortunately, even in the easiest case of the five point superamplitude this is very hard to achieve. As long as it is not understood 
how to obtain $f_n(\{x_i,\theta_i,\tilde\theta_i\})$ from the output of the BCFW recursion, \cref{eq:BCFW_6D_integrated} will be limited to numerical applications.

Indeed, in \cref{section:6damps,section:uplift_huang} we will extensively use a \texttt{Mathematica} implementation of  the integrated BCFW recursion \eqref{eq:BCFW_6D_integrated}. Independence of  $X_{a\dot{a}}$ and the arbitrary reference momenta entering $\mathcal{B}_{2}$ and $\mathcal{B}_{n-2}$ provides a nontrivial check of the numerical results obtained from the implementation. In fact, taking the four point amplitude \eqref{eq:A4_6d} as initial data, independence of the six-point component amplitudes on $X_{a\dot{a}}$  requires the explicit minus sign appearing in the BCFW recursion relation \cref{eq:BCFW_6D}.

\subsection{Proof of dual conformal symmetry of $\mathcal{N}=(1,1)$ superamplitudes}\label{section:ProofDual}
With the help of the BCFW recursion and the inversion rules \eqref{eq:inversion6d_first} to \eqref{eq:inversion6d_last} it is straightforward to inductively prove the dual conformal covariant inversion of the $\mathcal{N}=(1,1)$ superamplitudes by showing that each term $\mathcal{B}_{i}$ in the integrated BCFW recursion \cref{eq:BCFW_6D_integrated} inverts as
\begin{equation}
 I[\mathcal{B}_{j}]=\left(\prod_i x_i^2\right)\mathcal{B}_{j}\,.
\end{equation}
Since the BCFW diagrams involving three-point amplitudes $\mathcal{B}_{2}$, $\mathcal{B}_{n-2}$ are related by cyclic relabeling of the indices, we only need to consider  one of them as well as the general diagram $\mathcal{B}_{j}$ without three-point functions.

We start out with $\mathcal{B}_{2}$, \cref{eq:B26d}, and investigate the inversion of $\left({\bf u}_{\hat{1}} - {\bf u}_{2}\right)\left({{\bf\tilde u}}_{\hat{1}} - {{\bf\tilde u}}_{2}\right)$. Simply plugging in the inversion rules yields

\begin{align}
 I[{{\bf\tilde u}}_{2}-{{\bf\tilde u}}_{\hat{1}}]=\beta^{-1}\sqrt{\frac{x^2_2}{x_{\hat{1}}^2x_{3}^2}}{\bf\tilde u}_{2}-\beta^{-1}\sqrt{\frac{x_{\hat{1}}^2}{x_{2}^2 x_{3}^2}}{{\bf\tilde u}}_{\hat{1}}+\frac{\beta^{-1}}{\sqrt{x_{\hat{1}}^2x^2_2x^2_3}}\left(\tilde{u}_{\hat 1}^{\dot{a}}[\tilde{\theta}_{\hat 1}|x_{\hat 1}|\hat{1}_{\dot{a}}]-\tilde{u}_{2}^{\dot{a}}[\tilde{\theta}_{2}|x_{2}|2_{\dot{a}}]\right)
\end{align}
Using $\tilde{u}_{2}^{\dot{a}}[2_{\dot{a}}|=\tilde{u}_{\hat 1}^{\dot{a}}[{\hat 1}_{\dot{a}}|$ and $x_2|\hat{1}_{\dot{a}}]=x_{\hat{1}}|\hat{1}_{\dot{a}}]=\tfrac{1}{2}(x_{\hat{1}}+x_2)|\hat{1}_{\dot{a}}]$ the inhomogeneous term can be rewritten
\begin{align}
 \left(\tilde{u}_{\hat 1}^{\dot{a}}[\tilde{\theta}_{\hat 1}|x_{\hat 1}|\hat{1}_{\dot{a}}]-\tilde{u}_{2}^{\dot{a}}[\tilde{\theta}_{2}|x_{2}|2_{\dot{a}}]\right)&=\tfrac{1}{2}\,\tilde{u}_{\hat 1}^{\dot{a}}\,\tilde{\xi}_{\hat 1}^{\dot b}\;[1_{\dot b}|x_{\hat 1}+x_2|1_{\dot a}]=\tfrac{1}{4}\,{{\bf\tilde u}}_{\hat{1}}\;\Tr\left[\,(x_{\hat 1}+x_2)(x_{\hat 1}-x_2)\;\right]\notag\\
&={{\bf\tilde u}}_{\hat{1}}(x_{\hat 1}^2-x_2^2)
\end{align}
and leads to the result
\begin{equation}
  I[{{\bf\tilde u}}_{2}-{{\bf\tilde u}}_{\hat{1}} ]=\beta^{-1}\sqrt{\frac{x^2_2}{x_{\hat{1}}^2x_{3}^2}}\left({{\bf\tilde u}}_{2}-{{\bf\tilde u}}_{\hat{1}}\right)\,.
\end{equation}
Similarly we find
\begin{align}
 I[{{\bf u}}_{2}-{{\bf u}}_{\hat{1}} ]&=\beta\sqrt{\frac{x^2_2}{x_{\hat{1}}^2x_{3}^2}}{\bf u}_{2}-\beta\sqrt{\frac{x_{\hat{1}}^2}{x_{2}^2 x_{3}^2}}{{\bf u}}_{\hat{1}}+\frac{\beta}{\sqrt{x_{\hat{1}}^2x^2_2x^2_3}}\left({u}_{\hat{1}\,a}\langle \theta_{\hat 1}|x_{\hat 1}|\hat{1}^{a}\rangle-u_{2\,a}\langle{\theta}_{2}|x_{2}|2^{a}\rangle\right)\notag\\
&=\beta\sqrt{\frac{x^2_2}{x_{\hat{1}}^2x_{3}^2}}\left({{\bf u}}_{2}-{{\bf u}}_{\hat{1}}\right)\,,
\end{align}
which together with
\begin{equation}
 I\left[\frac{f_{n-1} \left(x_{\hat 1},x_3,\dots,x_n\right)}{x_{13}^2} \right]=\frac{x_{\hat 1}^2x_{ 3}^2}{x_{ 2}^2}\left(\prod_{i=1}^n x_{i}^2\right)\;\frac{f_{n-1} \left(x_{\hat 1},x_3,\dots,x_n\right)}{x_{13}^2}\,,
\end{equation}
proves the desired inversion of  $\mathcal{B}_{2}$. What remains is to check the inversion of $\mathcal{B}_{j}$ given in \cref{eq:Bj}. Again inserting the inversion rules we obtain
\begin{align}
 I\bigl[\,\langle \theta_{j+1\,\hat{1}}|x_{\hat{1}\,j+1}|\theta_{j+1\,\hat{1}}\rangle\,\bigr]&=\left(\langle \theta_{j+1}|x^{-1}_{j+1}-\langle\theta_{\hat{1}}|x^{-1}_{\hat{1}}\right)\,x_{\hat{1}}^{-1}x_{\hat{1}\,j+1}x_{j+1}^{-1}\,\left(x^{-1}_{j+1}|\theta_{j+1}\rangle-x^{-1}_{\hat{1}}|\theta_{\hat{1}}\rangle\right)\notag\\
&=\frac{1}{x_{\hat{1}}^2x_{j+1}^2}\,\langle \theta_{j+1\,\hat{1}}|x_{\hat{1}\,j+1}|\theta_{j+1\,\hat{1}}\rangle\,,\label{eq:inversion1}
\end{align}
where we have used $x_{\hat{1}\,j+1}^2=0$. The inversion of $[\tilde\theta_{j+1\,\hat{1}}|x_{\hat{1}\,j+1}|\tilde\theta_{j+1\,\hat{1}}]$ can be obtained by chiral conjugation\footnote{The relative minus sign in the inversion of $\theta$, $\tilde\theta$ drops out.} of \eqref{eq:inversion1} and together with
\begin{multline}
  I\left[ \frac{f_{j + 1}\left(x_{\hat{1}},\dots,x_{j+1}\right)f_{n - j + 1}\left(x_{\hat{1}},x_{j+1},\dots,x_n\right)}{x_{1\,j+1}^2} \right]=\\
x_{\hat{1}}^4\,x_{j+1}^4\,\left(\prod_{i=1}^n x_{i}^2\right)\;\frac{f_{j + 1}\left(x_{\hat{1}},\dots,x_{j+1}\right)f_{n - j + 1}\left(x_{\hat{1}},x_{j+1},\dots,x_n\right)}{x_{1\,j+1}^2}
\end{multline}
this concludes the proof of the dual conformal symmetry of the tree superamplitudes.

\section{Tree-level superamplitudes of $\mathcal{N}=(1,1)$ SYM theory}\label{section:6damps}

In four dimensions the supersymmetric BCFW recursion together with the dual conformal invariance 
allowed for the construction of analytical formulae for all superamplitudes of $\cN=4$ SYM theory \cite{Drummond:2008cr}. The key to this remarkable result was the use of dual conformal invariant functions for the construction of a manifest dual conformal covariant solution to the BCFW recursion. Of similar importance was the MHV decomposition \eqref{eq:MHV_decomposition} of the superamplitudes, allowing to successively solve the recursion for the increasingly complex N${}^p$MHV superamplitudes. Although the non-chiral superamplitudes of $\cN=(1,1)$ SYM do not possess a conformal symmetry and an analogue of the helicity violation decomposition of the 4d theory,
 they still have a dual conformal symmetry and obey a supersymmetric BCFW recursion relation. Hence, it is natural to try to find dual conformal invariant functions suitable to construct a solution to the super-BCFW recursion of $\mathcal{N}=(1,1)$ SYM. Unfortunately, the six-dimensional 
BCFW recursion, as reviewed in \cref{section:BCFW6d}, is ill suited to produce compact analytical expressions. In contrast to four dimensions the shift \eqref{eq:shift6d} is not uniquely fixed and contains auxiliary spinor variables $x_a$, $x_{\dot a}$. Although the amplitudes are independent of these variables, their removal is non-trivial. The main obstacle is that the individual BCFW diagrams are in general not independent of $x_a$, $x_{\dot a}$ but only their sum, denying any obvious elimination of the auxiliary variables. In spite of its limitations the six-dimensional BCFW recursion is a powerful tool to obtain numerical values for arbitrary tree amplitudes of $\cN=(1,1)$ SYM theory. As we will explain in what follows, this can be exploited to determine manifest dual conformal covariant representations of superamplitudes.
\subsection{Analytical superamplitudes from numerical BCFW}\label{section:IdeaBCFW}
The general idea is to fix a sufficiently large set of dual conformal covariant functions $\Omega_{n,i}$ which are invariant under the dual symmetries $\{P_{AB},M^A_{\;\;B}, Q_A,\tilde{Q}^A,B,\tilde{B}\}$, covariant under the dual dilatation $D$, and are symmetric under chiral conjugation. In other words the $\Omega_{n,i}$ are Lorentz invariant functions of differences of dual variables, have Grassmann degree $\Omega_{n,i}=\mathcal{O}(\theta^{n-4}\tilde{\theta}^{n-4})$, are of dimension $-n$, and invert in the same way as $f_n$
\begin{equation}
I[\Omega_{n,j}]=\left(\prod_i x_i^2\right) \Omega_{n,j}\,.
\end{equation}
On the support of the momentum and supermomentum conserving delta functions, the $\Omega_{n,i}$ possess all continuous symmetries of $f_n$. Note that the invariance under the supersymmetry generators $\overline{q}^A$ and $\overline{\widetilde{q}}_A$ follows from the invariance under  $Q_A$, $\tilde{Q}^A$ and the covariance under dual conformal boosts $K_{AB}$, compare \cref{eq:S_dual,eq:Dualconformal6d}. Besides chiral symmetry, we could equally enforce the other discrete symmetries, which are cyclic invariance and the reflection symmetry. As will become clear in what follows enforcing symmetry under chiral conjugation is essential.

Given a set of functions $\{\Omega_{n,j}\}$, we can make the ansatz
\begin{equation}\label{eq:ansatz6d}
f_n=\sum_i \alpha_i \Omega_{n,i}\,,
\end{equation}
By construction, the coefficients $\alpha_{i}$ are dimensionless, dual conformal invariant functions of differences $x_{ij}$ of the region momenta $x_{i}$. 
The only dual conformal covariant objects that can be built from the $x_{ij}$ are the traces
\begin{align}
 \widetilde{\Tr}(i_1\dots i_{2k})&:=\left(x_{i_1i_2}\right)_{A_1A_2}\left(x_{i_2i_3}\right)^{A_2A_3}\dots\left(x_{i_{2k-1}i_{2k}}\right)_{A_{2k-1}A_{2k}}\left(x_{i_{2k}i_1}\right)^{A_{2k}A_1}\,.
\end{align}
However, these traces contain six-dimensional Levi-Civita tensors if six linear independent momenta are present, i.\,e.~if $k>3$ and $n>6$. Since  $\cN=(1,1)$ SYM is a non-chiral theory, all of its component amplitudes should be free of Levi-Civita tensors. Consequently, all  Levi-Civita tensors present in the coefficients $\alpha_i$ have to cancel out if we project the ansatz \cref{eq:ansatz6d} onto any component amplitude. The functions $\Omega_{n,i}$ are symmetric under chiral conjugation and therefore cannot produce Levi-Civita tensors. Hence, we conclude that only the chiral symmetric traces
\begin{equation}
 \Tr(i_1\dots i_{2k})=\tfrac{1}{2}\left(\widetilde{\Tr}(i_1\dots i_{2k})+\widetilde{\Tr}(i_2\dots i_{2k}i_1)\right)
\end{equation}
can appear in the coefficients. These traces are given by
\begin{equation}
 \begin{aligned}
\Tr(i\, j\, k\, l)&=2(x_{ij}^2x_{kl}^2-x_{ik}^2x_{jl}^2+x_{il}^2x_{jk}^2) \,\\
\Tr(i_1\,\dots \, i_{2k})&=-\tfrac{1}{2}\sum_{\alpha=2}^{2k}(-1)^\alpha x_{i_1i_{\alpha}}^2\Tr(i_2\,\dots\, i_{\alpha-1}\,i_{\alpha+1}\,\dots\, i_{2k})\,.
\end{aligned}
\end{equation}
We draw the important conclusion that the  coefficients $\alpha_i$ are rational functions of dual conformal invariant cross ratios
\begin{align}
 u_{ijkl}&=\frac{x_{ij}^2x_{kl}^2}{x_{il}^2x_{kj}^2}\,,&&\text{with}&x_{ij}^2&\neq0\,,\;x_{kl}^2\neq0\,,\;x_{il}^2\neq0\,,\;x_{kj}^2\neq0\,.
\end{align}
At multiplicity $n$, only $\nu_n=\tfrac{1}{2}n(n-5)$ of these cross ratios are independent. Since there are no cross ratios at four and five points, the $\alpha_i$ will be rational numbers in these cases. Unless the choice of the $\Omega_{n,i}$ has been extremely good, the $\alpha_i$ will depend on the cross-ratios for multiplicities greater than five. Nevertheless, it is straightforward to determine them using a numerical implementation of the BCFW recursion relation.  Evaluating both sides of \cref{eq:ansatz6d} for a given phase space point $\pi_j$ on a sufficiently large number of component amplitudes, the resulting linear equations can be solved for $\alpha_i(\pi_j)$. Numbering the cross ratios $\{u_1,u_2,\dots,u_{\nu_n}\}$ we make an ansatz for each of the coefficients
\begin{equation}\label{eq:ansatzAlpha_i}
 \alpha_i=\frac{\displaystyle a_0+\sum\limits_{i=1}^k\sum\limits_{\{n_j\}_k}a_{n_1\,\dots\, n_{\nu_n}}\prod\limits_{\sigma=1}^{\nu_n} u_{\sigma}^{n_{\sigma}}}{\displaystyle b_0+\sum\limits_{i=1}^k\sum\limits_{\{n_j\}_k}b_{n_1\,\dots\, n_{\nu_n}}\prod\limits_{\sigma=1}^{\nu_n} u_{\sigma}^{n_{\sigma}}}\,,
\end{equation}
where $\{n_j\}_k$ are all different distributions of $k$ powers among the cross ratios. Inserting the values of the cross ratios and the calculated values of the coefficients $\alpha_i(\pi_j)$ for a sufficiently large number of phase space points, the resulting linear equations can be solved for $\{a_I,b_I\}$. 

Some remarks are in order here. It is very important to randomly choose the set of component amplitudes used to calculate the $\alpha_i(\pi_j)$. As will be demonstrated later, 
picking only amplitudes of a particular sector, like e.\,g.~only gluon amplitudes, can lead to dual conformal extensions of this particular sector that are not equal to the full superamplitude. In practice one will successively increase the rank $k$ of the polynomials in \cref{eq:ansatzAlpha_i} until a solution is found. In order to not have to worry about numerical uncertainties or instabilities, we chose to use rational phase space points. Using momentum twistors it is straightforward to generating four-dimensional rational phase space points which can be used to obtain rational six-dimensional phase space points of the form $p_i^\mu=\{p_i^0,0,p_i^2,p_i^3,0,p_i^5\}$. Although these phase space points only have four non-zero components, they are sufficiently complex to yield non-zero results for all massive amplitudes \footnote{The only flaw in using them would have been the ruled out six-dimensional Levi-Civita tensors.}. The obvious benefit of the rational phase space points is that all found solutions to 
the 
ansatz \cref{eq:ansatz6d} are exact. An important property of the described method for the determination of the superamplitudes is that the obtained representations will contain only linear independent subsets of the basis functions $\Omega_{n,i}$. This may become an obstacle when looking for nice solutions with very simple coefficients $\alpha_i$ or ultimately for master formulae valid for arbitrary multiplicities since these not necessarily consist only of linear independent $\Omega_{n,i}$.

Essential for making the ansatz \cref{eq:ansatz6d} is the knowledge of the possible dual conformal covariant objects involving dual fermionic momenta $\theta_i$, $\tilde\theta_i$. Therefore we recall the inversion of the dual coordinates, compare \eqref{eq:inversion6d_first}-\eqref{eq:inversion6d_last},
\begin{align}
 I[x_{ij}^{AB}]&=-(x_i^{-1}x_{ij}x_j^{-1})_{AB}\,,& I[(x_{ij})_{AB}]&=-(x_i^{-1}x_{ij}x_j^{-1})^{AB}\,,\\
I[\theta_{i}^{A}]&=\theta_{i}^{B}(x^{-1}_{i})_{BA}\,,&I[\tilde\theta_{i\,A}]&=(x^{-1}_{i})^{AB}\tilde\theta_{i\,B}\,.
\end{align}
Clearly the objects
\begin{equation}
 \begin{gathered}\label{eq:Basis}
 \begin{aligned}
&\langle \theta_{i_1}|x_{i_1i_2}\dots x_{i_{2k-1}i_{2k}}|\theta_{i_{2k}}\rangle&\qquad\quad&[\tilde\theta_{i_1}|x_{i_1i_2}\dots x_{i_{2k-1}i_{2k}}|\tilde\theta_{i_{2k}}]\\ 
\end{aligned}\\
\langle \theta_{i_1}|x_{i_1i_2}\dots x_{i_{2k}i_{2k+1}}|\tilde\theta_{i_{2k+1}}]
\end{gathered}
\end{equation}
have inversion weight minus one on each of the appearing dual points but lack a translation invariance in $\theta$ and $\tilde\theta$. Fortunately there is a unique way to obtain manifest dual translation invariant objects from the dual conformal covariants \cref{eq:Basis}. We define the dual translation invariant objects
\begin{align}\label{}
 \langle B_{ijk}|&=\langle \theta_{ij}|x_{jk}x_{ki}|+\langle \theta_{ik}|x_{kj}x_{ji}|\,,&[ \widetilde{B}_{ijk}|&=[ \tilde\theta_{ij}|x_{jk}x_{ki}|+[ \tilde\theta_{ik}|x_{kj}x_{ji}|\,.
\end{align}
Because of 
$\langle B_{ijk}|=-|x_{ij}x_{jk}|\theta_{ki}\rangle-|x_{ik}x_{kj}|\theta_{ji}\rangle$
\footnote{In the sense of e.g.~$
\langle \theta_{ij}| x_{jk}x_{ki}|^{C}=\theta^{A}_{ij}\, x_{jk\, AB}\, x_{ki}^{BC}=
-x_{ik}^{CB}\, x_{kj\, BA}\, \theta_{ji}^{A}=-{}^{C}|x_{ik}\,x_{kj}|\theta_{ji}\rangle
$.} we define  $|B_{ijk}\rangle=-\langle B_{ijk}|$ and similar for the chiral conjugate. The dual conformal inversion properties become obvious if we expand them in $\theta$ and $\tilde\theta$, leading to
\begin{equation}
 \langle B_{ijk}|=-x_{jk}^2\langle \theta_i|+\langle \theta_{j}|x_{jk}x_{ki}|+\langle \theta_{k}|x_{kj}x_{ji}|\,.
\end{equation}
Hence, the dual conformal covariant, dual translation invariant building blocks for the superamplitudes are
\begin{align}
\langle  B_{i_1i_2i_3}|m_1\, \dots \,m_{2k}|B_{j_{1}j_{2}j_{3}}\rangle&=\langle  B_{i_1i_2i_3}|x_{i_1 m_1}x_{m_1 m_2} \dots x_{m_{2k}j_1}|B_{j_{1}j_{2}j_{3}}\rangle\,,\label{eq:BxB}\\
[  \widetilde{B}_{i_1i_2i_3}|m_1\, \dots \,m_{2k}|\widetilde{B}_{j_{1}j_{2}j_{3}}]&=[  \widetilde{B}_{i_1i_2i_3}|x_{i_1 m_1}x_{m_1 m_2}\dots x_{m_{2k}j_1}|\widetilde{B}_{j_{1}j_{2}j_{3}}]\,,\label{eq:BtxBt}
\intertext{and}
 \langle  B_{i_1i_2i_3}|m_1\, \dots \,m_{2k+1}|\widetilde{B}_{j_{1}j_{2}j_{3}}]&=\langle  B_{i_1i_2i_3}|x_{i_1 m_1}x_{m_1 m_2}\dots x_{m_{2k+1}j_1}|\widetilde{B}_{j_{1}j_{2}j_{3}}]\,.\label{eq:BBt}
\end{align}
They all have inversion weight minus one on every appearing dual point, e.\,g.
\begin{equation}
 I\left(\,\langle  B_{i_1i_2i_3}|m_1\, \dots \,m_{2k+1}|\widetilde{B}_{j_{1}j_{2}j_{3}}]\,\right)=\frac{\langle  B_{i_1i_2i_3}|m_1\, \dots \,m_{2k+1}|\widetilde{B}_{j_{1}j_{2}j_{3}}]}{x_{i_1}^2x_{i_2}^2x_{i_3}^2x_{m_1}^2\dots x_{m_{2k+1}}^2x_{j_1}^2x_{j_2}^2x_{j_3}^2}
\end{equation}
Keeping in mind that the degree in both $\theta$ and $\tilde\theta$ always increases by one if we successively increase the multiplicity, the last of the building blocks appears most natural. The first two building blocks necessarily appear in pairs and lead to a partial decoupling of the chiral and anti-chiral supermomenta. Consequently the building blocks \cref{eq:BxB,eq:BtxBt} alone cannot be sufficient to construct an even multiplicity amplitude. Furthermore they are very unfavorable from the four-dimensional perspective as the massless projection of amplitudes containing them has an obscured $R$ symmetry, for details we refer to \cref{section:uplift_huang}.  Although we found solutions to \cref{eq:ansatz6d} containing all three types of building blocks, we will neglect the building blocks \cref{eq:BxB,eq:BtxBt} in what follows. 

To be more precisely we will try to find representations of the superamplitudes with the general form
\begin{equation}\label{eq:MasterFormula}
 f_n=\sum_{I\,J\,K}\beta_{IJK}\prod_{i=1}^{n-4}\langle B_{I_i}|J_i|\widetilde{B}_{K_i}]\,,
\end{equation}
where the coefficients $\beta_{IJK}$ are functions of the dual conformal covariants $x_{ij}^2$ with the correct mass dimension and the correct inversion weights on each of the dual points in the multi-indices $I$, $J$, $K$. Manifest symmetry under chiral conjugation implies $\beta_{IJK}=(-1)^{n-4}\beta_{KJI}$.

Clearly not all of the building blocks \eqref{eq:BBt} are independent. All simple relations follow from
\begin{gather}\label{eq:propertiesB}
\begin{aligned}
  \langle B_{i\,j\,k}|&=\langle B_{i\,k\,j}|\,,\qquad&\langle B_{i\,i+1\,k}|&=-\langle B_{i+1\,i\,k}|\,,\\
\langle B_{i\,j\,j+1}|&=0\,,\qquad&\langle B_{i\,i+1\,j}|x_{i\,i+1}&=0\,,
\end{aligned}\\
\intertext{and}
\langle B_I|\dots\, i\, j\, k\, j\, l\,\dots |\widetilde{B}_J]=-x_{jk}^2 \langle B_I|\dots\, i\, l\,\dots |\widetilde{B}_J]\,.
\end{gather}

\subsection{Compact analytical results}
\subsubsection{The four and five-point amplitudes}\label{section:A5}
As an instructive illustration of the severe restrictions the dual conformal covariance, \cref{eq:inversionA6d}, puts on the functional form of the superamplitudes, we consider the four point amplitude. Indeed, dual conformal covariance fixes the four point amplitude up to a constant and the only possible ansatz is
\begin{equation}\label{eq:ansatzA4}
f_{4} = \frac{\alpha}{x^2_{1 3} x^2_{2 4}}\,.
\end{equation}
The constant can be fixed by performing the dimensional reduction onto any massless four-dimensional amplitude. For the MHV gluon amplitude with negative helicity gluons at positions three and four we obtain
\begin{align}
 A_4(1^1_{\;\;\dot{1}},2^1_{\;\;\dot{1}},3^2_{\;\;\dot{2}},4^2_{\;\;\dot{2}}) &= \frac{\alpha}{x^2_{1 3} x^2_{2 4}} \langle 1^{1} 2^{1} 3^{2} 4^{2}\rangle \left[ 1_{\dot{1}} 2_{\dot{1}} 3_{\dot{2}} 4_{\dot{2}}\right]\notag\\ 
&= \frac{\alpha}{x^2_{1 3} x^2_{2 4}}
\det\begin{pmatrix} 0 & 0 & \lambda_{3\alpha} & \lambda_{4\alpha}\\  \tilde{\lambda}_1^{\dot{\alpha}} & \tilde{\lambda}_2^{\dot{\alpha}} & 0 & 0\end{pmatrix} 
\det\begin{pmatrix}0 & 0 & \lambda_3^{\alpha} & \lambda_4^{\alpha}\\  -\tilde{\lambda}_{1\dot{\alpha}}  & -\tilde{\lambda}_{2\dot{\alpha}}  & 0 & 0\end{pmatrix}\notag\\
&= \frac{\alpha \left<3 4\right>^2 \left[1 2\right]^2}{\left<1 2\right> \left[2 1\right] \left<2 3\right> \left[3 2\right]} = - \alpha\frac{\left<3 4\right>^4}{\left<1 2\right> \left<2 3\right> \left<3 4\right> \left<4 1\right>} \,.
\end{align}
Comparison with the well known Parke-Taylor formula yields $\alpha=-i$. This trivial calculation should be compared to the comparably complicated calculation using the BCFW recursion in references \cite{Cheung:2009dc,Dennen:2009vk}.

Recalling the known result for the five point amplitude, \cref{eq:5pkt_6d}, we want to find the most simple representation of $f_5$ that is manifest dual conformal covariant. Hence we are searching for dual translation invariant functions of mass dimension minus five, that are of degree one in both $\theta$ and $\tilde\theta$ and invert as
\begin{equation}\label{Inversion_5Pkt}
I[f_{5}] =  x^{2}_{1} x^{2}_{2} x^{2}_{3} x^{2}_{4} x^{2}_{5} f_{5}\,.
\end{equation}
The most simple dual conformal covariant building blocks invariant under chiral conjugation are given by
\begin{align}\label{eq:def_Omega}
\Omega_{i\,j\,k\,l\,m}&:=\tfrac{1}{2}\left(\langle B_{ijl}|\widetilde{B}_{ikm}]-\langle B_{ikm}|\widetilde{B}_{ijl}]\right)\,,&&\text{with}&I[\Omega_{i\,j\,k\,l\,m}]&=\frac{\Omega_{i\,j\,k\,l\,m}}{x_i^2x_j^2x_k^2x_l^2x_m^2}\,.
\end{align}
Obviously $\Omega_{ijklm}$ is  zero if less than three of its indices are distinct. From the properties of $\langle B_{ijk}|$, \cref{eq:propertiesB}, and its definition above follow the properties
\begin{equation}\label{eq:propertiesOmega}
\begin{aligned}
 \Omega_{i\,j\,k\,l\,m}&=\Omega_{i\,l\,k\,j\,m}\,,\qquad&\Omega_{i\,j\,k\,l\,m}&=-\Omega_{i\,k\,j\,m\,l}\,,\qquad&\Omega_{i\,i+1\,k\,l\,m}&=-\Omega_{i+1\,i\,k\,l\,m}\,,\\
\Omega_{i\,j\,k\,j+1\,m}&=0\,.& \Omega_{i\,j\,k\,j\,m}&=0\,&\Omega_{i\,i\,k\,j\,m}&=0\,.
\end{aligned}
\end{equation}
At five point level the indices of $\Omega_{i\,j\,k\,l\,m}$ need to be a permutation of $\{1,2,3,4,5\}$. Applying the symmetry properties \eqref{eq:propertiesOmega} to all these permutations of the indices reveal that they are either zero or up to a sign equal to $\Omega_{1\,2\,3\,4\,5}$. Furthermore $\Omega_{1\,2\,3\,4\,5}$ is cyclically symmetric
\begin{equation}
 \Omega_{1\,2\,3\,4\,5}=\Omega_{1\,2\,3\,4\,5\,1}=\Omega_{3\,4\,5\,1\,2}=\Omega_{4\,5\,1\,2\,3}=\Omega_{5\,1\,2\,3\,4}\,,
\end{equation}
and has the reflection symmetry
\begin{equation}
\Omega_{1\,2\,3\,4\,5}=-\Omega_{5\,4\,3\,2\,1}\,.
\end{equation}
Therefore the simplest possible structure for the five point amplitude is
\begin{align}\label{eq:ansatz_f5}
 \frac{\Omega_{1\,2\,3\,4\,5}}{x_{13}^2x_{24}^2x_{35}^2x_{41}^2x_{52}^2}&&&\text{with}&I\left[\frac{\Omega_{1\,2\,3\,4\,5}}{x_{13}^2x_{24}^2x_{35}^2x_{41}^2x_{52}^2}\right]&=x_1^2x_2^2x_3^2x_4^2x_5^2\frac{\Omega_{1\,2\,3\,4\,5}}{x_{13}^2x_{24}^2x_{35}^2x_{41}^2x_{52}^2}\,.
\end{align}
Since there are no dual conformal invariant cross ratios at five point level, we know that \cref{eq:ansatz_f5} is either up to a constant equal to $f_5$ or  we need to make a more complicated ansatz including the building blocks $\langle  B_{ijk}|x_{il}x_{lk}|\widetilde{B}_{kmn}]$. Comparing this ansatz with the numerical BCFW recursion we indeed find the beautiful result
\begin{equation}\label{eq:f_5symmetric}
f_{5} = -i \frac{\Omega_{1\,2\,3\,4\,5}}{x^{2}_{13} x^{2}_{24} x^{2}_{35} x^{2}_{4 1} x^{2}_{5 2}}\,.
\end{equation}
This remarkably compact representation of the five point amplitude makes all continuous and discrete symmetries of the superamplitude manifest. Interestingly it can be simplified even more if we do not require manifest symmetry under chiral conjugation. On the support of the momentum and supermomentum conserving delta functions $\langle B_{124}|\widetilde{B}_{135}]$ is symmetric under chiral conjugation
\begin{equation}\label{eq:selfconjugate}
 \langle B_{124}|\widetilde{B}_{135}]=-\langle B_{135}|\widetilde{B}_{124}]
\end{equation}
and the five point amplitude is given by
\begin{equation}\label{eq:A5compact}
 \mathcal{A}_5=-i\delta^{(4)}(q)\delta^{(4)}(\tilde q)\frac{\langle B_{124}|\widetilde{B}_{135}]}{x^{2}_{13} x^{2}_{24} x^{2}_{35} x^{2}_{4 1} x^{2}_{5 2}}\,.
\end{equation}
This is the most compact dual conformal covariant expression of the five point amplitude 
available and should be compared to the form \eqref{eq:5pkt_6d} of  \cite{Bern:2010qa}. Making the dual conformal properties manifest led to a significant simplification.
Another manifest dual conformal covariant representation has been reported in \cite{Huang:2011um} by uplifting the four-dimensional five point amplitude of non-chiral superspace. We will discuss the potential uplift of massless four-dimensional amplitudes in \cref{section:uplift_huang}.
\subsubsection{The six-point amplitude}\label{section:A6}
As it turned out, the four and also the five point amplitudes were trivial examples of our general ansatz \cref{eq:ansatz6d}, since the coefficients $\alpha_i$ were constants. At six points they will in general no longer be constant but rational functions of the three  dual conformal invariant cross ratios
\begin{align}\label{crossrations_definition}
u_{1} &= \frac{x_{13}^{2} x_{4 6}^{2}}{x_{14}^{2} x_{3 6}^{2}}\,,& \qquad u_{2} &= \frac{x_{15}^{2} x_{2 4}^{2}}{x_{14}^{2} x_{2 5}^{2}}\,,& \qquad  u_{3} &= \frac{x_{2 6}^{2} x_{3 5}^{2}}{x_{2 5}^{2} x_{3 6}^{2}}\,.
\end{align}
Similar to the five point case we try to find a representation of the six point amplitude using only the simplest of the building blocks of \cref{eq:BBt}. To further reduce the resulting basis, we require chiral symmetry of the building blocks. Hence we only use the $\Omega_{i\,j\,k\,l\,m}$ defined in \cref{eq:def_Omega}. In contrast to five points the objects $\Omega_{i\,j\,j\,l\,m}$ are not all zero at multiplicity six. Nevertheless, we neglect them and stick to the $\Omega_{i\,j\,k\,l\,m}$ with distinct indices. What we are left with are the six building blocks
\begin{equation}\label{BB_Bloecke}
\begin{aligned}
\Omega_{1} &:= \Omega_{1\,2\,3\,4\,5}\,,\\[+0.2cm] \Omega_{2} &:= \Omega_{2\,3\,4\,5\,6}\\[+0.2cm]
\Omega_{3} &:= \Omega_{3\,4\,5\,6\,1}\,,\\[+0.2cm] \Omega_{4} &:= \Omega_{4\,5\,6\,1\,2}\,, \\[+0.2cm]
\Omega_{5} &:= \Omega_{5\,6\,1\,2\,3}\,,\\[+0.2cm] \Omega_{6} &:= \Omega_{6\,1\,2\,3\,4}
\end{aligned} 
\end{equation}
The basis of fifteen terms that we built from the $\Omega_{i}$ is
\begin{equation}
\Omega_{i\,j}=\frac{\beta_{ij}\Omega_i \Omega_j}{x^2_{13}x^2_{24}x^2_{35}x^2_{46}x^2_{51}x^2_{62}} 
\end{equation}
where the $\beta_{ij}$ cancel out the inversion weights of the four overlapping indices present in $\Omega_i \Omega_j$. Because of the existence of the three cross ratios, $\beta_{ij}$ are not uniquely fixed. One possible choice is
\begin{equation}\label{eq:betaij}
\beta_{ij}=\begin{pmatrix}
            0&(x_{24}^2 x_{35}^2)^{-1}&(x_{14}^2 x_{35}^2)^{-1}&(x_{15}^2 x_{24}^2)^{-1}&(x_{13}^2 x_{25}^2)^{-1}&(x_{13}^2 x_{24}^2)^{-1}\\
0&0&(x_{35}^2 x_{46}^2)^{-1}&(x_{25}^2x_{46}^2)^{-1}&(x_{26}^2 x_{35}^2)^{-1}&(x_{24}^2 x_{36}^2)^{-1}\\
0&0&0&(x_{15}^2 x_{46}^2)^{-1}&(x_{35}^2 x_{46}^2)^{-1}&(x_{13}^2 x_{46}^2)^{-1}\\
0&0&0&0&(x_{15}^2 x_{26}^2)^{-1}&(x_{14}^2  x_{26}^2)^{-1}\\
0&0&0&0&0&(x_{13}^2 x_{26}^2)^{-1}\\
0&0&0&0&0&0
\end{pmatrix}
\end{equation}
We exclude terms of the form $(\Omega_i)^2$ and make the following ansatz for the five point amplitude
\begin{equation}\label{eq:ansatz_f6}
 f_6=\sum_{i<j}\alpha_{ij}\Omega_{i\,j}\,.
\end{equation}
with $\alpha_{ij}=\alpha_{ij}(u_1,u_2,u_3)$ being a rational function of the cross ratios. Making an ansatz of the form \cref{eq:ansatzAlpha_i} it is straightforward to determine the $\alpha_{ij}$. The first observation is that out of our fifteen basis elements only eleven are linearly independent, leading to a large number of different representations of the form
\eqref{eq:ansatz_f6}. The highly nontrivial linear relations between the $\Omega_{i\,j}$ are only valid on the support of the momentum and supermomentum conserving delta functions and can be determined in the same way as the amplitude. The two ten-term and two eleven-term identities involving complicated functions of the cross ratios can be used to transform a particular solution to \cref{eq:ansatz_f6} to any other solution of this form. The complexity of the coefficients $\alpha_{ij}$ varies largely with the choice of linear independent $\Omega_{i\,j}$ in the solution, e.\,g.~some solutions involve rational functions of degree twelve in the cross rations $u_i$. The three simplest of the solutions involve nine $\Omega_{i\,j}$ and rational functions of degrees less than three. One particular of these simple solutions is
\begin{equation}\label{eq:solutionf6}
\left(\alpha_{ij}\right)=\frac{1}{1+u_1-u_2-u_3}
 \begin{pmatrix}
  0&u_2 u_3&-u_3&u_2
   \left(u_3-u_1\right)&0&0\\
0&0&0&0&u_3\left(u_2-u_1\right)&-u_2\\
0&0&0&0&-u_2&u_1
   \left(u_2+u_3\right)\\
0&0&0&0&u_2 u_3&-u_3\\
0&0&0&0&0&0\\
0&0&0&0&0&0
 \end{pmatrix}\,.
\end{equation}
Inserting the coefficients, the definitions of $\Omega_{ij}$ and the cross rations $u_i$, as well as the identity
\begin{equation}
\Tr \left(1\, 2\, 3\, 5\, 6\, 4\right) = x^{2}_{1 4} x^{2}_{2 5} x^{2}_{3 6} \left(1 + u_1 - u_2 - u_3\right)\,,
\end{equation}
into the ansatz \cref{eq:ansatz_f6}, the six point amplitude reads
\begin{equation}\label{eq:f_6}
\begin{aligned}
f_{6} &= \frac{1}{x^{2}_{1 3} x^{2}_{2 4} x^{2}_{3 5} x^{2}_{4 6} x^{2}_{5 1} x^{2}_{6 2}} \frac{i}{\Tr \left(1\, 2\, 3\, 5\, 6\, 4\right)}\Biggl(- x^{2}_{26} \Omega_{1} \Omega_{3} - x^{2}_{1 5} \Omega_{2} \Omega_{6} - x^{2}_{2 4} \Omega_{3} \Omega_{5}- x^{2}_{35}  \Omega_{4} \Omega_{6} \\
&\={} \hspace{3.5cm} +\frac{x_{26}^2 x_{15}^2}{x_{25}^{2}} \Omega_{1} \Omega_{2}+\frac{x_{24}^2 x_{35}^2}{x_{25}^{2}} \Omega_{4} \Omega_{5}+  \left(\frac{x_{15}^{2} x_{2 4}^{2}}{x_{2 5}^{2}} - \frac{x_{13}^{2} x_{4 6}^{2}}{x_{3 6}^{2}}\right) \Omega_{2} \Omega_{5}\\
&\= {} \hspace{3.5cm}+ \left(\frac{x_{2 6}^{2} x_{3 5}^{2}}{x_{2 5}^{2}} - \frac{x_{13}^{2} x_{4 6}^{2}}{x_{14}^{2}}\right) \Omega_{1} \Omega_{4} + \left(\frac{x_{15}^{2} x_{2 4}^{2}}{x_{14}^{2}} + \frac{x_{2 6}^{2} x_{3 5}^{2}}{x_{3 6}^{2}}\right) \Omega_{3} \Omega_{6}  \Biggr)\,.
\end{aligned} 
\end{equation}
Note that this representation of the six point amplitude has an unphysical pole $ u_2 + u_3 - u_1=1$, contained in the trace. From the analysis of the solutions to \cref{eq:ansatz_f6} we conclude that unphysical poles are a general feature for representations in terms of the $\Omega_{ij}$. Of course, all the unphysical poles are only spurious and cancel out if we project onto any component amplitude. The other two nine term solutions can be obtained by cyclic permutations of \cref{eq:f_6}.

Albeit all continuous symmetries and the symmetry under chiral conjugation of the six point amplitude are manifest in the solutions to \cref{eq:ansatz_f6}, the cyclic\footnote{\Cref{eq:f_6} is manifest invariant under the cyclic permutation $i\rightarrow i+3$.} and reflection symmetry are not obvious. However, there is no obstacle in finding manifest cyclically symmetric representations by constructing manifest cyclically symmetric basis elements  from the  $\Omega_i$. As a consequence of the manifest cyclic invariance of the basis, the coefficients in the general ansatz \cref{eq:ansatz6d} are cyclically symmetric as well, i.\,e.~are rational functions of symmetric polynomials of the cross ratios. 

There are three types of such manifest cyclically symmetric basis elements
\begin{equation}
\begin{aligned}
&g_1(u_1,u_2,u_3)\Omega_{12}+\text{five cyclic rotations}\\
&g_2(u_1,u_2,u_3)\Omega_{13}+\text{five cyclic rotations}\\
&g_3(u_1,u_2,u_3)\Omega_{14}+f_3(u_2,u_3,u_1)\Omega_{24}+f_3(u_3,u_1,u_2)\Omega_{36}
\end{aligned}
\end{equation}
The functions $g_i$ are arbitrary rational functions of the cross ratios leaving a lot of freedom to define a cyclic basis. Looking at the solution \cref{eq:solutionf6}, reasonable choices are $g_1\in \{u_1u_2,u_1u_3,u_2u_3\}$,    $g_2\in \{u_1,u_2,u_3\}$, and $g_3\in \{u_1(u_2\pm u_3),u_2(u_3\pm u_1),u_3(u_1\pm u_2)\}$. Indeed, this leads to a solution involving only three cyclically symmetric basis elements. Choosing $g_1=u_2u_3$,    $g_2=u_3$, and  $g_3=u_2(u_1+ u_3)$ we find
\begin{equation}
(\alpha_i)=\frac{1}{3-u_1-u_2-u_3}\begin{pmatrix}
1&-2&1
\end{pmatrix}
\end{equation}
or equivalently
\begin{multline}\label{eq:f6cyclic}
f_{6} = \frac{1}{x^{2}_{1 3} x^{2}_{2 4} x^{2}_{3 5} x^{2}_{4 6} x^{2}_{5 1} x^{2}_{6 2}} \frac{i}{x^{2}_{1 4} x^{2}_{2 5} x^{2}_{3 6}(3 - u_{1} - u_{2} - u_{3})}\left(\frac{x^{2}_{15}x^{2}_{26}}{x^{2}_{25}}\,\Omega_{1} \Omega_{2} - 2\, x^{2}_{26}\, \Omega_{1} \Omega_{3} + \right. \\
+ \left. \left(\frac{x^{2}_{13}x^{2}_{46}}{x^{2}_{14}} + \frac{x^{2}_{26}x^{2}_{35}}{x^{2}_{25}} \right) \Omega_{1} \Omega_{4} + \mbox{cyclic permutations}\right)\,.
\end{multline}
Clearly this representation is not minimal as it consists of all fifteen $\Omega_{ij}$. The contained unphysical pole at $u_{1} + u_{2} + u_{3} = 3$, might be expressed by the traces
\begin{equation}
\begin{gathered}
x^{2}_{1 4} x^{2}_{2 5} x^{2}_{3 6} \left(3 - u_{1} - u_{2} - u_{3}\right) = \tfrac{1}{2}\left(\;\Tr \left(1\, 2\, 3\, 5\, 6\, 4\right) + \text{cyclic permutations}\;\;\right)\,.
\end{gathered} 
\end{equation}

As emphasized in \cref{section:IdeaBCFW} it is very important to randomly choose the component amplitudes which are used to calculate the coefficients $\alpha_i$ in the general ansatz \cref{eq:ansatz6d}. Since we are dealing with a maximally supersymmetric theory one might wonder if it would not be sufficient to consider e.\,g.~only gluon amplitudes and let supersymmetry care for all other amplitudes. Indeed this is a widespread claim within the literature which can be easily disproved. In fact, only eight of the fifteen $\Omega_{ij}$ are linear independent on gluon amplitudes compared to eleven on all component amplitudes. Consequently, supersymmetrizing gluon amplitudes as has been done in reference \cite{Dennen:2009vk} for the three, four and five point amplitudes will not yield the correct superamplitude for multiplicities greater than five.
Having said that, it is nevertheless interesting to investigate how such a supersymmetrization of the gluon amplitudes looks like. Therefore we try to find a dual conformal invariant extension of the gluon amplitudes, that is a solution to \cref{eq:ansatz6d} valid on all gluon amplitudes. At six points we do not have to worry about six-dimensional Levi-Civita tensors and it is not necessary to use chiral self-conjugate building blocks. Instead of the $\Omega_i$ we use the building blocks
\begin{align}\label{eq:OmegaUpDown}
\Omega_{i,j}^u&:=\langle B_{i\,i+1\,i+3}|\widetilde{B}_{i\,i+2\,i+4}]\,,&\Omega_{i,j}^d&:=\langle B_{i\,i-1\,i-3}|\widetilde{B}_{i\,i-2\,i-4}]\,
\end{align}
where the label $j$ indicates that the indices $\{i,i\pm1,i\pm2,\pm3,i\pm4\}$  in $\Omega_{i,j}^{u/d}$ have to be taken modulo $j$. Whenever the label $j$ is equal to the multiplicity $n$, we will usually drop it. 
 The $\Omega_{i}^{u/d}$ are related to the chiral self-conjugate $\Omega_i$ by
\begin{equation}\label{eq:relationOmegaUD}
\Omega_i=\tfrac{1}{2}\left(\Omega_i^u-\Omega_{i+4}^d\right)\,.
\end{equation}
The resulting ansatz for the dual conformal extension of the gluon sector is
\begin{align}\label{eq:ansatzf6gluon}
f_6\bigr\rvert_{\text{gluons}}=\frac{i}{x^{2}_{1 3} x^{2}_{2 4} x^{2}_{3 5} x^{2}_{4 6} x^{2}_{5 1} x^{2}_{6 2}x^{2}_{1 4} x^{2}_{2 5} x^{2}_{3 6}}\Bigl(\sum_{i,j}\begin{aligned}[t]&\alpha_{ij}x_{i-1\,j-1}^2\Omega_i^{u}\Omega_j^{u}\\&\quad+\beta_{ij}x_{i-1\,j+1}^2\Omega_i^{u}\Omega_j^{d}\\&\quad\quad+\gamma_{ij}x_{i+1\,j+1}^2\Omega_i^{d}\Omega_j^{d}\Bigr)\end{aligned}
\end{align}
Since the gluon sector is not closed under dual conformal symmetry, the massless coefficients $\alpha_{ij}$, $\beta_{ij}$, $\gamma_{ij}$ are in general rational functions of the Lorentz invariants $x_{kl}^2$. As expected not all of the $\Omega_i^{u/d}\Omega_j^{u/d}$ are linear independent on the gluon amplitudes. A good indication that we will find dual conformal covariant solutions to \cref{eq:ansatzf6gluon} is the fact that all two term identities that the $\Omega_i^{u/d}\Omega_j^{u/d}$ fulfill on the gluon amplitudes are in fact dual conformal covariant. On the support of the momentum and supermomentum conserving delta functions we have for example the six identities
\begin{equation}\label{eq:2termIds}
\begin{aligned}
x_{i-1\,i+1}^2\Omega^{u}_i\Omega^{d}_i\bigr\rvert_{\text{gluons}}&=x_{i+2\,i+4}^2\Omega^{u}_{i+3}\Omega^{d}_{i+3}\bigr\rvert_{\text{gluons}}\,,\\
x_{i-1\,i+3}^2\Omega^{u}_i\Omega^{d}_{i+2}\bigr\rvert_{\text{gluons}}&=x_{i+2\,i}^2\Omega^{u}_{i+3}\Omega^{d}_{i+5}\bigr\rvert_{\text{gluons}}\,.
\end{aligned}
\end{equation}
Indeed there are 24  nice three term solutions to \cref{eq:ansatzf6gluon} that are all dual conformal covariant. One of these solutions is
\begin{equation}\label{eq:f6gluon}
f_6\bigr\rvert_{\text{gluons}}=\frac{i}{x^{2}_{1 3} x^{2}_{2 4} x^{2}_{3 5} x^{2}_{4 6} x^{2}_{5 1} x^{2}_{6 2}x^{2}_{1 4} x^{2}_{2 5} x^{2}_{3 6}} \left(x_{24}^2\Omega^d_3 \Omega^u_3 - x_{13}^2\Omega^d_2 \Omega^d_6-x_{26}^2\Omega^u_1 \Omega^u_3 \right)\,.
\end{equation}
Unfortunately, none of the found dual conformal extensions of the gluon sector were equal to the superamplitude. However, they all gave the correct ultra helicity violating (UHV) amplitudes.
\subsubsection{Towards higher multiplicities}\label{section:HigherMultiplicities}
Inspired by the compact representations \cref{eq:f_5symmetric,eq:f_6,eq:f6cyclic} the next logical step is to try to find a nice representation of the seven point amplitude with the ultimate goal of a master formula valid for arbitrary multiplicities. The main difficulty at higher multiplicities is to make a good choice for the basis $\Omega_{n,i}$ used to express the amplitude, compare \cref{eq:ansatz6d}. Going from five to six partons the number of terms in the amplitudes increased roughly by a factor of ten. Hence, the number of terms in the seven point amplitude is expected to be of order 100, making systematic studies of the solutions to \cref{eq:ansatz6d} impossible for multiplicities $n>6$. Furthermore, the generic solution $\alpha_i$  to \cref{eq:ansatz6d} contains complicated rational functions of the $\nu_n=\tfrac{1}{2}n(n-5)$ cross ratios which require a huge calculational effort to be obtained from the BCFW recursion using \cref{eq:ansatzAlpha_i}.

At seven points, a natural starting point is to use a basis constructed from products of the chiral self-conjugate $\Omega_{i\,j\,k\,l\,m}$ defined in \cref{eq:def_Omega}. Hence, the ansatz for the seven point amplitude reads
\begin{equation}\label{eq:ansatz_f7}
f_{7} = \sum_{I, J,K} \alpha_{I J K} \beta_{I J K} \Omega_{I} \Omega_{J} \Omega_{K}\,
\end{equation} 
where the coefficient $\beta_{I J K}$ are functions of the covariants $x_{ij}^2$ compensating the negative inversion weights in the dual points present in $\{I,J,K\}$.  The $\beta_{I J K}$ have mass dimension -22 and are straightforward to obtain by counting the inversion weights in $\{I,J,K\}$, compare \cref{eq:betaij}. The dimensionless $\alpha_{I J K}$ are rational functions of the seven cross ratios
\begin{equation}
\begin{gathered}
u_{i} = \frac{x^2_{i\, i + 2} x^2_{i+3\, i+6}}{x^2_{i\, i + 3} x^2_{i + 2\, i+6}}\,.
\end{gathered} 
\end{equation}
Even if we restrict the basis to products of distinct $\Omega_{I}$ and only consider $\Omega_{I}=\Omega_{i_1\,i_2\,i_3\,i_4\,i_5}$ with distinct indices we end up with more than $10^4$ basis elements. Solving for $\alpha_{I J K}(\pi_i)$ at different phase space points reveals that approximately $70$ basis elements are required to obtain a representation of the seven point amplitude. Analyzing different choices of a linear independent subset of basis elements, for none of them  the complexity of the coefficients $\alpha_{I J K}$ turned out to be sufficiently low to justify the computational effort necessary to determine their analytical form. Due to the astronomical number of different solutions to \cref{eq:ansatz_f7} it is impossible to decide whether or not simple solutions to it exist or if the restriction to use only the building blocks $\Omega_{ijklm}$ needs to be relaxed. 

Looking at the representations \cref{eq:f_5symmetric,eq:f_6,eq:f6cyclic} found for the five and six point amplitude we observe the prefactor
\begin{align}
&\frac{1}{\prod x^2_{ij}}\,,&&\text{with}&I\left[\frac{1}{\prod x^2_{ij}}\right]&=\frac{\left(\prod x_i^2\right)^{n-3}}{\prod x^2_{ij}}\,,
\end{align}
containing the product of all $\tfrac{1}{2}n(n-3)$ physical poles. It seems natural to expect this prefactor in a potential master formula for arbitrary multiplicities. With the definition
\begin{equation}\label{eq:defOmegaIJK}
\Omega_{I;J;K}=\tfrac{1}{2}\left(\langle B_I|J|\widetilde{B}_K]-\langle B_K|J|\widetilde{B}_I]\right)
\end{equation}
of the chiral self-conjugate building blocks $\Omega_{I;J;K}$ we can easily write down a nice ansatz valid for arbitrary multiplicities
\begin{align}\label{eq:MasterAnsatz}
f_n=\frac{1}{\prod x^2_{ij}}\sum_{I,J,K,L}\alpha_{I,J,K,L}\;\Omega_{I}\;\prod_{i=1}^{n-5}\Omega_{J_i;K_i;L_i}\,.
\end{align}
Here the sum goes over all multi-indices $I=\{i_1,i_2,i_3,i_4,i_5\}$, $J=\{J_1,\dots,J_{n-5}\}$, $K=\{K_1,\dots,K_{n-5}\}$, $L=\{L_1,\dots,L_{n-5}\}$ with $|J_i|=|L_i|=3$, $|K_i|=2i-1$, where $\{I,J,K,L\}$ is a permutation of $\{\{1\}^{n-4},\dots,\{n\}^{n-4}\}$. By construction the $\alpha_{I,J,K,L}$ are dimensionless and dual conformal invariant. Clearly the representation of the five point amplitude \cref{eq:f_5symmetric} is a solution to \cref{eq:MasterAnsatz} whereas the representations of the six point amplitude \cref{eq:f_6,eq:f6cyclic} are no solution to it. We leave it to future work to investigate whether there exist nice solutions to the master ansatz \cref{eq:MasterAnsatz} for multiplicities greater than five.

\subsection{The little group decomposition of the superamplitudes}\label{section:littleGroupDecomposition}
With regard to the MHV decomposition of the massless amplitudes of $\cN=4$ SYM it would be nice to have a similar decomposition of the massless six-dimensional amplitudes of $\mathcal{N} = (1,1)$ into sectors of varying complexity. Here we propose a decomposition of the amplitudes according to the violation of the $SU(2)\times SU(2)$ little group which is the six-dimensional analog of the MHV-band decomposition introduced for massive amplitudes on the Coulomb branch in 
$\mathcal{N} = 4$ SYM in \cite{Craig:2011ws}.

Our starting point is the decomposition of the six-dimensional superamplitude into the component amplitudes $A_{n}$
\begin{equation}\label{eq:decomposition6d}
\mathcal{A}_{n} = \sum_{I,J} \xi_{i_1a_1} \xi_{i_2a_2}\dots \xi_{i_na_n} \tilde{\xi}^{\dot{b}_1}_{j_1} \tilde{\xi}^{\dot{b}_2}_{j_2}\dots\tilde{\xi}^{\dot{b}_n}_{j_n} A_{n}\left(i_1^{a_1},i_2^{a_2},\dots ,i_n^{a_n},j_{1 \dot{b}_1}, j_{2 \dot{b}_2},\dots ,j_{n \dot{b}_n} \right)\,.
\end{equation}
Recall the connection of the six dimensional Grassmann variables to the non-chiral
four dimensional ones of section \cref{section:dimensional_reduction} under
dimensional reduction
\begin{align}\label{eq:mapEta}
\xi_{a} &= \left(\tilde{\eta}_{3},\eta^{1} \right)\,,&  \tilde{\xi}^{\dot{a}}& = \left(\tilde{\eta}_{2}, -\eta^{4}\right)\,,
\end{align}
which implied the reduction for the supermomenta
\begin{align}
q^{A} &= \left(\begin{array}{c} q_\alpha^1 \\  \tilde{q}^{\dot{\alpha}}_{3} \end{array}\right) \,,&
\tilde{q}_{A} & = \left(\begin{array}{cc} -q^{\alpha\,4} \, , & -\tilde{q}_{\dot{\alpha}\,2} 
 \end{array}\right)\,.
\end{align}

It is instructive to translate the MHV decomposition of the massless four-dimensional superamplitudes into six-dimensional language. Because of the $SU(4)_R$ symmetry the N${}^p$MHV superamplitude in chiral superspace has the Grassmann dependence
\begin{align}
\text{4d chiral superspace:}&&\mathcal{A}_n^{\text{N${}^p$MHV}}&=\mathcal{O}\left((\eta^1)^{p+2}(\eta^2)^{p+2}(\eta^3)^{p+2}(\eta^4)^{p+2}\right)
\end{align}
According to \cref{eq:relationOfsuperfields}, the chiral super field $\mathcal{A}_n( \varPhi_1,\dots,\varPhi_1)$ is related to the non-chiral superfield $\mathcal{A}_n( \varUpsilon_1,\dots,\varUpsilon_n)$ by the half Fourier transformation
\begin{align}\label{eq:FT}
\mathcal{A}_n( \varUpsilon_1,\dots,\varUpsilon_n) = \prod_{i}\int d\eta_i^{3} d\eta_i^{2} \;e^{ \eta_i^{2}\tilde{\eta}_{i\,2} +  \eta_i^{3}\tilde{\eta}_{i\,3}} \mathcal{A}_n( \varPhi_1,\dots,\varPhi_1)\,.
\end{align}
Consequently, the N${}^p$MHV superamplitude in non-chiral superspace has the Grassmann dependence
\begin{align}
\text{4d non-chiral superspace:}&&\mathcal{A}_n^{\text{N${}^p$MHV}}&=\mathcal{O}\left((\eta^1)^{p+2}(\tilde\eta_2)^{n-p-2}(\tilde\eta_3)^{n-p-2}(\eta^4)^{p+2}\right)\,.
\end{align}
With the help of the map \cref{eq:mapEta} between the four-dimensional and six-dimensional Grassmann variables we can deduce which of the six-dimensional component amplitudes $A_{n}\left(i_1^{a_1},\dots ,i_n^{a_n},j_{1 \dot{b}_1}, \dots ,j_{n \dot{b}_n} \right)$, defined in \cref{eq:decomposition6d}, correspond to massless four-dimensional N${}^p$MHV amplitudes
\begin{align}
\text{6d non-chiral superspace:}&&\mathcal{A}_n\bigr\rvert_{\text{N${}^p$MHV}}&=\mathcal{O}\left((\xi_1)^{n-p-2}(\xi_2)^{p+2}(\tilde\xi^{\dot{1}})^{n-p-2}(\tilde\xi^{\dot{2}})^{p+2}\right)\,.
\end{align}
Hence, the $SU(4)_R$ symmetry of the massless chiral superamplitudes in four dimensions leads to a Grassmann dependence of the form  $(\xi_1)^{n-a}(\xi_2)^{a}(\tilde\xi^{\dot{1}})^{n-a}(\tilde\xi^{\dot{2}})^{a}$ in six dimensions. From the six-dimensional perspective the Grassmann dependence of the superamplitudes in the massless four-dimensional limit is a consequence of  breaking the $SU(2) \times SU(2)$ little group to a $U(1)$ little group in four dimensions because  on the four-dimensional subspace the chiral and anti-chiral spinors $\lambda^A$ and $\tilde\lambda_A$ are equal.

In the case of the massive four dimensional amplitudes the $SU(4)_R$ symmetry is broken and the Grassmann dependence of the  corresponding six-dimensional superamplitude is no longer restricted, i.e.~all terms of the form $(\xi_1)^{n-a}(\xi_2)^{a}(\tilde\xi^{\dot{1}})^{n-b}(\tilde\xi^{\dot{2}})^{b}$ are appearing except the ones with $a,\,b\,\in\{0,n\}$. We then propose the following little group decomposition of the superamplitudes of 
$\cN=(1,1)$ SYM
\begin{align}
\mathcal{A}_n&=\sum_{a=1}^{n-1}\sum_{b=1}^{n-1}\mathcal{A}_n^{a\times b}\,,&&\text{with}&\mathcal{A}_n^{a\times b}&=\mathcal{O}\left((\xi_1)^{n-a}(\xi_2)^{a}(\tilde\xi^{\dot{1}})^{n-b}(\tilde\xi^{\dot{2}})^{b}\right)\,.
\end{align}
\begin{figure}[t]\label{Baender_6D_pic}
\centering
\includegraphics[width=0.8\linewidth]{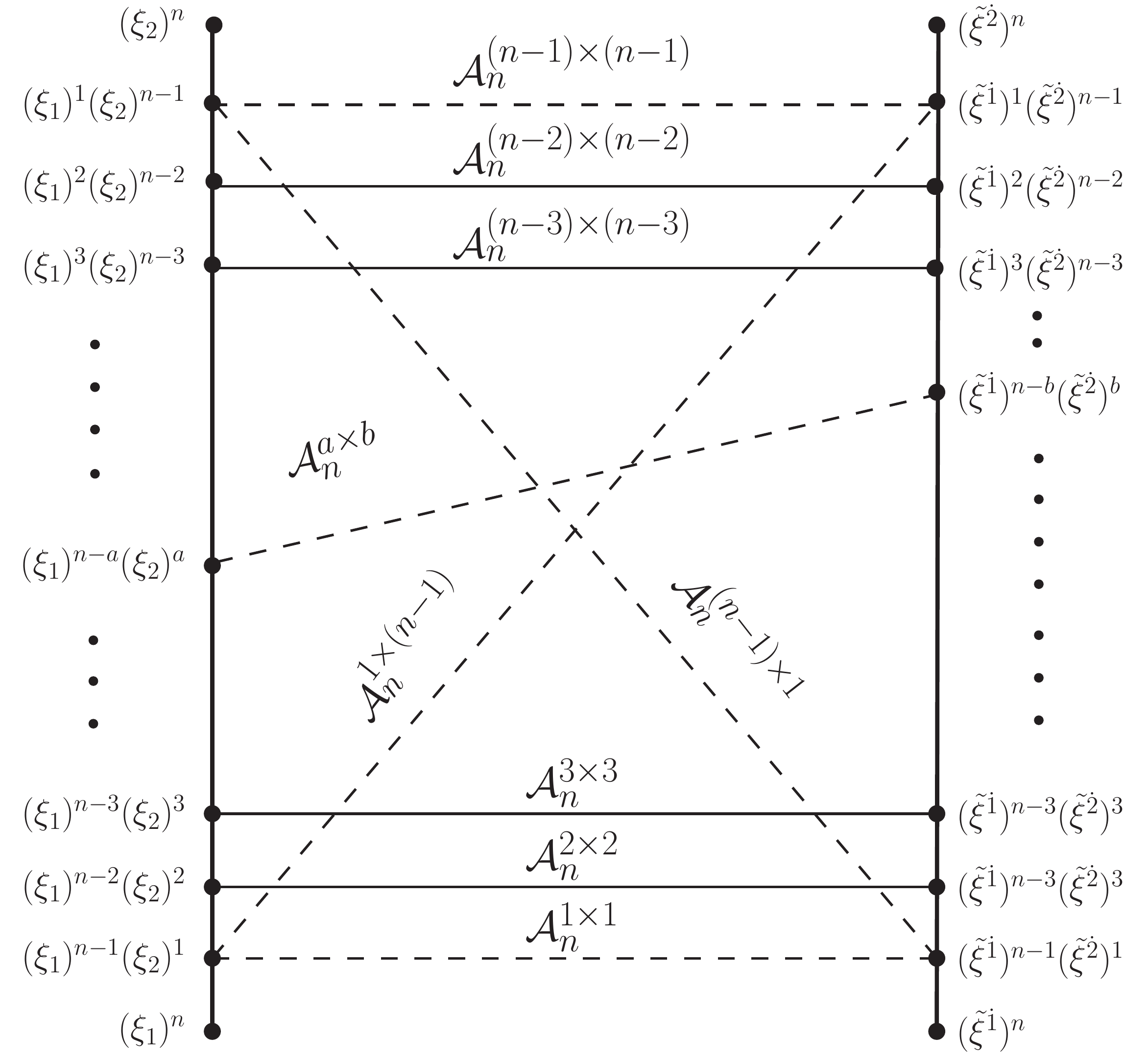}
\caption[Little group decomposition of the  $\mathcal{N} = (1,1)$ SYM amplitudes.]{Little group decomposition of the  $\mathcal{N} = (1,1)$ SYM amplitudes. The general  amplitude $\mathcal{A}^{ a \times b}_n$ has the Grassmann dependence $(\xi_1)^{n-a} (\xi_2)^{a} (\tilde{\xi}^{\dot{1}})^{n-b}(\tilde{\xi}^{\dot{2}})^{b}$. In the massless four-dimensional limit only $\mathcal{A}^{(p+2) \times (p+2)}_n$ with $p = 0,1,\dots,n-4$ are non-zero and give the N${}^p$MHV amplitudes (continuous horizontal lines). Some examples of amplitudes that are vanishing in the massless limit to four dimensions are represented by dashed lines.}
\end{figure}
This decomposition can be further motivated by translating the Grassmann dependence of $\mathcal{A}_n^{a\times b}$ back to chiral superspace using \cref{eq:mapEta,eq:FT}
\begin{align}
\text{6d:}&&&\mathcal{A}_n^{a\times b}&&\qquad\longrightarrow\qquad &&\text{4d:}&&\mathcal{O}\left((\eta^1)^{a}(\eta^3)^{a}(\eta^{2})^{b}(\eta^4)^{b}\right)
\end{align}
Hence the little group decomposition in six dimensions corresponds to breaking the four-dimensional $SU(4)_R$ symmetry to a $SU(2)_R\times SU(2)_R$ symmetry.

For the little group decomposition to be of use, the complexity of the $\mathcal{A}_n^{a\times b}$ should vary with the values of $a$ and $b$. In the massless  four-dimensional theory the simplest amplitudes are the MHV amplitudes.  In the massive case, gluon amplitudes with helicity configurations of the form $(+\,-\,\dots\,-)$ or $(-\,+\,\dots\,+)$ are no longer zero and belong to the ultra helicity violating (UHV) amplitudes. The UHV amplitudes are the simplest of the massive amplitudes and vanish in the massless limit.
Within the little group decomposition of the six-dimensional superamplitudes the UHV amplitudes are given by
\begin{align}\label{eq:defUHV}
\text{UHV amplitudes:}&&\mathcal{A}^{ 1 \times 1}_n\,,& &\mathcal{A}^{(n - 1) \times 1}_n\,,&&\mathcal{A}^{1 \times (n - 1)}_n\,,&&\mathcal{A}^{(n - 1) \times (n - 1)}_n\,.
\end{align}
Now that the simplest  parts of the superamplitude is identified, the numerical BCFW recursion relation can be used to investigate their analytical form.
\subsection{The UHV amplitudes in $\mathcal{N} = (1,1)$ SYM theory}\label{section:UHV}
Since the UHV amplitudes are not closed under the dual conformal symmetry, we cannot expect the coefficients $\alpha_i$ in our general ansatz \cref{eq:ansatz6d} to be dual conformal invariant. In general the coefficients $\alpha_i$ will be rational functions of the $\rho_n=\tfrac{1}{2}n(n-3)$ Lorentz invariants $\{x_{13}^2,x_{24}^2,\dots\}=:\{s_1,s_2,\dots,s_{\rho_n}\}$ and similar to \cref{eq:ansatzAlpha_i} they can be obtained by solving the linear equations derived from the ansatz
\begin{equation}\label{eq:ansatzXij}
  \alpha_i=\frac{\displaystyle \sum\limits_{\{n_i\}_k}a_{n_1\dots n_{\rho_n}}\prod\limits_{\sigma=1}^{\rho_n} s_{\sigma}^{n_\sigma}}{\displaystyle \sum\limits_{\{n_i\}_k}b_{n_1\dots n_{\rho_n}}\prod\limits_{\sigma=1}^{\rho_n} s_{\sigma}^{n_\sigma}}\,,
\end{equation}
where $\{n_j\}_k$ are all different distributions of $k$ powers among the Lorentz invariants. In contrast to the dual conformal invariant case, \cref{eq:ansatzAlpha_i}, numerator and denominator need to be homogeneous polynomials of equal degree $k$. 

\subsubsection{Six-point case}

To get an idea of the complexity of the UHV amplitudes we turn to the six point case and make the same ansatz as in \cref{eq:ansatzf6gluon} for the gluon sector 
\begin{align}\label{eq:ansatzf6UHV}
f_6^{\text{UHV}}=\frac{i}{x^{2}_{1 3} x^{2}_{2 4} x^{2}_{3 5} x^{2}_{4 6} x^{2}_{5 1} x^{2}_{6 2}x^{2}_{1 4} x^{2}_{2 5} x^{2}_{3 6}}\Bigl(\sum_{i,j}\begin{aligned}[t]&\alpha_{ij}x_{i-1\,j-1}^2\Omega_i^{u}\Omega_j^{u}\\&\quad+\beta_{ij}x_{i-1\,j+1}^2\Omega_i^{u}\Omega_j^{d}\\&\quad\quad+\gamma_{ij}x_{i+1\,j+1}^2\Omega_i^{d}\Omega_j^{d}\Bigr)\,,
\end{aligned}
\end{align}
where the $\Omega^{u/d}_{i}=\Omega^{u/d}_{i,n}$ were defined in \cref{eq:OmegaUpDown}.
Looking for solutions to \cref{eq:ansatzf6UHV}, the first observation is that only two of the $\Omega_i^{u/d}\Omega_j^{u/d}$ are linear independent. Further restricting to either the little group sectors $1\times 1\,\cup\,(n-1)\times (n-1)$ or  $1\times (n-1)\,\cup\,(n-1)\times 1$ every single term in \cref{eq:ansatzf6UHV} gives a solution. This is an impressive display of the simplicity of the Grassmann dependence of the UHV amplitudes as well as a belated justification of the use of the dual conformal covariant building blocks $\Omega_j^{u/d}$. Unfortunately, all the two term solutions to \cref{eq:ansatzf6UHV}
 have very complicated coefficients. In order to cure this fact we try to find non-minimal and ideally dual conformal covariant solutions with simple coefficients. Due to \cref{eq:relationOmegaUD} we already know four dual conformal covariant solutions to \cref{eq:ansatzf6UHV}, namely \cref{eq:f_6} and its cyclic rotations and \cref{eq:f6cyclic}. 
A key observation towards simple dual conformal covariant representations of the UHV amplitudes is that on the UHV amplitudes the basis elements $\Omega_i^{u/d}\Omega_j^{u/d}$ obey the same dual conformal covariant two term identities \eqref{eq:2termIds} as on the gluon amplitudes. Hence it is natural to look for nice three term solutions similar to ones obtained for the gluon sector. A very basic way to to find three term solutions to \cref{eq:ansatzf6UHV} is to fix one of the coefficients to some simple function of the cross ratios, e.\,g. $\alpha_{13}=g(u_1,u_2,u_3)$  and solve for the remaining coefficients. With regard to the nice representations found for the gluons \eqref{eq:f6gluon}, we start with the most simple choices possible $g=\pm1$. Indeed for $\alpha_{13}=-1$ we find four simple dual conformal covariant solutions 
\begin{equation}\label{eq:f6UHV}
\begin{aligned}
-i \,\left(x^{2}_{1 3} x^{2}_{2 4} x^{2}_{3 5} x^{2}_{4 6} x^{2}_{5 1} x^{2}_{6 2}x^{2}_{1 4} x^{2}_{2 5} x^{2}_{3 6}\right)\,f_6^{\text{UHV}}&=x_{24}^2\Omega^d_3 \Omega^u_3 - x_{13}^2\Omega^d_2 \Omega^d_6-x_{26}^2\Omega^u_1 \Omega^u_3 \\
&=x_{46}^2\Omega^d_3 \Omega^u_1 - x_{15}^2\Omega^d_4 \Omega^d_6-x_{26}^2\Omega^u_1 \Omega^u_3\\
&=x_{13}^2\Omega^d_6 \Omega^u_4 - x_{15}^2\Omega^d_4 \Omega^d_6-x_{26}^2\Omega^u_1 \Omega^u_3\\
&=x_{24}^2\Omega^d_6 \Omega^u_6 - x_{13}^2\Omega^d_2 \Omega^d_6-x_{26}^2\Omega^u_1 \Omega^u_3\,.\end{aligned}
\end{equation}
As it turns out, the 24 three-term solutions that can be obtained in this way exactly match the 24 gluon representations found in \cref{section:A6}, i.\,e.~the dual conformal extension of the UHV sector includes the gluon sector. This observation is highly nontrivial. At this point it is not clear whether this is a special feature of the six point amplitude or a multiplicity independent statement.

\subsubsection{UHV amplitudes with two massive legs at arbitrary multiplicities}\label{section:UHVtwoMass}
Motivated by compact formulae obtained in reference \cite{Craig:2011ws} for massive $\mathcal{N}=4$ SYM amplitudes with two neighboring massive legs, we investigate the UHV sector in the special kinematics where only the first two legs are massive from the four-dimensional point of view. By cyclic permutations of the indices this is straightforward to translate to the case were  another pair of consecutive legs is massive. In six-dimensional language this is equivalent to the restriction to phase space points of the form 
\begin{align}\label{eq:M1M2}
 p^{5}_{1} &= - p^{5}_{2}\,,&  p^{6}_{1} &= - p^{6}_{2}&&\text{and}& p^{5}_{i} &= p^{6}_{i} = 0&& \mbox{ for }&  i &= 3,\dots,n\,.
\end{align}
Similar to the four-dimensional calculation in reference \cite{Craig:2011ws} we are searching for a formula valid for all multiplicities. Therefore we make the recursive ansatz
\begin{equation}\label{eq:recursionAnsatz}
 f_{n}=f_{n-1}\left(\sum_{i}\alpha_i\Omega_{i,n}^u+\beta_i\Omega_{i,n}^d\right)\,,
\end{equation}
where at each recursion step we only use the $2n$ dual conformal covariant building blocks $\Omega_{i,n}^u$ defined in \cref{eq:OmegaUpDown}. Due to the special kinematics \cref{eq:M1M2} we do not have to worry about six-dimensional Levi-Civita tensors for multiplicities larger than six, hence there is no need for chiral self-conjugate building blocks. The coefficients $\alpha_i$, $\beta_i$ have mass dimension minus six and their functional dependence on the Lorentz invariants $x_{ij}^2$ can be obtained by modifying the ansatz \cref{eq:ansatzXij} accordingly. We successively determine the solutions to \cref{eq:recursionAnsatz} and at each multiplicity we keep all one term solutions and feed them back into the recursive ansatz \cref{eq:recursionAnsatz}. As initial data we take the ten equivalent representations of the full five point amplitude
following from \cref{eq:A5compact}, \cref{eq:selfconjugate} and the cyclic invariance of
the amplitude
\begin{align}
  f_5&=- \frac{i}{x^{2}_{13} x^{2}_{24} x^{2}_{35} x^{2}_{41} x^{2}_{52}}\Omega_{i,5}^u\,,&f_5&= \frac{i}{x^{2}_{13} x^{2}_{24} x^{2}_{35} x^{2}_{41} x^{2}_{52}}\Omega_{i,5}^d\,.
\end{align}
%
%obtained in \cref{section:A5}. 
Note that the discrete symmetries making the above 10 representations identical only
hold within five-point kinematics. Only the two $f_{n}$ of this set proportional to $\Omega_{1,5}^u$ or $\Omega_{5,5}^d$ yield one term solutions in the recursive construction of $f_6$ and out of the four one term solutions they produce again only two, namely
\begin{align}
f_6^{\text{UHV}}&=-\frac{i}{x^{2}_{13} x^{2}_{24} x^{2}_{35} x^{2}_{46} x^{2}_{51}x^{2}_{62}}\frac{\Omega_{1,5}^u\Omega_{4,6}^u}{x^{2}_{14}x^{2}_{25}}\\
\intertext{and}
f_6^{\text{UHV}}&=\frac{i}{x^{2}_{13} x^{2}_{24} x^{2}_{35} x^{2}_{46} x^{2}_{51}x^{2}_{62}}\frac{\Omega_{5,5}^d\Omega_{5,6}^u}{x^{2}_{13}x^{2}_{25}} \,.
\end{align}
lead to one term solutions for $f_7$. Interestingly both solutions are dual conformal covariant with inversion weight one on each dual point just like the full amplitude. 
Both solutions for $f_6^{\text{UHV}}$ nicely evolve through all subsequent recursion steps. Looking at the two representations they yield for the UHV amplitudes of multiplicity seven
\begin{align}
 f_7^{\text{UHV}}&=-\frac{i}{x^{2}_{13} x^{2}_{24} x^{2}_{35} x^{2}_{46} x^{2}_{57}x^{2}_{61}x^{2}_{72}}\Omega_{1,5}^u\frac{\Omega_{4,6}^u}{x^{2}_{14}x^{2}_{25}}\frac{\Omega_{5,7}^u}{x^{2}_{15}x^{2}_{26}}\\
\intertext{and}
f_7^{\text{UHV}}&=\frac{i}{x^{2}_{13} x^{2}_{24} x^{2}_{35} x^{2}_{46} x^{2}_{57}x^{2}_{61}x^{2}_{72}}\Omega_{5,5}^d\frac{\Omega_{5,6}^u}{x^{2}_{13}x^{2}_{25}}\frac{\Omega_{6,7}^u}{x^{2}_{13}x^{2}_{26}}\,,
\end{align}
it is straightforward to generalize them to arbitrary multiplicities. We conjecture the formulae
\begin{align}\label{eq:UHV1}
 f_n^{\text{UHV}}&=-\frac{i}{\prod x_{i\,i+2}}\Omega_{1,5}^u\prod_{i=6}^n\frac{\Omega_{i-2,i}^u}{x^{2}_{1\,i-2}x^{2}_{2\,i-1}}\\
\intertext{and}\label{eq:UHV2}
f_n^{\text{UHV}}&=\frac{i}{\prod x_{i\,i+2}}\Omega_{5,5}^d\prod_{i=6}^n\frac{\Omega_{i-1,i}^u}{x^{2}_{13}x^{2}_{2\,i-1}}\,,
\end{align}
to be valid representations for UHV amplitudes of multiplicities greater than four. Up to multiplicity $n=13$ both formulae have been checked by determining the solutions to the recursive ansatz \cref{eq:recursionAnsatz} which seems sufficient to us to consider \cref{eq:UHV1,eq:UHV2} to be proven. 

With regard to the three term solutions \eqref{eq:f6UHV} for all gluon and UHV amplitudes on general kinematics, we expect the formulae \cref{eq:UHV1,eq:UHV2} to be valid for other sectors as well. The natural guess is of course that the dual conformal extensions of the UHV amplitudes on the special kinematics \cref{eq:M1M2} produce the correct gluon amplitudes. However, this is not the case. The reason might be that the gluon sector does not undergo the same significant simplifications as the UHV sector if we specialize the kinematics. Fortunately the found dual conformal extensions of \cref{eq:UHV1,eq:UHV2} yield an even bigger class of amplitudes. We find the remarkable results that \cref{eq:UHV1} is equal to the superamplitude on all little group sectors of the form $1\times a$, $(n-1)\times a$, whereas \cref{eq:UHV2} is correct for the chiral conjugate little group sectors $a\times 1$, $a\times  (n-1)$. We indicate this by writing
\begin{align}\label{eq:master1}
 f_n^{1\times a\;(n-1)\times a}&=-\frac{i}{\prod x_{i\,i+2}}\Omega_{1,5}^u\prod_{i=6}^n\frac{\Omega_{i-2,i}^u}{x^{2}_{1\,i-2}x^{2}_{2\,i-1}}\\
\intertext{and}\label{eq:master2}
f_n^{a\times 1\;a\times  (n-1)}&=\frac{i}{\prod x_{i\,i+2}}\Omega_{5,5}^d\prod_{i=6}^n\frac{\Omega_{i-1,i}^u}{x^{2}_{13}x^{2}_{2\,i-1}}\,.
\end{align}
Clearly the chiral conjugate of the formula for $f_n^{1\times a\;(n-1)\times a}$ is an alternative representation of $f_n^{a\times 1\;a\times  (n-1)}$ and vice versa.
\section{From massless 4d to massless 6d superamplitudes }\label{section:uplift_huang}
A very exciting question, first discussed in \cite{Huang:2011um}, is whether or not it is possible to obtain the massless tree-level superamplitudes of the six-dimensional $\cN=(1,1)$ SYM by uplifting the massless non-chiral  superamplitudes of the four-dimensional $\cN=4$ SYM. If so, as claimed by the author of \cite{Huang:2011um}, this implies that the massive four-dimensional amplitudes of $\cN=4$ SYM can be obtained from the massless ones. Since the non-chiral superamplitudes of $\cN=4$ are straightforward to obtain using the well behaved non-chiral BCFW recursion, described in \cref{section:BCFWnonChiral}, such a correspondence could provide an easy way to obtain the tree amplitudes of $\cN=(1,1)$ SYM.

In this section we want to thoroughly investigate the potential uplift of the massless four-dimensional amplitudes and thereby clarify some points in \cite{Huang:2011um}.

\subsection{Dimensional reduction of $\mathcal{N} = (1,1)$ SYM revisited}\label{uplift_huang_diskussion}
As explained in detail in \cref{section:dimensional_reduction} performing the dimensional reduction of the superamplitudes of $\mathcal{N} = (1,1)$ to massless four dimensions yields the non-chiral superamplitudes of $\mathcal{N} = (1,1)$. The symmetries of the six-dimensional and four-dimensional superamplitudes have been discussed in detail in \cref{section:symmetries_N=4,generators_max_6d}. The most relevant in this discussion are the discrete symmetry under chiral conjugation and the $R$-symmetry of the four-dimensional superamplitudes. In particular the invariance under the $R$-symmetry generators $\mathpzc{m}_{\,n m}$ and $\widetilde{\mathpzc{m}}_{\,n' m'}$ implies that all $R$-symmetry indices within a superamplitude are contracted. 

With the help of the maps between the six-dimensional on-shell variables $\{\lambda_i^A$, $\tilde\lambda_{i\,A}$, $\xi_i^a$, $\tilde\xi_i^{\dot{a}}\}$ and the massless four-dimensional on-shell variables $\{\lambda_i^\alpha$, $\tilde\lambda_i^{\dot{\alpha}}$, $\eta_i^m$, $\tilde\eta_i^{m'}\}$ \cref{eq:grassmann_map,eq:MapSpinors6d4d} it is straightforward to obtain the projection of every six-dimensional object. Since there is a one-to-one map between the supermomentum conserving delta functions \eqref{eq:projektion_delta} we neglect them straight away and investigate the correspondence
\begin{align}\label{eq:uplift}
f_n^{\text{6d}}&&&\begin{gathered}[c]\underrightarrow{\quad\text{projection}\quad}\\\overleftarrow{\quad\;\;\,\text{uplift?}\;\;\,\quad}\end{gathered} && f_n^{\text{4d}}\,.
\end{align}
 The tree-level amplitudes of $\mathcal{N} = (1,1)$ SYM theory consist of  Lorentz invariant contractions of momenta $p_i$ and supermomenta $q_i$, $\tilde{q}_i$. The only purely bosonic Lorentz invariants are traces of an even number of  momenta $(k_i)_{AB}$, $(k_i)^{AB}$. However chiral conjugate traces project to the same four-dimensional  traces
\begin{align}
\begin{aligned}[c]\text{Tr}^{6d}(\Gamma_{+}\slashed{k}_1 \slashed{k}_2 \dots \slashed{k}_{2 n})&=(k_1)_{A_1A_2}\dots(k_{2n})^{A_{2n}A_1}\\\text{Tr}^{6d}(\Gamma_{-}\slashed{k}_1 \slashed{k}_2 \dots \slashed{k}_{2 n})&=(k_1)^{A_1A_2}\dots(k_{2n})_{A_{2n}A_1}\end{aligned}&&\longrightarrow & &\Tr{}^{4d}\left(\slashed{k}_1 \slashed{k}_2 \dots \slashed{k}_{2 n}\right) 
\end{align}
where $\slashed{k}_i$ denotes the contraction of the momentum $k_i$ with either the six-dimensional or the four-dimensional gamma matrices and $\Gamma_{\pm}=\tfrac{1}{2}(1\pm\gamma^7)$. Hence, the presence of traces in $f_n^{\text{6d}}$ that are not chiral self-conjugate would already spoil the uplift. The chiral conjugate traces differ by terms containing the six-dimensional Levi-Civita tensor. Since $\mathcal{N} = (1,1)$ SYM is a non-chiral theory it is symmetric under chiral conjugation $(p_i)_{AB}\leftrightarrow (p_i)^{AB}$, $q_i\leftrightarrow \tilde{q}_i$ and therefore free of six-dimensional Levi-Civita tensor. In conclusion, the only purely bosonic invariants in $f_n^{\text{6d}}$ are chiral self-conjugate traces whose projections can be uniquely uplifted from four dimensions
\begin{align}
\tfrac{1}{2}\Tr{}^{6d}\left(\slashed{k}_1 \slashed{k}_2 \dots \slashed{k}_{2 n}\right) &&\LR & & \Tr{}^{4d}\left(\slashed{k}_1 \slashed{k}_2 \dots \slashed{k}_{2 n}\right) 
\end{align}
 Inserting the definition of the gamma matrices, the four-dimensional trace may be written as the sum of two chiral conjugate traces of four-dimensional Pauli matrices
 \begin{equation}
 \Tr{}^{4d}\left(\slashed{k}_1 \slashed{k}_2 \dots \slashed{k}_{2 n}\right) =(k_1)_{\alpha_1\dot{\alpha}_2}\dots(k_{2n})^{\dot{\alpha}_{2n}\alpha_1}+(k_1)^{\dot{\alpha}_1\alpha_2}\dots(k_{2n})_{\alpha_{2n}\dot{\alpha}_1}
 \end{equation}
There are three possible Lorentz invariants containing supermomenta. All of them have a unique projection to four dimensions
\begin{align}
\langle q_{i}|k_1 \dots k_{2 n}|\tilde{q}_{j}] &&\longrightarrow & & \langle q^{1}_{i}|{k}_1  \dots k_{2 n}|q^{4}_{j} \rangle+ [\tilde{q}_{i 3}|k_1  \dots k_{2 n}|\tilde{q}_{j 2}]\label{eq:type1}\\
\langle q_{i}|k_1 \dots k_{2 n + 1}|q_{j}\rangle &&\longrightarrow && \langle q^{1}_{i}|k_1  \dots k_{2 n + 1}|\tilde{q}_{j 3}] - [\tilde{q}_{i 3}|k_1  \dots k_{2 n + 1}|q_{j}^{1}\rangle\label{eq:type2}\\
[\tilde{q}_{i}|k_1  \dots k_{2 n + 1}|\tilde{q}_{j}] &&\longrightarrow && \langle q^{4}_{i}|k_1 \dots k_{2 n + 1}|\tilde{q}_{j 2}] - [\tilde{q}_{i 2}|k_1  \dots k_{2 n + 1}|q_{j}^{4}\rangle\label{eq:type3}
\end{align}
Non-chirality of the four-dimensional superamplitudes implies their invariance under the exchanges $q^{1}_{i}\leftrightarrow \tilde{q}_{i 3}$ and  $q^{4}_{i}\leftrightarrow \tilde{q}_{i 2}$. Since Lorentz invariants of the last two types, \cref{eq:type2,eq:type3}, can only occur pairwise in a six-dimensional superamplitude, it follows that the projection of a six-dimensional superamplitude has always a manifest chiral symmetry in four dimensions. Apparently none of these three six-dimensional Lorentz invariants leads to a manifest $R$-symmetry in four dimensions. However, any reasonable representation of $f_n^{\text{4d}}$ has a manifest $R$-symmetry. In conclusion, a potential uplift of $f_n^{\text{4d}}$ to six-dimensions can only consist of building blocks whose projection to four dimensions is $R$-symmetric. From the investigation of the three types of six-dimensional Lorentz invariants and their projections, \cref{eq:type1,eq:type2,eq:type3},  it immediately follows that there is only one such object
\begin{equation} \label{eq:correspondence6d4d}
\begin{aligned}
 \langle q_{i}|k_1 \dots k_{2 n}|\tilde{q}_{j}]+[\tilde{q}_{i}|k_1 \dots k_{2 n}|q_{j}\rangle &&\LR &&  \langle q^{m}_{i}|{k}_1  \dots k_{2 n}|q_{j\,m} \rangle+ [\tilde{q}_{i m'}|k_1  \dots k_{2 n}|\tilde{q}_{j}^{m'}]
\end{aligned}
\end{equation}
Unlike the claim in \cite{Huang:2011um} there is no combination of six-dimensional Lorentz invariants  of the second and third type, \cref{eq:type2,eq:type3},  that has a $R$ invariant projection to four dimensions. For further details see \cref{sec:Connection_six_four_Invariants}.
 We conclude that if a correspondence of the form \cref{eq:uplift} exists, then the involved representations of $f_n^{\text{6d}}$ and $f_n^{\text{4d}}$ only contain the building blocks \cref{eq:correspondence6d4d}. As will be explained in the next section, for multiplicities larger than five this is a severe constraint on the representations of $f_n^{\text{6d/4d}}$.
\subsection{Uplifting massless superamplitudes from four to six dimensions}\label{section:check_uplift}
We want to discuss the implications of \cref{eq:correspondence6d4d}. At four point level $f_4^{\text{4d}}$ is purely bosonic and the uplift is trivial
\begin{align}
f_4^{\text{4d}}=-\frac{i}{x_{13}^2x_{24}^2} &&\Longrightarrow & &f_4^{\text{6d}}=-\frac{i}{x_{13}^2x_{24}^2}\,.
\end{align}
At five points, any representation of $f_5^{\text{4d}}$ that has a manifest $R$-symmetry and a manifest symmetry under chiral conjugation automatically only consists of the building blocks \cref{eq:correspondence6d4d}. Since any reasonable representation of $f_5^{\text{4d}}$ has a manifest $R$-symmetry and the chiral symmetry can be made manifest by replacing e.\,g.~the MHV part by the chiral conjugate of the $\overline{\text{MHV}}$ part, any representation of $f_5^{\text{4d}}$ can be uplifted to six dimensions. By uplifting the representation, \cref{eq:f_5_4d}, 
\begin{equation}
f_5^{\text{4d}}=
 \frac{i}{x_{1 3}^2 x_{2 4}^4 x_{2 5}^4 x_{3 5}^2 x_{4 1}^2}\left( \langle B_{5 4 2}|\, 1\, 2\, 3\,| B_{5 4 2}\rangle + 
   [ \tilde B_{5 4 2}|\, 1\, 2\, 3\,| \tilde B_{5 4 2}]\right)
\end{equation}
obtained from the BCFW recursion yields the following representation of the six-dimensional amplitude
\begin{align}\label{eq:f_5uplift}
 f_5^{\text{6d}}&=
 \frac{i}{x_{1 3}^2 x_{2 4}^4 x_{2 5}^4 x_{3 5}^2 x_{4 1}^2}\tfrac{1}{2}\left( \langle B_{5 4 2}|\, 1\, 2\, 3\,| \tilde{B}_{5 4 2}] + 
   [ \tilde B_{5 4 2}|\, 1\, 2\, 3\,|  B_{5 4 2}\rangle\right) \\
   &=i\frac{\Omega_{542;123;542}}{x_{1 3}^2 x_{2 4}^4 x_{2 5}^4 x_{3 5}^2 x_{4 1}^2}\,,
\end{align}
where the factor of $\tfrac{1}{2}$ originates from the definition \eqref{eq:defBB} and we inserted the definition of $\Omega_{I;J;K}$, \cref{eq:defOmegaIJK}. We checked numerically that \cref{eq:f_5uplift} is indeed equal to the five-point amplitude in six dimensions.

Unfortunately the uplift starts to be non-trivial already at multiplicity six.  Let $\{\Omega_i\}$ denote a set  of the chiral self-conjugate building blocks \eqref{eq:correspondence6d4d} for the six-dimensional superamplitudes
\begin{equation}
\Omega_i \qquad\LR \qquad  \omega_i+\tilde{\omega}_i
\end{equation}
where $\omega_i=\mathcal{O}(q^2)$ and $\tilde{\omega}_i=\mathcal{O}(\tilde{q}^2)$ are the chiral conjugates in the projection of $\Omega_i $. As a consequence of \cref{eq:correspondence6d4d} an uplift able representation of the six-point amplitudes has the form
\begin{equation}\label{eq:f_6uplift}
f_6^{\text{4d}}=\sum_{i,j}\alpha_{ij}(\omega_i+\tilde{\omega}_i)(\omega_j+\tilde{\omega}_j)
\end{equation}
and uplifts to
\begin{equation}
f_6^{\text{6d}}=\sum_{i,j}\alpha_{ij}\Omega_i \Omega_j\,.
\end{equation}
From \cref{eq:f_6uplift} it follows
\begin{align}
\left(f_6^{\text{4d}}\right)^{\text{MHV}}\!\!\!&=\sum_{i,j}\alpha_{ij}\tilde{\omega}_i\tilde{\omega}_j\,,\,\,\,&\left(f_6^{\text{4d}}\right)^{\text{NMHV}}\!\!\!&=\sum_{i,j}\alpha_{ij}(\omega_i\tilde{\omega}_j+\tilde\omega_i\omega_j)\,,\,\,\,&\left(f_6^{\text{4d}}\right)^{\overline{\text{MHV}}}\!\!\!&=\sum_{i,j}\alpha_{ij}\omega_i\omega_j\,.
\end{align}
Comparing this with the representation \cref{eq:f_6_4d} obtained for $f_6^{\text{4d}}$ from the BCFW recursion it is apparent that a generic representation of $f_6^{\text{4d}}$ does not have the form \cref{eq:f_6uplift} required for an uplift. In contrast to the five point case, making the chiral symmetry manifest does not solve the problem because the minimal helicity violating (minHV) NMHV amplitudes are independent of the MHV and $\overline{\text{MHV}}$ amplitudes. As a consequence, it is straightforward to turn a generic representation into the form
\begin{equation}
f_6^{\text{4d}}=\sum_{i,j}\beta_{ij}\omega_i\omega_j+\gamma_{ij}(\omega_i\tilde{\omega}_j+\tilde\omega_i\omega_j)+\beta_{ij}\tilde{\omega}_i\tilde{\omega}_j\,,
\end{equation}
but in general the coefficients $\beta_{ij}$ and $\gamma_{ij}$ are unrelated. This is the key issue, that to our mind has been overlooked in reference \cite{Huang:2011um}. As a result, finding any representation of $f_n^{\text{4d}}$ is not sufficient to obtain the six-dimensional amplitude. In fact, under the assumption that the uplift works, obtaining $f_n^{\text{6d}}$ is equivalent to finding a representation of the form 
\begin{equation}\label{eq:f_nuplift}
f_n^{\text{4d}}=\sum_{|I|=n-4}\alpha_{I}\prod_{k=1}^{n-4}(\omega_{i_k}+\tilde{\omega}_{i_k})\,,
\end{equation}
for the four-dimensional amplitude. While such a representation trivially uplifts to
\begin{equation}\label{eq:f_nlifted}
f_n^{\text{6d}}=\sum_{|I|=n-4}\alpha_{I}\prod_{k=1}^{n-4}\Omega_{i_k}\,,
\end{equation}
obtaining it is non-trivial and a rigorous proof that \cref{eq:f_nlifted} is always a valid representation of the six-dimensional superamplitude is still missing. Of course we could use a numerical implementation of the non-chiral BCFW recursion relation to determine a solution to an ansatz of the form \cref{eq:f_nuplift} but this is not easier than determining $f_n^{\text{6d}}$ directly, using the methods described in \cref{section:IdeaBCFW}.

Albeit it seems save to say that the uplift is of no practical relevance for the determination of the six-dimensional superamplitudes, it is still very fascinating from the theoretical point of view. It is intriguing that the correct representation of the MHV superamplitude 
\begin{equation}\label{eq:f_nMHV}
(f_n^{\text{4d}})^{\text{MHV}}=\sum_{|I|=n-4}\alpha_{I}\prod_{k=1}^{n-4}\tilde{\omega}_{i_k}\,,
\end{equation}%
might be sufficient to get the whole six-dimensional  superamplitude, or equivalently all massive four-dimensional amplitudes.

One thing that would immediately invalidate the uplift are identities of the $\omega_{i}+\tilde{\omega}_{i}$ that do not uplift to identities of the $\Omega_i$. Though we do not have a concrete counterexample for the uplift, there are indeed four-dimensional identities of strings of momenta $k_i$ that do not have a six-dimensional counterpart, i.\,e.
\begin{align}\label{eq:identity4d}
\text{4d:}&& (p_ik_1\dots k_{2n}p_i)_{\a}^{\;\;\b} &=\left(\,|i\rangle[i|k_1\dots k_{2n}|i]\langle i|\,\right)_{\a}^{\;\;\b}=-(p_ik_{2n}\dots k_1p_i)_{\a}^{\;\;\b}\\
\intertext{but}
\text{6d:}&& (p_ik_1\dots k_{2n}p_i)_{A}^{\;\;B} &=\left(\,|i_{\dot{a}}][i^{\dot{a}}|k_1\dots k_{2n}|i^b\rangle\langle i_b|\,\right)\!\!{}_{A}^{\;\;B}\neq-(p_ik_{2n}\dots k_1p_i)_{A}^{\;\;B}
\end{align}
At this point we do not see how such identities could not spoil the uplift without restricting  the  allowed four-dimensional building blocks.
 
Using the numerical implementation of the six-dimensional BCFW recursion it is possible to numerically check the uplift. The easiest way to do so is to make an ansatz \eqref{eq:ansatz6d} for $f_n^{\text{6d}}$ using only the minimal building blocks $\Omega_{ijkl}$ defined in \cref{eq:BBt} and determine a solution $\alpha_i(\pi)$ for a massless phase space point with momenta of the form $\{p_i^1,p_i^2,p_i^3,p_i^4,0,0\}$. Since the coefficients are functions of the Lorentz invariants $x_{ij}^2$ they have identical numerical values on the 'massive' phase space point  $\{p_i^1,0,p_i^3,p_i^4,0,p_i^2\}$ and we can check whether the obtained coefficients $\alpha_i(\pi)$ provide a solution to the massive amplitudes as well. In fact, we checked that up to multiplicity eight that representation of the massless non-chiral amplitudes containing only the minimal building blocks $\langle B_{ijk}|B_{ilm}\rangle+[ \tilde{B}_{ijk}|\tilde{B}_{ilm}]$ did always uplift to six dimensions.  Since the eight-point amplitude is 
already 
very complicated, there is no reason to believe that the uplift of a representation containing only the minimal building blocks will fail beyond eight points. In case of more complicated building blocks the identities \eqref{eq:identity4d} might become an issue even at multiplicities lower than eight.

\section{Conclusion and outlook}

A central motivation for this work was to take first steps towards a generalization of the 
analytic construction of massless QCD amplitudes from $\mathcal{N}=4$ SYM ones of \cite{Dixon:2010ik,Schuster:2013aya,Melia:2013epa} to massive QCD amplitudes by employing $\cN=4$ SYM superamplitudes 
on the Coulomb branch. For this we constructed all standard and hidden symmetries of the massless six-dimensional superamplitudes of $\cN=(1,1)$ SYM theory thereby correcting small mistakes in the proof of the dual conformal symmetry given in \cite{Dennen:2010dh}. We exploited the symmetries of the six-dimensional amplitudes to derive the symmetries of massive tree amplitudes in $\mathcal{N}=4$ SYM theory and showed that the five dimensional dual conformal symmetry of the massive amplitudes leads to the presence of non-local Yangian-like generators $m^{(1)}$, $p^{(1)}$ associated to the masses and momenta in on-shell superspace. An interesting open question is whether or not there exist level-one supermomenta as 
well.

Furthermore, we explained how  analytical formulae for tree-level superamplitudes of $\mathcal{N}=(1,1)$ SYM can be obtained from a numerical implementation of the BCFW recursion relation. The developed method is very general and can be applied to other theories as well. We used it to derive compact manifest dual conformally covariant representations of the five- and six-point superamplitudes. To facilitate the investigation of the six-dimensional superamplitudes we proposed a little group decomposition of them. The little group decomposition is the six-dimensional analog of the MHV-band decomposition in 4d introduced in \cite{Craig:2011ws}. It  allows a separation into parts of varying complexity as well as the identification of those 
pieces of the superamplitude that survive in the massless limit to four-dimensions. We exploited the little group decomposition to study UHV amplitudes leading to arbitrary multiplicity formulae valid for large classes of component amplitudes with two consecutive 
massive legs. 

We demonstrated that within a maximally supersymmetric theory it is not always sufficient to consider only gluon amplitudes and the remaining amplitudes follow by supersymmetry. Indeed, the supersymmetrization of the six-dimensional gluon amplitudes, as has been done in reference \cite{Dennen:2009vk} for the three, four and five point amplitudes,  will not necessarily yield the correct superamplitude for multiplicities greater than five. We  derived examples of supersymmetric, dual conformally covariant representations of the gluon sector which do not coincide with the superamplitude. Nevertheless, we observed that dual conformal extensions and consequently supersymmetrizations of subsets of amplitudes reproduce at least part of the other component amplitudes. It would be interesting to investigate this in more detail in the future since finding dual conformal extensions of subsets of amplitudes is much simpler than finding the whole superamplitude.

In \cite{Huang:2011um} it has been claimed that all superamplitudes of $\mathcal{N}=(1,1)$ SYM can be obtained by uplifting massless tree-level superamplitudes of $\mathcal{N}=4$ SYM in non-chiral superspace. In our work we derived the superconformal and dual superconformal symmetries of the non-chiral superamplitudes and used the non-chiral BCFW recursion to prove the dual conformal symmetry as well as to derive the five and six-point superamplitudes.  We thoroughly investigated the implications of a potential uplift by identifying the correct four- and six-dimensional Lorentz invariants that should appear in such a correspondence. By performing numerical checks we confirmed the uplift of representations containing only a restricted set of dual conformal covariant and chiral self-conjugate building blocks up to multiplicity eight. However, we proved that finding a representation of the massless non-chiral superamplitudes of $\mathcal{N}=4$ SYM that can be uplifted is non-trivial for 
multiplicities larger than five. One possible flaw of the uplift are identities of the four-dimensional building blocks that do not uplift to identities of the corresponding six-dimensional building blocks. We gave  examples of such identities that need to be avoided by restricting the allowed building blocks in order to not spoil the uplift. Despite being of no practical relevance for the determination of the six-dimensional superamplitudes or the massive four-dimensional amplitudes at this point, it is still very fascinating to note that the correct representation of the non-chiral MHV superamplitudes in four dimensions 
could be sufficient to obtain all six-dimensional superamplitudes, or equivalently all massive four-dimensional amplitudes on the Coulomb branch of $\mathcal{N}=4$ SYM theory.

\acknowledgments

We thank H.~Elvang and M.~Kiermaier for discussions. JP thanks the Pauli Center for Theoretical 
Studies Z\"urich
and the Institute for Theoretical Physics at the ETH Z\"urich for hospitality and
support in the framework of a visiting professorship.

\appendix
\section{Spinor Conventions}
%%%%%%%%%%%%%%%%%%%%%%%%%%%%%%%%%%%%%%%%%%%%%%%%%%%%%%%%%%%%%%%%%%%%%%%%%%%%%%%%
\label{appendix:Spinors}
In this appendix we summarize our convention for the four- and six-dimensional spinors and provide the identities relevant for calculations within the spinor helicity formalism.  
\subsection{Four-Dimensional Spinors}
Raising and lowering of spinor indices is defined by left multiplication with the $\epsilon$ symbol and its inverse:
\begin{align}
\lambda_\alpha&=\epsilon_{\alpha\beta}\lambda^\beta\,,&\lambda^\alpha&=\epsilon^{\alpha\beta}\lambda_\beta\,,\\
\tilde{\lambda}_{\dot{\alpha}}&=\epsilon_{\dot{\alpha}\dot{\beta}}\tilde{\lambda}^{\dot{\beta}}\,,&\tilde{\lambda}^{\dot{\alpha}}&=\epsilon^{\dot{\alpha}\dot{\beta}}\tilde{\lambda}_{\dot{\beta}}\,,
\end{align}
where the antisymmetric $\epsilon$ symbol is defined as
\begin{align}
\epsilon&=i\sigma_2&
\epsilon_{12}&=\epsilon_{\dot{1}\dot{2}}=-\epsilon^{12}=-\epsilon^{\dot{1}\dot{2}}=1\,
\end{align}
and is obeying the equations
\begin{equation}
\begin{aligned}\label{Schouten_4D_1}
\epsilon_{\alpha\beta}\epsilon^{\beta\gamma}&=\delta_\alpha^\gamma\,,&\epsilon_{\dot{\alpha}\dot{\beta}}\epsilon^{\dot{\beta}\dot{\gamma}}&=\delta_{\dot{\alpha}}^{\dot{\gamma}}\,,\\
\epsilon_{\beta \gamma} \delta^{\alpha}_{\delta} + \epsilon_{\gamma \delta} \delta^{\alpha}_{\beta} +\epsilon_{\delta \beta} \delta^{\alpha}_{\gamma} &= 0&    \epsilon^{\dot{\beta}\dot{\gamma}} \delta^{\dot{\delta}}_{\dot{\alpha}} +\epsilon^{\dot{\gamma}\dot{\delta}} \delta^{\dot{\beta}}_{\dot{\alpha}} + \epsilon^{\dot{\delta}\dot{\beta}} \delta^{\dot{\gamma}}_{\dot{\alpha}} &= 0
\end{aligned}
\end{equation}
For the spinor products we choose the conventions
\begin{align}
\ang{\lambda}{\mu}&=\lambda^\alpha\mu_\alpha&&\text{and}	& [\tilde{\lambda}\,\tilde{\mu}]&=\tilde{\lambda}_{\dot{\alpha}}\tilde{\mu}^{\dot{\alpha}}\,,
\end{align}
which implies
\begin{align}
\lambda_\alpha\mu_\beta-\lambda_\beta\mu_\alpha&=\epsilon_{\alpha\beta}\,\ang{\lambda}{\mu}&
\tilde\lambda_{\dot\alpha}\tilde\mu_{\dot\beta}-\tilde\lambda_{\dot\beta}\tilde\mu_{\dot\alpha}&=-\epsilon_{\dot\alpha\dot\beta}\,[\tilde\lambda\,\tilde\mu]
\end{align}
The four-dimensional sigma matrices are defined as
\begin{align}
\sigma^\mu_{\phantom{\mu}\,\alpha\dot{\alpha}}&=(1,\vec{\sigma})_{\alpha\dot{\alpha}}&&\text{and}&\bar{\sigma}^{\mu\,\dot{\alpha}\alpha}&=(1,-\vec{\sigma})^{\dot{\alpha}\alpha}
\end{align}
and have the properties
\begin{align}
\sigma^{\mu}\bar{\sigma}^{\nu}+\sigma^{\nu}\bar{\sigma}^{\mu}&=2\eta^{\mu\nu}\,,&\bar{\sigma}^{\mu}\sigma^{\nu}+\bar{\sigma}^{\nu}\sigma^{\mu}&=2\eta^{\mu\nu}\,,\\
\sigma^\mu_{\alpha\dot{\alpha}}\bar{\sigma}^{\dot{\beta}\beta}_\mu&=2\delta_\alpha^\beta\delta_{\dot{\alpha}}^{\dot{\beta}}\,,&\sigma^{\mu\,\alpha\dot{\beta}}&=\bar{\sigma}^{\mu\,\dot{\beta}\alpha}\,,
\end{align}
which are consequences of the properties of the ordinary three-dimensional Pauli matrices $\vec{\sigma}=\begin{pmatrix}\sigma_1&\sigma_2&\sigma_3\end{pmatrix}$
\begin{align}
\sigma_1&=\begin{pmatrix}0&1\\1&0\end{pmatrix}\,,&\sigma_2&=\begin{pmatrix}0&-i\\i&0\end{pmatrix}\,,&\sigma_3&=\begin{pmatrix}1&0\\0&-1\end{pmatrix}\,.
\end{align}
Raising and lowering of spinor indices on derivatives with respect to a spinor leads to an additional minus sign
\begin{align}
\frac{\partial}{\partial \lambda_\alpha}=\frac{\partial \lambda^\beta}{\partial \lambda_\alpha}\frac{\partial}{\partial \lambda^\beta}=-\epsilon_{\alpha\beta}\frac{\partial}{\partial \lambda^\beta}\,,
\end{align}
which is a general feature of derivatives carrying $su(2)$ indices.
\subsection{Six dimensional Spinors}
The six-dimensional Pauli matrices fulfill the algebra
\begin{equation}\label{eq:Sigma}
\Sigma^\mu\widetilde\Sigma^\nu+\Sigma^\nu\widetilde\Sigma^\mu=2\eta^{\mu\nu}\,.
\end{equation}
We choose the antisymmetric representation
\begin{align}\label{eq:Pauli6d}
\Sigma^0&=i\sigma_1\otimes\sigma_2\,,&\widetilde\Sigma^0&=-\Sigma^0\,,\\
\Sigma^1&=i\sigma_2\otimes\sigma_3\,,&\widetilde\Sigma^1&=\Sigma^1\,,\\
\Sigma^2&=-\sigma_2\otimes\sigma_0\,,&\widetilde\Sigma^2&=-\Sigma^2\,,\\
\Sigma^3&=-i\sigma_2\otimes\sigma_1\,,&\widetilde\Sigma^3&=\Sigma^3\,,\\
\Sigma^4&=-\sigma_3\otimes\sigma_2\,,&\widetilde\Sigma^4&=-\Sigma^4\,,\\
\Sigma^5&=i\sigma_0\otimes\sigma_2\,,&\widetilde\Sigma^5&=\Sigma^5\,.
\end{align}
They satisfy the following identities
\begin{align}
\Sigma^\mu_{AB}&=\tfrac{1}{2}\epsilon_{ABCD}\widetilde\Sigma_{\mu}^{CD}\,,&\widetilde\Sigma_{\mu}^{AB}&=\tfrac{1}{2}\epsilon^{ABCD}\Sigma^\mu_{CD}\\
\Sigma^\mu_{AB}\Sigma_{\mu\,CD}&=-2\epsilon_{ABCD}\,, &
\widetilde\Sigma^{\mu\,AB}\widetilde\Sigma_{\mu}^{CD}&=-2\epsilon^{ABCD}\,,\\
\widetilde\Sigma_\mu^{AB}\Sigma^{\mu}_{CD}&=-2(\delta_C^A\delta_D^B-\delta_C^B\delta_D^A)\,,&\Tr(\widetilde\Sigma^\mu\Sigma^\nu)&=4\eta^{\mu\nu}\,.
\end{align}
The six dimensional Shouten identity reads
\begin{equation}
\delta_A^F\epsilon_{BCDE}+\delta_B^F\epsilon_{CDEA}+\delta_C^F\epsilon_{DEAB}+\delta_D^F\epsilon_{EABC}+\delta_E^F\epsilon_{ABCD}=0\,,
\end{equation}
and contractions of epsilon tensors may be deduced from
\begin{multline}
\epsilon_{ABCD}\epsilon^{EFGD}=\delta_A^E\delta_B^F\delta_C^G+\delta_A^F\delta_B^G\delta_C^E+\delta_A^G\delta_B^E\delta_C^F
-\delta_C^E\delta_B^F\delta_A^G-\delta_C^F\delta_B^G\delta_A^E-\delta_C^G\delta_B^E\delta_A^F
\end{multline}
The first four of the six dimensional sigma matrices are simply related to the Weyl representation of the four dimensional gamma matrices
\begin{align}\label{eq:SigmaGamma}
\Sigma^\mu&= 1\otimes\epsilon\cdot\gamma^\mu=\begin{pmatrix}
                                        0&-\sigma^{\mu\,\alpha}_{\phantom{\mu\,\alpha}\dot{\beta}}\\
\bar{\sigma}^{\mu\,\phantom{\dot\alpha}\beta}_{\phantom{\mu\,}\dot\alpha}&0
                                       \end{pmatrix}\,,&\widetilde\Sigma^\mu&= \gamma^\mu\cdot1\otimes\epsilon^{-1}=\begin{pmatrix}
                                        0&-\sigma_{\phantom{\mu}\alpha}^{\mu\,\phantom{\alpha}\dot\beta}\\
\bar\sigma^{\mu\,\dot\alpha}_{\phantom{\mu\,\dot\alpha}\beta}&0
                                       \end{pmatrix}\,.
\end{align}
\subsubsection{Three-Point Kinematics}\label{appendix:threePoint}
The three-point kinematics
\begin{align}
 p_1+p_2+p_3&=0\,, &p_i^2=0
\end{align}
imply the vanishing of all invariants
\begin{equation}
 p_1\cdot p_2=p_1\cdot p_3=p_2\cdot p_3=0\,. 
\end{equation}
As a consequence of \cref{eq:invariants}, the spinor products $\langle i|j]$ have rank one and posses a bispinor representation. A consistent set of spinors $\{u_i,\tilde u_i\}$ associated to the external legs has been introduced by C. Cheung and D. O'Connell in \cite{Cheung:2009dc} and reads
\begin{align}
 \langle i_a|j_{\dot{a}}]&=u_{i\,a}\tilde{u}_{j\,\dot{a}}\,,& \langle j_a|i_{\dot{a}}]&=-u_{j\,a}\tilde{u}_{i\,\dot{a}}\,,&&\text{for $\{i,j\}$ cyclic.}
\end{align}
Due to momentum conservation, these spinors are subject to the constraints
\begin{align}\label{eq:P1}
 u_1^a\langle 1_a|=u_2^a\langle 2_a|=u_3^a\langle 3_a|\,,&\tilde{u}_1^{\dot a}[ 1_{\dot a}|&=\tilde{u}_1^{\dot a}[ 1_{\dot a}|=\tilde{u}_1^{\dot a}[ 1_{\dot a}|\,.
\end{align}
Furthermore pseudoinverses of the spinors can be introduced
\begin{align}\label{eq:P2}
 u_aw_b-u_bw_a&=\epsilon_{ab}\,,&\tilde{u}_{\dot{a}}\tilde{w}_{\dot{b}}-\tilde{u}_{\dot{b}}\tilde{w}_{\dot{a}}&=\epsilon_{\dot{a}\dot{b}}\,.
\end{align}
In order to reduce the redundancy in the definition of the spinors $w_i$ and $\tilde{w}_i$ it is convenient to impose the constraints
\begin{align}\label{eq:P3}
 w_1^a\langle 1_a|+ w_2^a\langle 2_a|+ w_3^a\langle 3_a|&=0\,,& \tilde{w}_1^{\dot{a}}[ 1_{\dot{a}}|+\tilde{w}_2^{\dot{a}}[ 2_{\dot{a}}|+\tilde{w}_3^{\dot{a}}[ 3_{\dot{a}}|&=0\,.
\end{align}

\section{The Non-Chiral Superconformal Algebra} \label{sec:Algebra_Non_Chiral}
The $su(2)\times su(2)$ Lorentz generators $\mathds{M}_{\alpha \beta}$, $\overline{\mathds{M}}_{\dot{\alpha} \dot{\beta}}$ and the $su(2) \times su(2)$ $R$-symmetry generators $\mathfrak{M}_{n m}$, $\widetilde{\mathfrak{M}}_{n'm'}$ act canonically on the remaining generators carrying Lorentz and $R$-symmetry indices: 
\begin{align}
 \left[\mathds{M}_{\alpha \beta},\mathds{M}^{\gamma \delta}\right] &= \delta_{(\beta}^{\;(\gamma} \mathds{M}_{\alpha)}^{\;\;\;\delta)} &[\overline{\mathds{M}}_{\dot{\alpha} \dot{\beta}}, \overline{\mathds{M}}^{\dot{\gamma} \dot{\delta}}] &= \delta_{(\dot{\beta}}^{\;(\dot{\gamma}} \overline{\mathds{M}}_{\dot{\alpha})}^{\;\;\;\dot{\delta})} \\ 
 [\mathds{M}_{\alpha \beta}, \mathds{P}^{\gamma \dot{\delta}}]  & = \delta_{(\beta}^{\;\gamma}  \mathds{P}_{\alpha)}^{\;\dot{\delta}} & [\overline{\mathds{M}}_{\dot{\alpha} \dot{\beta}}, \mathds{P}^{\gamma \dot{\delta}}] &= \delta_{(\dot{\beta}}^{\;\dot{\delta}}  \mathds{P}_{\;\dot{\alpha})}^{\gamma}  \\
 [\mathds{M}_{\alpha \beta}, \mathds{K}^{\gamma \dot{\delta}}] &= -\delta_{(\beta}^{\;\gamma}  \mathds{K}_{\alpha)}^{\;\dot{\delta}} & [\overline{\mathds{M}}_{\dot{\alpha} \dot{\beta}}, \mathds{K}^{\gamma \dot{\delta}}] &= -\delta_{(\dot{\beta}}^{\;\dot{\delta}}  \mathds{K}_{\;\dot{\alpha})}^{\gamma} \\
[\mathds{M}_{\alpha \beta}, \mathds{Q}^{\gamma n}] &= \delta_{(\beta}^{\gamma} \mathds{Q}^{n}_{\alpha)} & [\overline{\mathds{M}}_{\dot{\alpha} \dot{\beta}}, \overline{\mathds{Q}}^{\dot{\gamma}}_{n}] &= \delta_{(\dot{\beta}}^{\dot{\gamma}} \overline{\mathds{Q}}_{\dot{\alpha}) n} \\
 [\mathds{M}_{\alpha \beta}, \overline{\widetilde{\mathds{Q}}}^{\gamma n'}] &= \delta_{(\beta}^{\gamma} \overline{\widetilde{\mathds{Q}}}^{n'}_{\alpha)}&  [\overline{\mathds{M}}_{\dot{\alpha} \dot{\beta}}, \widetilde{\mathds{Q}}^{\dot{\gamma}}_{n'}] &= \delta_{(\dot{\beta}}^{\dot{\gamma}} \widetilde{\mathds{Q}}_{\dot{\alpha}) n'}\\
[\mathds{M}_{\alpha \beta}, \mathds{S}^{\gamma}_{n}] &= - \delta^{\gamma}_{(\beta} \mathds{S}_{\alpha) n} & [\overline{\mathds{M}}_{\dot{\alpha} \dot{\beta}}, \overline{\mathds{S}}^{\dot{\gamma} n}] &= - \delta^{\dot{\gamma}}_{\;(\dot{\beta}} \overline{\mathds{S}}^{n}_{\dot{\alpha})} \\
[\mathds{M}_{\alpha \beta}, \overline{\widetilde{\mathds{S}}}^{\gamma}_{n'}]&= - \delta^{\gamma}_{(\beta} \overline{\widetilde{\mathds{S}}}_{\alpha) n'} & [\overline{\mathds{M}}_{\dot{\alpha} \dot{\beta}}, \widetilde{\mathds{S}}^{\dot{\gamma} n'}] &= - \delta^{\dot{\gamma}}_{\;(\dot{\beta}} \widetilde{\mathds{S}}^{n'}_{\dot{\alpha})}\\
 \left[\mathfrak{M}_{n m},\mathfrak{M}^{k l}\right] &= \delta_{(m}^{\;(k} \mathfrak{M}_{n)}^{\;l)} & [\widetilde{\mathfrak{M}}_{n' m'}, \widetilde{\mathfrak{M}}^{k' l'}] &= \delta_{(m'}^{\;(k'} \widetilde{\mathfrak{M}}_{n')}^{\;l')} \\
 [\mathfrak{M}_{n m}, \mathfrak{P}^{k l'}]   &= \delta_{(m}^{\;k}  \mathfrak{P}_{n)}^{\;l'} & [\widetilde{\mathfrak{M}}_{n' m'}, \mathfrak{P}^{k l'}] &= \delta_{(m'}^{\;l'}  \mathfrak{P}_{\;n')}^{k} \\
 [\mathfrak{M}_{n m}, \mathfrak{K}^{k l'}] &= - \delta_{(m}^{\;k}  \mathfrak{K}_{n)}^{\;l'} & [\widetilde{\mathfrak{M}}_{n' m'}, \mathfrak{K}^{k l'}] &= - \delta_{(m'}^{\;l'}  \mathfrak{K}_{\;n')}^{k} \\
[\mathfrak{M}_{n m}, \mathds{Q}^{\gamma k}] &= \delta_{(m}^{k} \mathds{Q}^{\gamma}_{n)} & [\widetilde{\mathfrak{M}}_{n' m'}, \widetilde{\mathds{Q}}^{k'}_{\dot{\alpha}}] &= \delta_{(n'}^{k'} \widetilde{\mathds{Q}}_{\dot{\alpha} m')} \\
 [\mathfrak{M}_{n m}, \overline{\mathds{S}}_{\dot{\gamma}}^{k}] &= \delta_{(m}^{k} \overline{\mathds{S}}_{\dot{\gamma} n)} & [\widetilde{\mathfrak{M}}_{n' m'}, \overline{\widetilde{\mathds{S}}}^{k'}_{\alpha}] &= \delta_{(n'}^{k'}\overline{\widetilde{\mathds{S}}}_{\alpha m')} \\
[\mathfrak{M}_{n m}, \mathds{S}_{\alpha}^{k}] &= - \delta^{k}_{(n} \mathds{S}_{\alpha m)} & [\mathfrak{M}_{n m}, \overline{\mathds{Q}}_{\dot{\alpha}}^{k}] &= - \delta^{k}_{(n} \overline{\mathds{Q}}_{\dot{\alpha} m)} \\ [\widetilde{\mathfrak{M}}_{n' m'}, \widetilde{\mathds{S}}_{\dot{\alpha}}^{k'}] &= - \delta_{(m'}^{\;k'} \widetilde{\mathds{S}}_{\dot{\alpha} n')}  & [\widetilde{\mathfrak{M}}_{n' m'}, \overline{\widetilde{\mathds{Q}}}_{\alpha}^{k'}] &= - \delta_{(m'}^{\;k'} \overline{\widetilde{\mathds{Q}}}_{\alpha n')}
\end{align}
The action of the dilatation $\mathds{D}$ and hypercharge $\mathds{B}$ on a generator $\mathds{G}$ is given by:
\begin{equation}
\left[\mathds{D}, \mathds{G} \right] =\dim\left(\mathds{G}\right) \mathds{G} \qquad \left[\mathds{B}, \mathds{G} \right] = \hyp\left(\mathds{G}\right) \mathds{G} 
\end{equation} 
The non-zero dimensions and hypercharges of the various generators are
\begin{equation}
\begin{gathered}
\begin{aligned}
\dim\left(\mathds{P}\right)& = 1\,, \qquad &\dim\left(\mathds{Q}\right) &= \dim \left(\widetilde{\mathds{Q}}\right) =  \dim\left(\overline{\mathds{Q}}\right) =  \dim (\overline{\widetilde{\mathds{Q}}}) = \tfrac{1}{2}\,,\\
\dim\left(\mathds{K}\right) &= -1\,, \qquad& \dim\left(\mathds{S}\right) &=  \dim(\widetilde{\mathds{S}}) = \dim\left(\overline{\mathds{S}}\right) =  \dim (\overline{\widetilde{\mathds{S}}}) = - \tfrac{1}{2}\,, 
\end{aligned}\\
\begin{aligned}
 \hyp\left(\mathds{Q}\right) &=  \hyp(\overline{\widetilde{\mathds{Q}}}) = \hyp\left(\overline{\mathds{S}}\right) = \hyp(\widetilde{\mathds{S}}) = \tfrac{1}{2}\,, \\ 
\hyp\left(\overline{\mathds{Q}}\right) &=  \hyp(\widetilde{\mathds{Q}}) = \hyp\left(\mathds{S}\right) = \hyp(\overline{\widetilde{\mathds{S}}}) = - \tfrac{1}{2} \,.
\end{aligned}
\end{gathered}
\end{equation}
The action of the $R$-dilatation $\mathfrak{D}$ on some generator $\mathds{G}$ is given by:
\begin{equation}
\left[\mathfrak{D}, \mathds{G} \right] =\ferm\left(\mathds{G}\right) \mathds{G}\,.
\end{equation} 
The non-zero fermionic dimensions of the superconformal generators are:
\begin{equation}
\begin{aligned}
\ferm\left(\mathfrak{P}\right) &= 1\,, \qquad& \ferm\left(\mathds{Q}\right) &= \ferm (\widetilde{\mathds{Q}}) =  \ferm\left(\overline{\mathds{S}}\right) = \ferm (\overline{\widetilde{\mathds{S}}}) = \tfrac{1}{2}\\
\ferm\left(\mathfrak{K}\right) &= -1 \qquad& \ferm\left(\overline{\mathds{Q}}\right) &= \ferm (\overline{\widetilde{\mathds{Q}}}) =  \ferm\left(\mathds{S}\right) = \ferm (\widetilde{\mathds{S}}) = -\tfrac{1}{2}\,.
\end{aligned}
\end{equation}
The remaining non-trivial commutation relations are
\begin{align}\label{eq:NCsuperconformal}
\{\mathds{Q}^{n}_\alpha , \overline{\mathds{Q}}_{\dot{\alpha}m}\} &= \delta^{n}_{m} \mathds{P}_{\alpha \dot{\alpha}} \hspace{3cm}& \{\widetilde{\mathds{Q}}_{\dot{\alpha}}^{n'},\overline{\widetilde{\mathds{Q}}}_{\alpha m'}\} &= \delta^{n'}_{m'} \mathds{P}_{\alpha \dot{\alpha}}\\
[\mathds{K}_{\alpha \dot{\alpha}}, \mathds{Q}^{\beta n}] &= \delta^{\beta}_{\alpha} \overline{\mathds{S}}_{\dot{\alpha}}^{n} 
& [\mathds{K}_{\alpha \dot{\alpha}}, \widetilde{\mathds{Q}}^{\dot{\beta} n'}] &= \delta^{\dot{\beta}}_{\dot{\alpha}} \overline{\widetilde{\mathds{S}}}_{\alpha}^{n'}\\
[\mathds{K}_{\alpha \dot{\alpha}},\overline{\mathds{Q}}^{\dot{\beta}}_{n}] &= \delta^{\dot{\beta}}_{\dot{\alpha}} \mathds{S}_{\alpha n} 
& [\mathds{K}_{\alpha \dot{\alpha}},\overline{\widetilde{\mathds{Q}}}^{\beta}_{n'}] &= \delta^{\beta}_{\alpha} \widetilde{\mathds{S}}_{\dot{\alpha} n'}\\
[\mathds{S}_{\alpha n}, \mathds{P}^{\beta \dot{\beta}}] &= \delta^{\beta}_{\alpha} \overline{\mathds{Q}}^{\dot{\beta}}_{n} & 
[\widetilde{\mathds{S}}_{\dot{\alpha} n'}, \mathds{P}^{\beta \dot{\beta}}] &= \delta^{\dot{\beta}}_{\dot{\alpha}} \overline{\widetilde{\mathds{Q}}}^{\beta}_{n'}\\
[\overline{\mathds{S}}_{\dot{\alpha}}^{n}, \mathds{P}^{\beta \dot{\beta}}] &= \delta^{\dot{\beta}}_{\dot{\alpha}} \mathds{Q}^{\beta n} & [\overline{\widetilde{\mathds{S}}}_{\alpha}^{n'}, \mathds{P}^{\beta \dot{\beta}}] &= \delta^{\beta}_{\alpha} \widetilde{\mathds{Q}}^{\dot{\beta} n'}\\
\{\mathds{S}_{\alpha n}, \overline{\mathds{S}}_{\dot{\alpha}}^{m}\} &= \delta^{m}_{n} \mathds{K}_{\alpha \dot{\alpha}} &  
\{\widetilde{\mathds{S}}_{\dot{\alpha} n'}, \overline{\widetilde{\mathds{S}}}_{\alpha}^{m'}\} &= \delta^{m'}_{n'} \mathds{K}_{\alpha \dot{\alpha}} \\
\{\mathds{S}_{\alpha n}, \overline{\widetilde{\mathds{Q}}}^{\beta}_{n'}\} &= \delta^{\beta}_{\alpha}  \mathfrak{K}_{n n'} & \{\widetilde{\mathds{S}}_{\dot{\alpha} n'}, \overline{\mathds{Q}}^{\dot{\beta}}_{n}\} &= - \delta^{\dot{\beta}}_{\dot{\alpha}}  \mathfrak{K}_{n n'}\\
\{\overline{\mathds{S}}_{\dot{\alpha}}^{n}, \widetilde{\mathds{Q}}^{\dot{\beta} n'}\} &= \delta^{\dot{\beta}}_{\dot{\alpha}}  \mathfrak{P}^{n n'} & \{\overline{\widetilde{\mathds{S}}}_{\alpha}^{n'}, \mathds{Q}^{\beta n}\} &= -  \delta^{\beta}_{\alpha} \mathfrak{P}^{n n'}\\
\left[\right.\overline{\mathds{Q}}^{\dot{\beta}}_{m}, \mathfrak{P}^{n n'}\left. \right] &= \delta^{n}_{m} \widetilde{\mathds{Q}}^{\dot{\beta} n'} & 
[\overline{\widetilde{\mathds{Q}}}^{\beta}_{m'}, \mathfrak{P}^{n n'}] &= - \delta^{n'}_{m'} \mathds{Q}^{\beta n} \\
[\mathds{S}_{\alpha m}, \mathfrak{P}^{n n'}] &= \delta^{n}_{m}\overline{\widetilde{\mathds{S}}}_{\alpha}^{n'} & [\widetilde{\mathds{S}}_{\dot{\alpha} m'}, \mathfrak{P}^{n n'}] &= - \delta^{n'}_{m'} \overline{\mathds{S}}_{\dot{\alpha}}^{n}\\
[\mathfrak{K}_{n n'}, \mathds{Q}^{\alpha m}] &= - \delta^{m}_{n} \overline{\widetilde{\mathds{Q}}}^{\alpha}_{n'} & 
[\mathfrak{K}_{n n'}, \widetilde{\mathds{Q}}^{\dot{\alpha} m'}] &= \delta^{m'}_{n'} \overline{\mathds{Q}}^{\dot{\alpha}}_{n} \\
[\mathfrak{K}_{n n'}, \overline{\mathds{S}}_{\dot{\alpha}}^{m}] &= - \delta^{m}_{n} \widetilde{\mathds{S}}_{\dot{\alpha} n'} & 
[\mathfrak{K}_{n n'}, \overline{\widetilde{\mathds{S}}}_{\alpha}^{n'}] &= \delta^{m'}_{n'} \mathds{S}_{\alpha n}
\end{align}
as well as
\begin{align}
 \{\mathds{S}_{\alpha n}, \mathds{Q}^{\beta m}\} &= \delta^{m}_{n} \mathds{M}^{\beta}_{\;\;\alpha}  - \delta^{\beta}_{\alpha} \mathfrak{M}^{m}_{\;\;n} + \tfrac{1}{2} \delta^{m}_{n} \delta^{\beta}_{\alpha} (\mathds{D} - \mathds{C} - \mathfrak{D})&\\
\{\widetilde{\mathds{S}}_{\dot{\alpha} n'}, \widetilde{\mathds{Q}}^{\dot{\beta} m'}\} &= \delta^{m'}_{n'} \overline{\mathds{M}}^{\dot{\beta}}_{\;\;\dot{\alpha}}  - \delta^{\dot{\beta}}_{\dot{\alpha}} \widetilde{\mathfrak{M}}^{m'}_{\;\;n'} + \tfrac{1}{2} \delta^{m'}_{n'} \delta^{\dot{\beta}}_{\dot{\alpha}} (\mathds{D} + \mathds{C} - \mathfrak{D})&\\
\{\overline{\mathds{S}}_{\dot{\alpha}}^{n}, \overline{\mathds{Q}}^{\dot{\beta}}_{m}\} &= \delta^{n}_{m} \overline{\mathds{M}}^{\dot{\beta}}_{\;\;\dot{\alpha}}  + \delta^{\dot{\beta}}_{\dot{\alpha}} \mathfrak{M}^{n}_{\;\;m} + \tfrac{1}{2} \delta^{n}_{m} \delta^{\dot{\beta}}_{\dot{\alpha}} (\mathds{D} + \mathds{C} + \mathfrak{D})&\\
\{\overline{\widetilde{\mathds{S}}}_{\alpha}^{n'}, \overline{\widetilde{\mathds{Q}}}^{\beta}_{m'}\} &= \delta^{n'}_{m'} \mathds{M}^{\beta}_{\;\;\alpha}  + \delta^{\beta}_{\alpha} \widetilde{\mathfrak{M}}^{n'}_{\;\;m'} + \tfrac{1}{2} \delta^{n'}_{m'} \delta^{\dot{\beta}}_{\dot{\alpha}} (\mathds{D} - \mathds{C} + \mathfrak{D})&\\
[\mathfrak{K}_{m m'}, \mathfrak{P}^{n n'}] &= \delta^{n'}_{m'} \delta^{n}_{m} \mathfrak{D} + \delta^{n}_{m} \widetilde{\mathfrak{M}}^{m'}_{\;\;n'} +  \delta^{n'}_{m'} \mathfrak{M}^{m}_{\;\;n}&\\
[ \mathds{K}_{\alpha \dot{\alpha}} ,  \mathds{P}^{\beta \dot{\beta}} ] &= \delta^{\beta}_{\alpha} \delta^{\dot{\beta}}_{\dot{\alpha}} \mathds{D} +  \delta^{\dot{\beta}}_{\dot{\alpha}} \mathds{M}_{\;\alpha}^{\beta}  +  \delta^{\beta}_{\alpha} \overline{\mathds{M}}_{\;\dot{\alpha}}^{\dot{\beta}}&
\end{align}
\subsection{The On-Shell Representation}\label{sec:on_shell_non_chiral}
We denote the generators of the on-shell representation of the non-chiral superconformal algebra by small letters $a,b,c$ and $\mathpzc{a}, \mathpzc{b}, \mathpzc{c}$. We introduce the following abbreviations
\begin{align}
\partial_{i \a} &= \frac{\partial}{\partial \lambda^{\a}_i}\,,&\partial_{i \da} &= \frac{\partial}{\partial \tilde\lambda^{\da}_i}\,,&\partial_{i n} &= \frac{\partial}{\partial \eta^{n}_i}\,,&\partial_{i n'} &= \frac{\partial}{\partial \tilde{\eta}_{i}^{n'}}
\end{align}
for derivatives with respect to the on-shell variables. The on-shell generators are
\begin{align}
p^{\dot{\alpha} \alpha} &= \sum_{i} \lambda_{i}^{\alpha} \tilde{\lambda}_{i}^{\dot{\alpha}} & k_{\dot{\alpha} \alpha} &= \sum_{i} \partial_{i\alpha}  \partial_{i \dot{\alpha}}\\
m_{\alpha \beta} &= \sum_{i}  \lambda_{i (\alpha}  \partial_{i \beta)} & \overline{m}_{\dot{\alpha} \dot{\beta}} &= \sum_{i}  \tilde{\lambda}_{i (\dot{\alpha}}  \partial_{i \dot{\beta})} \\
q^{\alpha n} &= \sum_{i}  \lambda^{\alpha}_{i} \eta^{n}_{i} &  \tilde{q}^{\dot\alpha n'} &= \sum_{i}  \tilde{\lambda}^{\dot\alpha}_{i} \tilde{\eta}_{i}^{n'} \\
\bar{q}^{\dot{\alpha}}_{n} &= \sum_{i}  \tilde{\lambda}^{\dot{\alpha}}_{i} \partial_{i n}
& \bar{\tilde{q}}^{\alpha}_{ n'} &= \sum_{i}  \lambda^{\alpha}_{i} \partial_{i n'}\\
s_{\alpha n} &= \sum_{i} \partial_{i \alpha} \partial_{i n} &\tilde{s}_{\dot{\alpha} n'} &= \sum_{i}  \partial_{i \dot{\alpha}}\partial_{i n'}  \\
\bar{s}^{n}_{\dot{\alpha}} &= \sum_{i}  \eta^{n}_{i} \partial_{i \dot{\alpha}} &\bar{\tilde{s}}_{\alpha}^{n'} &= \sum_{i} \tilde{\eta}_{i}^{n'} \partial_{i \alpha}  \\
d  &= \tfrac{1}{2} \sum_{i} \left( \lambda^{\alpha}_{i} \partial_{i \alpha} + \tilde{\lambda}^{\dot{\alpha}}_{i} \partial_{i \dot{\alpha}} + 2 \right)& b &=  \tfrac{1}{2} \sum_{i}\left(\eta^{n}_{i} \partial_{i n} - \tilde{\eta}^{n'}_{i} \partial_{i n'}\right)\\
c &=  \tfrac{1}{2} \sum_{i}\left(-\lambda^{\alpha}_{i} \partial_{i \alpha} + \tilde{\lambda}^{\dot{\alpha}}_{i} \partial_{i \dot{\alpha}} + \eta^{n}_{i} \partial_{i n} - \tilde{\eta}^{n'}_{i} \partial_{i n'}\right)\hspace{-1cm}&\\
\mathpzc{p}^{n n'} &= \sum_{i} \eta_{i}^{n} \tilde{\eta}_{i}^{n'} &\mathpzc{k}_{\,\;n n'} &= \sum_{i} \partial_{i n}  \partial_{i n'}\\
\mathpzc{m}_{\,n m} &= \sum_{i} \eta_{i(n} \partial_{i m)} & \widetilde{\mathpzc{m}}_{\,n' m'} &= \sum_{i} \tilde{\eta}_{i (n'} \partial_{i m')}\\
\mathpzc{d} &= \tfrac{1}{2} \sum_{i} \left(\eta^{n}_{i} \partial_{i n} + \tilde{\eta}^{n'}_{i} \partial_{i n'} - 2\right)&
\end{align}
\subsection{The Dual Representation} \label{sec:dual_non_chiral}
We denote the generators of the dual representation of the non-chiral superconformal algebra by capital letters $A,B,C$ and $\mathcal{A}, \mathpzc{B}, \mathpzc{C}$. We present the dual representation in dual non-chiral superspace $(x,y,\theta,\tilde{\theta})$ using the following abbreviations  
\begin{align}
\partial_{i \alpha \dot{\alpha}} &= \frac{\partial}{\partial x_{i}^{\dot{\alpha}\alpha }}=\tfrac{1}{2}\sigma^{\mu}_{\alpha\dot\alpha}\frac{\partial}{\partial x_i^\mu}\,,&
\partial_{i n n'} &= \frac{\partial}{\partial y_{i}^{n n'}}\,, &\partial_{i \alpha n} &= \frac{\partial}{\partial \theta_{i}^{\alpha n}}\,,& \partial_{i\dot{\alpha} n'} &= \frac{\partial}{\partial \tilde{\theta}_{i}^{\dot{\alpha} n'}} 
\end{align} 
for derivatives with respect to the dual variables.
In the dual superspace $\{x_i^{\dot{\alpha}\alpha},\theta_i^{m\,\alpha},\tilde\theta_i^{m'\,\dot\alpha}\}$ the generators of the dual non-chiral superconformal symmetry are given by
\begin{equation}\label{eq:dualConformalNC}
\begin{gathered}
 \begin{aligned} 
P_{\alpha \dot{\alpha}}& = \sum_{i} \partial_{i \alpha \dot{\alpha}} \qquad &\mathpzc{P}_{n n'} &= -\sum_{i} \partial_{i n n'} \\
Q_{\alpha n} &= -\sum_{i} \partial_{i \alpha n} \qquad& \widetilde{Q}_{\dot{\alpha} n'} &= -\sum_{i} \partial_{i \dot{\alpha} n'} \\
\overline{Q}^{n}_{\dot{\alpha}} &= \sum_{i} (\theta^{\alpha n}_{i} \partial_{i \alpha \dot{\alpha}} + y_{i}^{n n'} \partial_{i\dot{\alpha} n' }) \qquad& \overline{\widetilde{Q}}^{n'}_{\alpha} &= \sum_{i} (\tilde{\theta}^{\dot{\alpha} n'}_{i} \partial_{i \alpha \dot{\alpha}}-y_{i}^{n n'} \partial_{i \alpha n }  ) \\
M_{\alpha \beta} &= \sum_{i} \left( \theta^{n}_{i (\alpha} \partial_{i \beta) n} + x_{i (\alpha}^{\dot{\alpha}} \partial_{i \beta) \dot{\alpha}} \right)&
\overline{M}_{\dot{\alpha} \dot{\beta}} &= \sum_{i} \left( \tilde{\theta}^{n'}_{i (\dot{\alpha}} \partial_{i \dot{\beta}) n'} + x_{i (\dot{\alpha}}^{\alpha} \partial_{i \dot{\beta}) \alpha} \right)\\
\mathpzc{M}_{\,n m} &= \sum_{i} \left( \theta_{i \alpha(n} \partial_{i m)}^{\alpha} + y_{i (n}^{\;\;\;\;n'} \partial_{i m) n'} \right)&
\widetilde{\mathpzc{M}}_{\,n' m'} &= \sum_{i} \left(\tilde{\theta}_{i \dot{\alpha}(n'} \partial_{i  m')}^{\dot{\alpha}} + y_{i n (n'} \partial_{i  m')}^{n} \right)\\
\overline{S}^{\dot{\alpha}}_{n} &= -\sum_{i} (\tilde{\theta}_{i}^{\dot{\alpha} n'} \partial_{i n n'} + x_{i}^{\alpha \dot{\alpha}} \partial_{i \alpha n}) \qquad& 
\overline{\widetilde{S}}^{\alpha}_{n'} &= \sum_{i} ( \theta_{i}^{\alpha n} \partial_{i n n'} -x_{i}^{\alpha \dot{\alpha}} \partial_{i  \dot{\alpha} n'} )\end{aligned}\\
\begin{aligned}
S^{\alpha n} &= \sum_{i}  \left(- \theta_{i}^{\alpha m} \theta_{i}^{\beta n} \partial_{i \beta m} + x_{i}^{\alpha \dot{\beta}} \theta_{i}^{\beta n} \partial_{i \beta \dot{\beta}} - \theta_{i }^{\alpha m} y_{i}^{n m'} \partial_{im m'} + y_{i}^{n m'} x_{i}^{\alpha \dot{\alpha}} \partial_{i\dot{\alpha} m' }\right)\\
\widetilde{S}^{\dot{\alpha} n'} &= \sum_{i}  \left(-\tilde{\theta}_{i}^{\dot{\alpha} m'} \tilde{\theta}_{i}^{\dot{\beta} n'} \partial_{i \dot{\beta} m'} + x_{i}^{\dot{\alpha} \beta} \tilde{\theta}_{i}^{\dot{\beta} n'} \partial_{i \beta \dot{\beta}}  -\tilde{\theta}_{i}^{ \dot{\alpha} m'} y_{i}^{m n'} \partial_{i m m'} - y_{i}^{m n'} x_{i}^{\dot{\alpha} \alpha} \partial_{i \alpha m}\right)\\
K_{\alpha \dot{\alpha}} &= \sum_{i}  \left( x_{i \alpha}^{\;\;\; \dot{\beta}} x_{i \dot{\alpha}}^{\;\;\; \beta} \partial_{i \beta \dot{\beta}} + x_{i \dot{\alpha}}^{\;\;\;\beta} \theta_{i \alpha}^{n} \partial_{i n \beta} + x_{i \alpha}^{\;\;\; \dot{\beta}} \tilde{\theta}_{i \dot{\alpha}}^{n'} \partial_{i n' \dot{\beta}} 
  + \theta_{i \alpha}^{n} \tilde{\theta}_{i\dot{\alpha}}^{n'} \partial_{i n n'}\right)\\
\mathpzc{K}^{n n'} &= \sum_{i} \left(- y_{i}^{n m'} y_{i}^{m n'} \partial_{i m m'} - \tilde{\theta}^{\dot{\alpha} n'}_{i} y_{i}^{n m'} \partial_{i \dot{\alpha} m'} - \theta^{\alpha n}_{i} y_{i}^{m n'} \partial_{i \alpha m}  +\theta_{i}^{n \alpha} \tilde{\theta}^{n' \dot{\alpha}} \partial_{\alpha \dot{\alpha}} \right)\\
D &=- \tfrac{1}{2} \sum_{i} \left(\theta_{i}^{n \alpha} \partial_{i  \alpha n} + \tilde{\theta}_{i}^{\dot{\alpha} n'} \partial_{i \dot{\alpha} n'} + 2 x^{\alpha \dot{\alpha}}_{i} \partial_{i \alpha \dot{\alpha}} \right)\\
\mathpzc{D} &= -\tfrac{1}{2} \sum_{i} \left( \theta_{i}^{n \alpha} \partial_{i  \alpha n} + \tilde{\theta}_{i}^{\dot{\alpha} n'} \partial_{i \dot{\alpha} n'} + 2 y^{n n'}_{i} \partial_{i n n'}\right)\\
B &= \tfrac{1}{2} \sum_{i} \left( \tilde{\theta}_{i}^{\dot{\alpha} n'} \partial_{i \dot{\alpha} n'}-\theta_{i}^{n \alpha} \partial_{i  \alpha n} \right)
\end{aligned} 
\end{gathered}
\end{equation} 
We note that there are seven other possibilities to choose the signs of the generators such that they fulfill the non-chiral superconformal algebra listed at the beginning of \cref{sec:Algebra_Non_Chiral}.  It is straightforward to obtain the generators in full non-chiral superspace $\{\lambda_i^\alpha,\tilde\lambda_i^{\dot\alpha},x_i^{\dot{\alpha}\alpha},\eta_{i}^m,\tilde\eta_{i}^{m'}\theta_i^{m\,\alpha},\tilde\theta_i^{m'\,\dot\alpha}\}$ by extending the action of the generators in dual non-chiral superspace such that they commute with the constraints \cref{eq:constraints_full_nonchiral}. Alternatively one could derive the action of the conformal and superconformal generators $K_{\a\da}$, $S^{\alpha n}$, $\widetilde{S}^{\dot{\alpha} n'}$,  $\overline{S}^{\dot{\alpha}}_{n}$ and $\overline{\widetilde{S}}^{\alpha}_{n'}$ in full superspace from their definition \eqref{eq:superconformalGenerators_NC} and the inversion rules \eqref{eq:inversion4dNC} of the onshell variables. The action of all remaining generators on the 
onshell variables can then be obtained from the non-chiral superconformal algebra. 
\section{Connection between 4d and 6d Lorentz invariants containing supermomenta} \label{sec:Connection_six_four_Invariants}
Similar to the 6d Lorentz invariants \eqref{eq:type1}, we try to find a combination of the invarnants \eqref{eq:type2} and \eqref{eq:type3} whose four dimensional projection is manifestly $R$-symmetry invariant. Because of the non-chiral nature of the six-dimensional amplitudes, the number of chiral and anti-chiral supermomenta are equal and the invariants \eqref{eq:type2} and \eqref{eq:type3} can only appear in the pairs
\begin{equation}
\begin{aligned} \label{eq:BadCorr6D4D}
 6d: \qquad \langle q_{i}|k_1 \dots k_{2 r + 1}|q_{j}\rangle [\tilde{q}_{k}| p_1  \dots p_{2 s + 1}|\tilde{q}_{l}]
\end{aligned}
\end{equation}
leading to the following four-dimensional projection
\begin{equation}\label{eq:4dProjectionBadCorr6D4D}
\begin{aligned}
 \Bigl( \langle q^{1}_{i}|k_1  \dots k_{2 r + 1}|\tilde{q}_{j 3}] - [\tilde{q}_{i 3}|k_1  \dots k_{2 r + 1}|q_{j}^{1}\rangle \Bigr) \Bigl( \langle q^{4}_{k}| p_1 \dots p_{2 s + 1}|\tilde{q}_{l 2}] - [\tilde{q}_{k 2}| p_1  \dots p_{2 s + 1}|q_{l}^{4}\rangle \Bigr)
\end{aligned}
\end{equation}
Unfortunately this combination is in general not $R$-symmetry invariant. There are two possibilities to make \cref{eq:4dProjectionBadCorr6D4D} $R$ invariant. First, we could impose restrictions on the supermomenta appearing in \cref{eq:BadCorr6D4D}. However, only the very special case $i=j=k=l$ of all supermomenta belonging to the same external leg, has a $R$ invariant projection. Since this strong restriction is completely unnatural from both the 6d and 4d perspective and does not allow for a construction of manifestly dual conformal covariant Lorentz invariants, it should be neglected. The second possibility is to antisymmetrize \cref{eq:4dProjectionBadCorr6D4D} in the indices $(1,4)$ and $(2,3)$, corresponding to $m$ and $m'$ $SU(2)$ contractions:
\begin{align}
 \left( \langle q^{[1}_{i}|k_1 ... k_{2 r + 1}|\tilde{q}_{j [3}] - [\tilde{q}_{i [3}|k_1 ... k_{2 r + 1}|q_{j}^{[1}\rangle  \right) \left( \langle q^{4]}_{k}| p_1 ... p_{2 s + 1}|\tilde{q}_{l 2]}] - [\tilde{q}_{k 2]}| p_1  ... p_{2 s + 1}|q_{l}^{4]}\rangle \right) \\
 = \Bigl( \langle q^{m}_{i}|k_1  ... k_{2 r + 1}|\tilde{q}_{j\,m'}] - [\tilde{q}_{i\,m'}|k_1  ... k_{2 r + 1}|q_{j}^{m}\rangle \Bigr) \left( \langle q_{k\,m}| p_1 ... p_{2 s + 1}|\tilde{q}_{l}^{m'}] - [\tilde{q}_{k}^{m'} | p_1  ... p_{2 s + 1}|q_{l\,m}\rangle \right)
\end{align}
This combination would require, among others, the following projection to arise from six dimensions
\begin{align} \label{eq:BadProjection}
 ... -  \langle q^{1}_{i}|k_1 ... k_{2 r + 1}|\tilde{q}_{j 2}] 
\langle q^{4}_{k}| p_1 \dots p_{2 s + 1}|\tilde{q}_{l 3}]  + ... \end{align}
However, from \cref{eq:type2,eq:type3} it follows that such a term cannot appear from a six dimensional projection. Even for the chiral self-conjugate case, with momenta being the same ($k = p$, $n = m$) and supermomenta being conjugate of each other ($i = k$, $j = l$), contributions of the form \eqref{eq:BadProjection} do not cancel. Consequently the blocks in \cref{eq:BadCorr6D4D} are irrelevant for a connection between the superamplitudes in six and four dimensions, since the latter are manifestly $R$-symmetry invariant. Therefore only the invariants of the type \cref{eq:correspondence6d4d} are natural objects for establishing such a bridge.

\bibliographystyle{JHEP}
\bibliography{bibliothek}
\end{document}